\documentclass[showpacs,preprintnumbers,amsmath,amssymb]{revtex4}
\usepackage{graphicx}
\usepackage{dcolumn}
\usepackage{bm}

\begin{document}

\title{HIGGS-INFLATON SYMBIOSIS, COSMOLOGICAL CONSTANT PROBLEM AND
  SUPERACCELERATION PHASE OF THE UNIVERSE
  IN TWO MEASURES FIELD THEORY
   WITH SPONTANEOUSLY BROKEN SCALE INVARIANCE}

\author
{E. I. Guendelman \thanks{guendel@bgu.ac.il} and A.  B. Kaganovich
\thanks{alexk@bgu.ac.il}}
\address{Physics Department, Ben Gurion University of the Negev, Beer
Sheva 84105, Israel}

\date{\today}

\begin{abstract}
We study the scalar sector of the Two Measures Field Theory (TMT)
model in the context of cosmological dynamics. The scalar sector
includes the dilaton $\phi$ and the Higgs $\upsilon$ fields. The
model possesses gauge and scale invariance. The latter is
spontaneously broken due to intrinsic features of the TMT
dynamics. The scalar sector dynamics represents an explicit
example of $k$-essence resulting from  first principles where
$\phi$ plays the role of the inflaton . In the model with the
inflaton $\phi$ alone, in different regions of the parameter space
the following different effects can take place {\it without fine
tuning of the parameters and initial conditions}: a) Possibility
of a power law inflation driven by the scalar field $\phi$ which
is followed by the late time evolution driven both by a small
cosmological constant and the scalar field $\phi$ with a
quintessence-like potential; smallness of the cosmological
constant can be achieved without fine tuning of dimensionfull
parameters. b) Possibility of {\it resolution of the old
cosmological constant problem}: this is done in a consistent way
hinted  by S. Weinberg in his comment concerning the question of
{\it how one can avoid his no-go theorem}. c) The power law
inflation without any fine tuning may end with damped oscillations
of $\phi$ around the state with zero cosmological constant. d)
There are regions of the parameters where the equation-of-state
$w=p/\rho$ in  the late time universe is $w<-1$ and $w$ {\it
asymptotically} (as $t\rightarrow\infty$) {\it approaches} $-1$
{\it from below}. This effect is achieved without any exotic term
in the action. In a model with both $\phi$ and $\upsilon$ fields,
a scenario which resembles the hybrid inflation is realized but
there are essential differences, for example: the Higgs field
undergos transition to a gauge symmetry broken phase
$<\upsilon>\neq 0$ soon {\it after} the end of a power law
inflation; there are
 two oscillatory regimes of $\upsilon$, one around $\upsilon =0$
 at 50 e-folding before the end of inflation, another - during
 transition to a gauge symmetry broken phase where the scalar dark
 energy density approaches zero without fine tuning; the gauge
 symmetry breakdown is achieved {\it without tachyonic mass term} in the
 action.

   \end{abstract}

\maketitle

\section{Introduction}

The cosmological constant problem \cite{Weinberg1}-\cite{CC}, the
accelerated expansion of the late time universe\cite{accel}, the
cosmic coincidence \cite{coinc} are challenges for the foundations
of modern physics (see also reviews on dark
energy\cite{de-review},\cite{Starobinsky}, dark matter
\cite{dm-review} and references therein). Numerous models have
been proposed with the aim to find answer to these puzzles, for
example: the quintessence\cite{quint},  coupled
quintessence\cite{Amendola}, $k$-essence\cite{k-essence}, variable
mass particles\cite{vamp}, interacting quintessence\cite{int-q},
Chaplygin gas\cite{Chapl}, phantom field\cite{phantom}, tachyon
matter cosmology\cite{tachyon}, brane world scenarios\cite{brane},
  etc.. These
puzzles have also motivated an interest in modifications and even
radical revisions of the standard gravitational theory (General
Relativity (GR))\cite{modified-gravity1},\cite{modified-gravity2}.
Although motivations for most of these models can be found in
fundamental theories like for example in brane
world\cite{extra-dim},  the questions concerning the Einstein GR
limit and relation to the regular particle physics, like standard
model, still remain unclear. The listed and a number of other
cosmological and astrophysical problems generate fundamental
questions which are directly addressed to particle physics, like:
what is the dark matter from the point of view  of particle
physics; what is the relation between dark energy and the vacuum
of particle physics after gauge symmetry breaking; how the latter
question is related to the cosmological constant problem; what is
the precise physics of the inflation of the early universe, etc..

On the other side, even without the cosmological input, particle
physics has its own  fundamental questions, like for example: what
is the origin of fermionic generations; whether  is it possible to
realize spontaneous gauge symmetry breaking without using
tachyonic mass term.

It is very hard to imagine that  it is possible  to propose ideas
 which are able to solve all the above mentioned problems  keeping
 at the same time unchanged the basis of fundamental physics, i.e.
 gravity and particle field theory. This paper may be regarded as
 {\it an attempt to find satisfactory answers at least to a part of the
 fundamental questions on the basis of first principles}, i.e. without
using semi-empirical models. In this paper we explore the so
called Two Measures Field Theory (TMT) where gravity (or more
exactly, geometry) and particle field theory intertwined in a very
non trivial manner, but the Einstein GR is restored when the
fermion matter energy density is much larger than the vacuum
energy density.

Here we {\it have no  purpose} of constructing a complete
realistic cosmological scenario. Instead, we concentrate our
attention on studying a number of {\it new surprising enough
effects} of TMT. We study the model which possesses gauge and
scale symmetry (the latter includes the shift symmetry of the
dilaton field $\phi$). The scale symmetry is spontaneously broken
due to intrinsic features of the TMT dynamics. Except for peculiar
properties of the TMT action, the latter does not contain any
exotic terms, i.e. {\it only terms presented in standard model
considered in curved space-time are present in our action}. The
scalar sector dynamics represents an explicit example of
$k$-essence resulting from first principles.

In context of cosmological dynamics, we study in this paper only
the scalar sector of the model which includes the dilaton $\phi$
and the Higgs $\upsilon$ fields. The Higgs contribution to the
action has no terms which could be treated as tachyonic mass
terms. The parameter space of the model is large enough that
allows to find different regions where the following different
effects can take place {\it without fine tuning of the parameters
and initial conditions}:

a) Possibility of a power law inflation driven by a scalar field
$\phi$ which is  followed by the late time evolution  driven both
by a small cosmological constant and a scalar field $\phi$ with a
quintessence-like potential; smallness of the cosmological
constant can be achieved {\it without fine tuning of dimensionfull
parameters}.

b) In another region of the parameters, there is a possibility of
the power law inflation  ended with damped oscillations of $\phi$
around the state with zero cosmological constant (we want to
emphasize again that this is realized without tuning of the
parameters and initial conditions). Thus this scenario includes
{\it a resolution of the old cosmological constant problem}, at
least at the classical level. This becomes possible because one of
the basic assumptions of the Weinberg no-go
theorem\cite{Weinberg1} is violated in our model.

d) There is a region of the parameters where solutions for the
late time universe possess the following unexpected features: the
dark energy density $\rho$ increases asymptotically (as
$t\rightarrow\infty$) approaching from below to a constant value;
the equation-of-state $w=p/\rho$ in the late time universe is
$w<-1$ and $w$ {\it asymptotically approaches} $-1$ {\it from
below}.
 This effect
is achieved without any exotic term in the action similar to what
is present in phantom models\cite{phantom}.

e) In the model with both dilaton $\phi$ and Higgs $\upsilon$
fields, the power law inflation consists of two stages: during the
first stage $\upsilon$ remains close to its initial value; about
60 e-folding before the end of inflation $\upsilon$ suffers a
transition to the state of damping oscillations around $\upsilon
=0$. After the end of the power law inflation a transition of the
Higgs field to a broken gauge symmetry  phase $<\upsilon>\neq 0$
occurs via oscillations of $\upsilon$ and $\phi$. It is
interesting that this early universe scenario is realized {\it
without tachyonic mass terms}. Another remarkable feature is that
{\it the energy density in the symmetry broken phase is zero
without tuning of the parameters and  initial conditions}.

Since we have not yet studied the particle creation that must be
due to oscillations of the Higgs field, we are not able to present
here a complete cosmological scenario including the described
above effects in a consistent way. Nevertheless it is evident that
the listed effects can be in principle be useful for a unified
resolution of some of the
 cosmological and particle physics fundamental problems.

The organization of the paper is the following: In Sec.II we
present a  review of the basic ideas of TMT and formulate a gauge
and scale invariant model. The remaining part of the main text of
the paper is devoted to study of the scalar sector dynamics of the
model including dilaton $\phi$ and Higgs $\upsilon$ fields.
However the way   the $\upsilon$  field  is introduced in our TMT
model requires a special consideration of its properties in order
to demonstrate that it is indeed able to perform all what one
should demand from the Higgs field, for example in the standard
model. This was the reason of the need to include also gauge
fields and fermions into the description of the model although
they are not the subject of this paper. This is done in Sec.II and
Appendixes A and B. In Appendix A we present equations of motion
in the original frame. Appendix B is devoted to the
spin-connections. Using results of Appendixes A and B, the
complete set of equations of motion in the Einstein frame is given
in Sec.IIC. In Sec.III we give a detailed formulation of the
scalar sector cosmological dynamics in the spatially flat FRW
universe. For pedagogical reasons there is a need to present the
results first without the Higgs field: in Secs.IV-VI for three
different shapes of the effective potential we study the attractor
behavior of the phase trajectories, the cosmological constant
problem and possibility of super-acceleration. Main features of
the cosmological dynamics driven both by the dilaton field and the
Higgs field are studied in Sec.VII. The Discussion section gives a
systematic analysis of the main ideas and results. In Appendix C
we shortly review our previous results of TMT\cite{GK6} concerning
the restoration of GR and resolution of the fifth force problem in
normal particle physics conditions, i.e. when the local fermionic
energy density is many orders of magnitude larger than the scalar
dark energy density in the space-time region occupied by the
fermion. Although this is not a subject of the present paper we
believe that it is important for reader to know that in normal
particle physics conditions TMT reproduces GR. In Appendix D we
shortly discuss what kind of model one would obtain when choosing
fine tuned couplings to measures $\Phi$ and$\sqrt{-g}$ in the
action. Some additional remarks concerning the relation between
the structure of TMT action and the cosmological constant problem
are given in Appendix E.

\section{Basis of Two Measures Field Theory and   Scale Invariant Model}
\subsection{Main ideas of the Two Measures Field Theory}

TMT is a generally coordinate invariant theory where {\it all the
difference from the standard field theory in curved space-time
consists only of the following three additional assumptions}:

1. The first assumption is the hypothesis that the effective
action at the energies below the Planck scale has to be of the
form\cite{GK1}-\cite{GK6}
\begin{equation}
    S = \int L_{1}\Phi d^{4}x +\int L_{2}\sqrt{-g}d^{4}x
\label{S}
\end{equation}
 including two Lagrangians $ L_{1}$ and $L_{2}$ and two
measures of integration $\sqrt{-g}$ and $\Phi$ or, equivalently,
two volume elements $\Phi d^{4}x$ and $\sqrt{-g}d^{4}x$
respectively. One is the usual measure of integration $\sqrt{-g}$
in the 4-dimensional space-time manifold equipped with the metric
 $g_{\mu\nu}$.
Another  is the new measure of integration $\Phi$ in the same
4-dimensional space-time manifold. The measure  $\Phi$ being  a
scalar density and a total derivative\footnote{For applications of
the measure $\Phi$ in string and brane theories see
Ref.\cite{Mstring}.} may be defined
\begin{itemize}

\item
either by means of  four scalar fields $\varphi_{a}$
($a=1,2,3,4$), cf. with the approach by Wilczek\footnote{See:
 F. Wilczek, Phys.Rev.Lett. {\bf 80}, 4851
(1998). Wilczek's goal was to avoid the use of a fundamental
metric, and for this purpose he needs five scalar fields. In our
case we keep the standard role of the metric from the beginning,
but enrich the theory with a new metric independent density.},
\begin{equation}
\Phi
=\varepsilon^{\mu\nu\alpha\beta}\varepsilon_{abcd}\partial_{\mu}\varphi_{a}
\partial_{\nu}\varphi_{b}\partial_{\alpha}\varphi_{c}
\partial_{\beta}\varphi_{d}.
\label{Phi}
\end{equation}

\item
or by means of a totally antisymmetric three index field
$A_{\alpha\beta\gamma}$
\begin{equation}
\Phi
=\varepsilon^{\mu\nu\alpha\beta}\partial_{\mu}A_{\nu\alpha\beta}.
\label{Aabg}
\end{equation}
\end{itemize}

To provide parity conservation in the case given by Eq.(\ref{Phi})
one can choose for example one of $\varphi_{a}$'s to be a
pseudoscalar; in the case given by Eq.(\ref{Aabg}) we must choose
$A_{\alpha\beta\gamma}$ to have negative parity. Some ideas
concerning the nature of the measure fields $\varphi_{a}$ are
discussed in Ref.\cite{GK6}. A special case of the structure
(\ref{S}) with definition (\ref{Aabg}) has been recently discussed
in Ref.\cite{hodge} in applications to supersymmetric theory and
the cosmological constant problem.

2. It is assumed that the Lagrangians $ L_{1}$ and $L_{2}$ are
functions of all matter fields, the dilaton field, the metric, the
connection  (or spin-connection )
 but not of the
"measure fields" ($\varphi_{a}$ or $A_{\alpha\beta\gamma}$). In
such a case, i.e. when the measure fields  enter in the theory
only via the measure $\Phi$,
  the action (\ref{S}) possesses
an infinite dimensional symmetry. In the case given by
Eq.(\ref{Phi}) these symmetry transformations have the form
$\varphi_{a}\rightarrow\varphi_{a}+f_{a}(L_{1})$, where
$f_{a}(L_{1})$ are arbitrary functions of  $L_{1}$ (see details in
Ref.\cite{GK3}); in the case given by Eq.(\ref{Aabg}) they read
$A_{\alpha\beta\gamma}\rightarrow
A_{\alpha\beta\gamma}+\varepsilon_{\mu\alpha\beta\gamma}
f^{\mu}(L_{1})$ where $f^{\mu}(L_{1})$ are four arbitrary
functions of $L_{1}$ and $\varepsilon_{\mu\alpha\beta\gamma}$ is
numerically the same as $\varepsilon^{\mu\alpha\beta\gamma}$. One
can hope that this symmetry should prevent emergence of a measure
fields dependence in $ L_{1}$ and $L_{2}$ after quantum effects
are taken into account.

3. Important feature of TMT that is responsible for many
interesting and desirable results of the field theory models
studied so far\cite{GK1}-\cite{GK6}
 consists of the assumption that all fields, including
also metric, connection (or vierbein and spin-connection) and the
{\it measure fields} ($\varphi_{a}$ or $A_{\alpha\beta\gamma}$)
are independent dynamical variables. All the relations between
them are results of equations of motion.  In particular, the
independence of the metric and the connection means that we
proceed in the first order formalism and the relation between
connection and metric is not necessarily according to Riemannian
geometry.

We want to stress again that except for the listed three
assumptions we do not make any changes as compared with principles
of the standard field theory in curved space-time. In other words,
all the freedom in constructing different models in the framework
of TMT consists of the choice of the concrete matter content and
the Lagrangians $ L_{1}$ and $L_{2}$ that is quite similar to the
standard field theory.

Since $\Phi$ is a total derivative, a shift of $L_{1}$ by a
constant, $L_{1}\rightarrow L_{1}+const$, has no effect on the
equations of motion. Similar shift of $L_{2}$ would lead to the
change of the constant part of the Lagrangian coupled to the
volume element $\sqrt{-g}d^{4}x $. In the standard GR, this
constant term is the cosmological constant. However in TMT the
relation between the constant
 term of $L_{2}$ and the physical cosmological constant is very non
trivial (see \cite{GK3}-\cite{K}).

In the case of the definition of $\Phi$ by means of
Eq.(\ref{Phi}), varying the measure fields $\varphi_{a}$, we get
\begin{equation}
B^{\mu}_{a}\partial_{\mu}L_{1}=0  \quad \text{where} \quad
B^{\mu}_{a}=\varepsilon^{\mu\nu\alpha\beta}\varepsilon_{abcd}
\partial_{\nu}\varphi_{b}\partial_{\alpha}\varphi_{c}
\partial_{\beta}\varphi_{d}.\label{varphiB}
\end{equation}
Since $Det (B^{\mu}_{a}) = \frac{4^{-4}}{4!}\Phi^{3}$ it follows
that if $\Phi\neq 0$,
\begin{equation}
 L_{1}=sM^{4} =const
\label{varphi}
\end{equation}
where $s=\pm 1$ and $M$ is a constant of integration with the
dimension of mass. In what follows we make the choice $s=1$.

In the case of the definition (\ref{Aabg}), variation of
$A_{\alpha\beta\gamma}$ yields
\begin{equation}
\varepsilon^{\mu\nu\alpha\beta}\partial_{\mu}L_{1}=0, \label{AdL1}
\end{equation}
that implies Eq.(\ref{varphi}) without the  condition $\Phi\neq 0$
needed in the model with four scalar fields $\varphi_{a}$.

 One should notice
 {\it the very important differences of
TMT from scalar-tensor theories with nonminimal coupling}: \\
 a) In general, the Lagrangian density $L_{1}$ (coupled to the measure
$\Phi$) may contain not only the scalar curvature term (or more
general gravity term) but also all possible matter fields terms.
This means that TMT modifies in general both the gravitational
sector  and the matter sector; b) If the field $\Phi$ were the
fundamental (non composite) one then instead of (\ref{varphi}),
the variation of $\Phi$ would result in the equation $L_{1}=0$ and
therefore the dimensionfull integration constant $M^4$ would not
appear in the theory.

Applying the Palatini formalism in TMT one can show (see for
example \cite{GK3} and Appendix C of the present paper)  that the
resulting relation between metric and  connection includes also
the gradient of the ratio of the two measures
\begin{equation}
\zeta \equiv\frac{\Phi}{\sqrt{-g}} \label{zeta}
\end{equation}
which is a scalar field. The gravity and matter field equations
obtained by means of the first order formalism contain both
$\zeta$ and its gradient. It turns out that at least at the
classical level, the measure fields affect the theory only through
the scalar field $\zeta$.

The consistency condition of equations of motion has the form of a
constraint which determines $\zeta (x)$ as a function of matter
fields. The surprising feature of the theory is
 that neither Newton constant nor curvature appear in this constraint
which means that the {\it geometrical scalar field} $\zeta (x)$
{\it is determined by the matter fields configuration}  locally
and straightforward (that is without gravitational interaction).

By an appropriate change of the dynamical variables which includes
a conformal transformation of the metric, one can formulate the
theory in a Riemannian (or Riemann-Cartan) space-time. The
corresponding conformal frame we call "the Einstein frame". The
big advantage of TMT is that in the very wide class of models,
{\it the gravity and all matter fields equations of motion take
canonical GR form in the Einstein frame}.
 All the novelty of TMT in the Einstein frame as compared
with the standard GR is revealed only
 in an unusual structure of the scalar fields
effective potential (produced in the Einstein frame), masses of
fermions  and their interactions with scalar fields as well as in
the unusual structure of fermion contributions to the
energy-momentum tensor: all these quantities appear to be $\zeta$
dependent. This is why the scalar field $\zeta (x)$ determined by
the constraint as a function of matter fields, has a key role in
dynamics of TMT models.

\subsection{$SU(2)\times U(1)$ gauge and scale invariant model}

The TMT models possessing a global scale
invariance\cite{G1,GK4,GK5} are of significant interest because
they demonstrate the possibility of  spontaneous breakdown of the
scale symmetry\footnote{The field theory models with explicitly
broken scale symmetry and their application to the quintessential
inflation type  cosmological scenarios have been studied in
Ref.\cite{K}. Inflation and transition to slowly accelerated phase
from higher curvature terms was studied in Ref.\cite{GKatz}. }. In
fact, if the action (\ref{S}) is scale invariant then this
classical field theory effect results from Eq.(\ref{varphi}),
namely  from solving the equation of motion (\ref{varphiB}) or
(\ref{AdL1}).
 One of the
interesting applications of the scale invariant TMT
models\cite{GK4} is a possibility to generate the exponential
potential for the scalar field $\phi$ by means of the mentioned
spontaneous symmetry breaking even  without introducing any
potentials for  $\phi$ in the Lagrangians  $ L_{1}$ and $L_{2}$ of
the action (\ref{S}). Some cosmological applications of this
effect have been also studied in
 Ref.\cite{GK4}.

 In order to show that TMT is able to provide a realistic
 results for gravity and particle physics,
 we present here a model with
  the $SU(2)\times U(1)$ gauge structure as in
the standard model (with standard content of the bosonic sector:
gauge vector fields $\vec{W}_{\mu}$ and $B_{\mu}$
 and Higgs doublet $H$). Although fermions (as well as gauge
 bosons) have no relation to
 the scalar sector dynamics studied in the present paper we
 include also fermions for two reasons:
 the first is to show that
 there exists a regime where the Einstein GR and regular particle
 physics are realized simultaneously (see Appendix D);
  the second reason is to show
 that the Higgs field implements all what it does in the standard
 model.
 We start from only one family of the so called
"primordial" fermionic fields \footnote{On first acquaintance the
theory looks bulky enough. Therefore, for pedagogical reason, we
do not include the primordial up and down quarks $U$ and
$D$}(exact definition of this term is explained in the last
paragraph of the next subsection):
 the primordial electron $E$ and neutrino  $N$.  Just as in the standard
$SU(2)\times U(1)$ gauge invariant model, we will proceed with the
following independent fermionic degrees of freedom:
 one primordial left lepton SU(2) doublet $L_{L}$
and right primordial singlets $N_{R}$ and $E_{R}$.

In addition, a dilaton field $\phi$ is needed in order to realize
a spontaneously broken global scale invariance\cite{G1}. It
governs the evolution of the universe on different stages: in the
early universe $\phi$ plays the role of inflaton and in the late
time universe it is transformed into a part of the dark energy.

According to the general prescriptions of TMT, we have to start
from studying the self-consistent system of gravity and matter
fields proceeding in the first order formalism. In the model
including fermions in curved space-time, this means that the
independent dynamical degrees of freedom are: all  matter fields,
 vierbein $e_{a}^{\mu}$, spin-connection $\omega_{\mu}^{ab}$ and the measure
$\Phi$ degrees of freedom, i.e.  $\varphi_{a}$ or
$A_{\alpha\beta\gamma}$. We postulate that in  addition to
$SU(2)\times U(1)$ gauge symmetry, the theory is invariant under
the global scale transformations:
\begin{eqnarray}
    e_{\mu}^{a}\rightarrow e^{\theta /2}e_{\mu}^{a}, \quad
\omega^{\mu}_{ab}\rightarrow \omega^{\mu}_{ab}, \quad
\varphi_{a}\rightarrow \lambda_{ab}\varphi_{b}\quad where \quad
\det(\lambda_{ab})=e^{2\theta}, \nonumber
\\
\phi\rightarrow \phi-\frac{M_{p}}{\alpha}\theta ,\quad
\Psi\rightarrow e^{-\theta /4}\Psi, \quad
\overline{\Psi}\rightarrow  e^{-\theta /4} \overline{\Psi}; \quad
\theta =const, \nonumber
\\
H\rightarrow H , \quad \vec{W}_{\mu}\rightarrow \vec{W}_{\mu}
\quad B_{\mu}\rightarrow B_{\mu}. \label{stferm}
\end{eqnarray}
If the definition (\ref{Aabg}) is used for the measure $\Phi$ then
 the transformations of $\varphi_{a}$ in
(\ref{stferm}) should be changed by
$A_{\alpha\beta\gamma}\rightarrow
e^{2\theta}A_{\alpha\beta\gamma}$. This global scale invariance
includes the shift symmetry\cite{Carroll} of the dilaton $\phi$
and this is the main factor why it is important for cosmological
applications of the theory\cite{G1,K,GK4,GK6}.

We choose an action which, except for the modification of the
general structure caused by the basic assumptions of TMT,
 {\it does not contain
 any exotic terms and  fields}.
Keeping the general structure (\ref{S}), it is convenient to
represent the action in the following form:
\begin{eqnarray}
&S=&\int d^{4}x e^{\alpha\phi /M_{p}} \left[(\Phi
+b_{g}\sqrt{-g})\left(-\frac{1}{\kappa}R(\omega ,e) +
\frac{1}{2}g^{\mu\nu}(D_{\mu}H)^{\dag}D_{\nu}H \right)+(\Phi
+b_{\phi}\sqrt{-g})\frac{1}{2}g^{\mu\nu}\phi_{,\mu}\phi_{,\nu}\right]
\nonumber\\
&& -\int d^{4}x e^{2\alpha\phi /M_{p}}[\Phi V_{1}(H)
+\sqrt{-g}V_{2}(H)] +\int d^{4}x\sqrt{-g}L_{gauge} +\int d^{4}x
e^{\alpha\phi /M_{p}}(\Phi +k\sqrt{-g})L_{fk}
\nonumber\\
     &&-\int d^{4}xe^{\frac{3}{2}\alpha\phi /M_{p}}
\left[(\Phi +h_{E}\sqrt{-g})f_{E}\overline{L}_{L}\,H\,E_{R}
+[(\Phi +h_{N}\sqrt{-g})f_{N}\overline{L}_{L}\,H^{c}\,N_{R}
+H.c.\right] \label{totaction}
\end{eqnarray}

The notations in (\ref{totaction}) are the following:
$g^{\mu\nu}=e^{\mu}_{a}e^{\nu}_{b}\eta^{ab}$; the scalar curvature
is $R(\omega ,V) =e^{a\mu}e^{b\nu}R_{\mu\nu ab}(\omega)$ where
\begin{equation}
R_{\mu\nu ab}(\omega)=\partial_{\mu}\omega_{\nu ab} +\omega_{\mu
a}^{c}\omega_{\nu cb} -(\mu\leftrightarrow\nu);
        \label{B}
\end{equation}
 \begin{equation}
D_{\mu}H\equiv \left(\partial_{\mu}
-\frac{i}{2}g\vec{\tau}\cdot\vec{W}_{\mu}
-\frac{i}{2}g^{\prime}B_{\mu}\right)H; \label{DHiggs}
 \end{equation}
\begin{equation}
L_{fk}=\frac{i}{2}\left[\overline{L}_{L}{\not\!\!D}L_{L}+
  \, \overline{E}_{R}{\not\!\!D}E_{R}+\overline{N}_{R}{\not\!\!D}N_{R}\right]
\label{Lfkin}
 \end{equation}
where
\begin{equation}
{\not\!\!D}\equiv e^{\mu}_{a}\gamma^{a}\overrightarrow{D}_{\mu}
-\overleftarrow{D}_{\mu} e^{\mu}_{a}\gamma^{a} \label{gamma-D}
\end{equation}
\begin{equation}
\overrightarrow{D}_{\mu}\equiv\vec{\partial}_{\mu}
+\frac{1}{2}\omega_{\mu}^{cd}\sigma_{cd}
-ig\vec{T}\cdot\vec{W}_{\mu} -ig^{\prime}\frac{Y}{2}B_{\mu} \qquad
 \overleftarrow{D}_{\mu}\equiv\overleftarrow{\partial}_{\mu}
-\frac{1}{2}\omega_{\mu}^{cd}\sigma_{cd}I
+i\,g\vec{T}\cdot\vec{W}_{\mu} +i\,g^{\prime}\frac{Y}{2}B_{\mu}
\label{D12L}
 \end{equation}
and, as usual, $\vec{T}=0$ for $SU(2)$ scalars and
$\vec{T}=\vec{\tau}/2$
 for $SU(2)$ spinors; $Y$ sets the standard hypercharges: $Y(L_{L})=-1$, \, $Y(E_{R}=-2)$, \,
$Y(N_{R}=0)$;
\begin{equation}
L_{gauge}=-\frac{1}{4}g^{\alpha\mu}g^{\beta\nu}
\left(B_{\alpha\beta}B_{\mu\nu}+W^{a}_{\alpha\beta}W^{a}_{\mu\nu}\right)
\label{Lgauge}
 \end{equation}
and finally $B_{\mu\nu}\equiv\partial_{\mu}B_{\nu}-
                               \partial_{\nu}B_{\mu}$, \quad
$W^{a}_{\mu\nu}\equiv\partial_{\mu}W^{a}_{\nu}
                               -\partial_{\nu}W^{a}_{\mu}
+g\varepsilon^{abc}W^{b}_{\mu}W^{c}_{\nu}$.

In (\ref{totaction}) there are two types of the gravitational
terms and
 of the "kinetic-like terms" (both
for the scalar fields and for the primordial fermionic ones) which
respect the scale invariance : the terms of the one type coupled
to the
 measure $\Phi$ and those of the other type
coupled to the measure $\sqrt{-g}$. For the same reason there are
two different sets of the Yukawa-like coupling terms of the
primordial fermions \footnote{We restrict ourselves here only with
the Dirac type neutrino mass terms}. Using the freedom in
normalization of the measure fields ($\varphi_{a}$  in the case of
using Eq.(\ref{Phi}) or $A_{\alpha\beta\gamma}$ when using
Eq.(\ref{Aabg})), we set the coupling constant of the scalar
curvature to the measure $\Phi$ to be  $-\frac{1}{\kappa}$.
Normalizing all the fields such that their couplings to the
measure $\Phi$ have no additional factors, we are not able in
general to provide the same in terms describing the appropriate
couplings to the measure $\sqrt{-g}$. This fact explains the need
to introduce the dimensionless real parameters $b_g, b_{\phi}, k,
h_{N}, h_{E}$ and we will only assume that {\it they have close
orders of magnitudes}. The real positive parameter $\alpha$ is
assumed to be of the order of unity; in all solutions presented in
this paper we set $\alpha =0.2$. For Newton constant we use
$\kappa =16\pi G $, \, $M_p=(8\pi G)^{-1/2}$.

One should  also point out the possibility of introducing two
different potential-like exponential functions of the dilaton
$\phi$ coupled to the measures $\Phi$ and $\sqrt{-g}$ with factors
$V_{1}(H)$ and $V_{2}(H)$. $V_{1}$ and $V_{2}$ must be
$\phi$-independent to provide the scale symmetry (\ref{stferm}).
However they can be Higgs-dependent and then they play the role of
the Higgs pre-potentials (we will see below how the effective
potential of the scalar sector is generated in the Einstein
frame).

The choice of the action (\ref{totaction}) needs a few additional
explanations:

1) With the aim to simplify the analysis of the results of the
model (containing too many free parameters) we have chosen the
coefficient $b_{g}$ in front of $\sqrt{-g}$ in the first integral
of (\ref{totaction}) to be a common factor of the gravitational
term
 $-\frac{1}{\kappa}R(\omega ,e)$ and of the kinetic term for the
 Higgs field $H$. Except for the simplification there are no reasons
 for such a fine tuned choice.

2) We choose the kinetic terms of the gauge bosons in the
conformal invariant form which is possible only if these terms are
coupled to the measure $\sqrt{-g}$. Introducing the coupling of
these terms to the measure $\Phi$ would lead to the nonlinear
field strength dependence  in the gauge fields equations of motion
as well as to non positivity of the energy.  Another consequence
is a possibility of certain unorthodox effects, like space-time
variations of the effective fine structure constant.

3) One can show that  for achieving the right chiral structure of
the fermion sector in the Einstein frame,
 one has to choose
in (\ref{totaction}) the coupling of
 the kinetic terms of all the left and right primordial fermions
to the measures $\sqrt{-g}$ and $\Phi$ to be universal. This
feature is displayed in the choice of the parameter $k$  to be the
common factor in front of the corresponding kinetic terms for all
primordial fermion degrees of freedom.

Except for these three items, Eq.(\ref{totaction}) describes {\em
the most general action of TMT satisfying the formulated above
symmetries}.

\subsection{Equations of motion in the Einstein
frame}

In Appendix A  we present the equations of motion resulting from
the action (\ref{totaction}) when using the original set of
variables. In Appendix B one can find the equation and solution
for the spin-connection. The common feature of all the equations
in the original frame is that the measure $\Phi$ degrees of
freedom appear only through dependence upon the scalar field
$\zeta$, Eq.(\ref{zeta}). In particular, all the equations of
motion and the solution for the spin-connection include terms
proportional to $\partial_{\mu}\zeta$, that makes space-time non
Riemannian and all equations of motion - noncanonical.

It turns out that with the set of the new variables ($\phi$, $H$,
$B_{\mu}$ and $\vec{W}_{\mu}$ remain the same)
\begin{equation}
\tilde{g}_{\mu\nu}=e^{\alpha\phi/M_{p}}(\zeta +b_{g})g_{\mu\nu},
\quad \tilde{e}_{a\mu}=e^{\frac{1}{2}\alpha\phi/M_{p}}(\zeta
+b_{g})^{1/2}e_{a\mu}, \quad
\Psi^{\prime}_{i}=e^{-\frac{1}{4}\alpha\phi/M_{p}} \frac{(\zeta
+k)^{1/2}}{(\zeta +b_{g})^{3/4}}\Psi_{i} , \quad i=N,E
\label{ctferm}
\end{equation}
which we call the Einstein frame,
 the spin-connections  become those of the
Einstein-Cartan space-time, see Appendix C. Since
$\tilde{e}_{a\mu}$, $N^{\prime}$ and $E^{\prime}$ are invariant
under the scale transformations (\ref{stferm}), spontaneous
breaking of the scale symmetry (by means of Eq.(4) which for our
model (\ref{totaction}) takes the form (\ref{app1})) is reduced in
the new variables to the {\it spontaneous breakdown of the shift
symmetry}
\begin{equation}
 \phi\rightarrow\phi +const.
 \label{phiconst}
\end{equation}
Notice that the Goldstone theorem generically is not applicable in
this sector of the theory\cite{G1}. The reason is the following.
In fact, the shift symmetry (\ref{phiconst}) leads to a conserved
dilatation current $j^{\mu}$. However, for example in the
spatially flat FRW universe the spatial components of the current
$j^{i}$ behave as $j^{i}\propto M^4x^i$ as
$|x^i|\rightarrow\infty$. Due to this anomalous behavior at
infinity, there is a flux of the current leaking to infinity,
which causes the non conservation of the dilatation charge. The
absence of the latter implies that one of the conditions necessary
for the Goldstone theorem is missing. The non conservation of the
dilatation charge is similar to the well known effect of
instantons in QCD where singular behavior in the spatial infinity
leads to the absence of the Goldstone boson associated to the
$U(1)$ symmetry.

 After the change of
variables  to the Einstein frame (\ref{ctferm}) and some simple
algebra, the gravitational equations (\ref{app12}) take the
standard GR form
\begin{equation}
G_{\mu\nu}(\tilde{g}_{\alpha\beta})=\frac{\kappa}{2}T_{\mu\nu}^{eff}
 \label{gef}
\end{equation}
where  $G_{\mu\nu}(\tilde{g}_{\alpha\beta})$ is the Einstein
tensor in the Riemannian space-time with the metric
$\tilde{g}_{\mu\nu}$; the energy-momentum tensor
$T_{\mu\nu}^{eff}$ is now
\begin{eqnarray}
T_{\mu\nu}^{eff}&=&\frac{\zeta +b_{\phi}}{\zeta +b_{g}}
\left(\phi_{,\mu}\phi_{,\nu}-\frac{1}{2}
\tilde{g}_{\mu\nu}\tilde{g}^{\alpha\beta}\phi_{,\alpha}\phi_{,\beta}\right)
-\tilde{g}_{\mu\nu}\frac{b_{g}-b_{\phi}}{2(\zeta +b_{g})}
\tilde{g}^{\alpha\beta}\phi_{,\alpha}\phi_{,\beta}
\nonumber\\
&+&(D_{\mu}H)^{\dag}D_{\nu}H
-\frac{1}{2}\tilde{g}_{\mu\nu}\tilde{g}^{\alpha\beta}(D_{\alpha}H)^{\dag}D_{\beta}H
+\tilde{g}_{\mu\nu}V_{eff}(\phi,\upsilon ;\zeta)
+T_{\mu\nu}^{(gauge)}
+T_{\mu\nu}^{(ferm,can)}+T_{\mu\nu}^{(ferm,noncan)}
 \label{Tmn}
\end{eqnarray}
The function $V_{eff}(\phi ;\zeta)$ has the form
\begin{equation}
V_{eff}(\phi,\upsilon ;\zeta)=
\frac{b_{g}\left[M^{4}e^{-2\alpha\phi/M_{p}}+V_{1}(\upsilon
)\right] -V_{2}(\upsilon)}{(\zeta +b_{g})^{2}}; \label{Veff1}
\end{equation}
$T_{\mu\nu}^{(gauge)}$ is the canonical energy momentum tensor for
the $SU(2)\times U(1)$ gauge fields sector;
$T_{\mu\nu}^{(ferm,can)}$ is the canonical energy momentum tensor
for (primordial) fermions $N^{\prime}$ and $E^{\prime}$ in curved
space-time including also their standard $SU(2)\times U(1)$ gauge
interactions. $T_{\mu\nu}^{(ferm,noncan)}$ is the {\em
noncanonical} contribution of the fermions into the energy
momentum tensor
\begin{equation}
 T_{\mu\nu}^{(ferm,noncan)}=-\tilde{g}_{\mu\nu}\Lambda_{dyn}^{(ferm)};
\qquad  \quad \Lambda_{dyn}^{(ferm)}\equiv
Z_{N}(\zeta)m_{N}(\zeta, \upsilon)
\overline{N^{\prime}}N^{\prime}+ Z_{E}(\zeta)m_{E}(\zeta,
\upsilon)\overline{E^{\prime}}E^{\prime} \label{Tmn-noncan}
\end{equation}
where $Z_{i}(\zeta)$ and $m_{i}(\zeta, \upsilon)$
($i=N^{\prime},E^{\prime}$) are respectively
\begin{equation}
Z_{i}(\zeta)\equiv \frac{(\zeta -\zeta^{(i)}_{1})(\zeta
-\zeta^{(i)}_{2})}{2(\zeta +k)(\zeta +h_{i})}, \qquad
\zeta_{1,2}^{(i)}=\frac{1}{2}\left[k-3h_{i}\pm\sqrt{(k-3h_{i})^{2}+
8b(k-h_{i}) -4kh_{i}}\,\right],
 \label{Zeta}
\end{equation}
\begin{equation}
m_{N}(\zeta,\upsilon)= \frac{\upsilon f_{N}(\zeta
+h_{N})}{\sqrt{2}(\zeta +k)(\zeta +b_{g})^{1/2}} \qquad
m_{E}(\zeta,\upsilon)= \frac{\upsilon f_{E}(\zeta
+h_{E})}{\sqrt{2}(\zeta +k)(\zeta +b_{g})^{1/2}}
 \label{muferm1}
\end{equation}

The structure of $T_{\mu\nu}^{(ferm,noncan)}$ shows that it
behaves as
 a sort of  variable cosmological constant\cite{var-Lambda}
but in our case it is originated by fermions. This is why we will
refer to it as {\it dynamical fermionic $\Lambda$
 term}. This fact is displayed explicitly
in Eq.(\ref{Tmn-noncan}) by defining $\Lambda_{dyn}^{(ferm)}$. One
has to emphasize the substantial difference of the way
$\Lambda_{dyn}^{(ferm)}$
 emerges here as compared to the models of the
condensate cosmology (see for example
Refs.\cite{cond1}-\cite{cond3}). In those models the dynamical
cosmological constant results from  bosonic or fermionic
condensates. In TMT model studied here,
 $\Lambda_{dyn}^{(ferm)}$ is
originated by fermions but there is no need for any condensate. In
Appendix C we show that $\Lambda_{dyn}^{ferm}$ becomes negligible
in gravitational experiments with observable matter. However it
may be  very important for some astrophysics and cosmology
problems\cite{GK6}.

The dilaton $\phi$ field equation (\ref{phi-orig}) in the new
variables reads
\begin{eqnarray}
&&\frac{1}{\sqrt{-\tilde{g}}}\partial_{\mu}\left[\frac{\zeta
+b_{\phi}}{\zeta
+b_{g}}\sqrt{-\tilde{g}}\tilde{g}^{\mu\nu}\partial_{\nu}\phi\right]
\nonumber\\
 &-&\frac{\alpha}{M_{p}(\zeta +b_{g})^{2}} \left[(\zeta
+b_{g})M^{4}e^{-2\alpha\phi/M_{p}}-(\zeta -b_{g})V_{1}(\upsilon)
-2V_{2}(\upsilon)-\delta b_{g}(\zeta
+b_{g})\frac{1}{2}\tilde{g}^{\alpha\beta}\phi_{,\alpha}\phi_{,\beta}\right]
\nonumber \\
 &=&-\frac{\alpha}{M_{p}}[Z_{N}(\zeta)m_{N}(\zeta,
\upsilon)\overline{N^{\prime}}N^{\prime}
+Z_{E}(\zeta)m_{E}(\zeta,\upsilon)\overline{E^{\prime}}E^{\prime}]
\equiv -\frac{\alpha}{M_{p}}\Lambda_{dyn}^{(ferm)}.
 \label{phief+ferm1}
\end{eqnarray}

The Higgs field equation in the unitary gauge (\ref{Higgs-orig})
 takes in the Einstein frame the following form
\begin{equation}
\Box \upsilon  +\frac{\zeta V_{1}^{\prime}+V_{2}^{\prime}}{(\zeta
+b_{g})^{2}}= -\frac{1}{\sqrt{2}(\zeta +b_{g})^{1/2}(\zeta +k)}
\left[f_{E}(\zeta +h_{E})\overline{E^{\prime}}E^{\prime} +
f_{N}(\zeta +h_{N})\overline{N^{\prime}}N^{\prime}\right],
\label{chief+ferm}
\end{equation}
where
\begin{equation}
\Box\upsilon =(-\tilde{g})^{-1/2}\partial_{\mu}
(\sqrt{-\tilde{g}}\tilde{g}^{\mu\nu}\partial_{\nu}\upsilon) \quad
and \quad V_{i}^{\prime}\equiv \frac{dV_i}{d\upsilon} \quad
(i=1,2). \label{Box}
\end{equation}
 We have omitted here interactions to gauge fields which are of
the canonical $SU(2)\times U(1)$ form due to Eq.(\ref{DHiggs}).

One can show that equations  for the primordial leptons in terms
of the variables (\ref{ctferm})
 take the standard form
of fermionic equations for $N^{\prime}$ and $E^{\prime}$ in the
Einstein-Cartan space-time  where the spin-connection
$\omega_{\mu}^{\prime ab}$ is determined by Eq.(\ref{C7}) and the
standard gauge interactions are also present. However the
non-Abelian structure of these interactions makes the resulting
equations very bulky.  It is more convenient to represent the
result of calculations for the lepton sector in the Einstein frame
in a form of the effective fermion action
$S_{eff}^{(ferm)}=\int\sqrt{-\tilde{g}}d^{4}xL_{eff}^{(ferm)}$
where
\begin{equation}
L_{eff}^{(ferm)}=
\frac{i}{2}\left[\overline{L}^{\prime}_{L}\tilde{\not\!\!D}L^{\prime}_{L}+
  \, \overline{E}^{\prime}_{R}\tilde{\not\!\!D}E^{\prime}_{R}
+\overline{N}^{\prime}_{R}\tilde{\not\!\!D}N^{\prime}_{R}\right] -
m_{N}(\zeta,\upsilon)\overline{N}^{\prime}N^{\prime} -
m_{E}(\zeta,\upsilon)\overline{E}^{\prime}E^{\prime}.
\label{Lf-eff-Ein}
 \end{equation}
Here $\tilde{\not\!\!D}\equiv\tilde{\overrightarrow{\not\!\!D}}
-\tilde{\overleftarrow{\not\!\!D}} $  and
\begin{equation}
\tilde{\overrightarrow{\not\!\!D}}\equiv
e^{\prime\mu}_{a}\gamma^{a}\left(\vec{\partial}_{\mu}
+\frac{1}{2}\omega_{\mu}^{\prime cd}\sigma_{cd}
-ig\vec{T}\cdot\vec{W}_{\mu}
-ig^{\prime}\frac{Y}{2}B_{\mu}\right); \quad
\tilde{\overleftarrow{\not\!\!D}}\equiv
\left(\overleftarrow{\partial}_{\mu}
-\frac{1}{2}\omega_{\mu}^{\prime cd}\sigma_{cd}I
+i\,g\vec{T}\cdot\vec{W}_{\mu} +i\,
g^{\prime}\frac{Y}{2}B_{\mu}\right)\gamma^{a}e_{a}^{\prime\mu}
\label{D12L-Ein}
 \end{equation}

All the novelty in (\ref{Lf-eff-Ein}), as compared with the
standard field theory approach to the $SU(2)\times U(1)$ unified
gauge theory, consists of the $\zeta$ dependence of the "masses",
Eq.(\ref{muferm1}), of the primordial fermions $N^{\prime}$,
$E^{\prime}$.

The scalar field $\zeta$ in Eqs.(\ref{Tmn})-(\ref{Lf-eff-Ein}) is
determined by means of the constraint (\ref{app10}) which in the
new variables (\ref{ctferm}) takes the form
\begin{eqnarray}
&&\frac{1}{(\zeta
+b_{g})^{2}}\left\{(b_{g}-\zeta)\left[M^{4}e^{-2\alpha\phi/M_{p}}+
V_{1}(\upsilon)\right]-2V_{2}(\upsilon)-\delta b_{g}(\zeta
+b_{g})\frac{1}{2}\tilde{g}^{\alpha\beta}\phi_{,\alpha}\phi_{,\beta}\right\}
\nonumber \\
&&=[Z_{N}(\zeta)m_{N}(\zeta)\overline{N^{\prime}}N^{\prime}
+Z_{E}(\zeta)m_{E}(\zeta)\overline{E^{\prime}}E^{\prime}]\quad
\equiv \quad \Lambda_{dyn}^{(ferm)}, \label{constraint2}
\end{eqnarray}
where
\begin{equation}
\delta =\frac{b_{g}-b_{\phi}}{b_{g}} \label{delta}
\end{equation}

One should point out the interesting and very important fact: the
same $\Lambda_{dyn}^{(ferm)}$ emerges in the following three
different places: in the noncanonical contribution of the fermions
into the energy momentum tensor (\ref{Tmn-noncan}), in the
effective coupling of the dilaton to fermions (the right hand side
of Eq.(\ref{phief+ferm1})) and in the constraint
(\ref{constraint2}).

Notice the very important fact that as a result of our choice of
the conformal invariant form for the kinetic terms of the gauge
bosons in the action(\ref{totaction}), the gauge fields do not
enter into the constraint.

The gauge fields equations in the Einstein frame become exactly
the same as in the standard field theory approach to the
$SU(2)\times U(1)$ unified gauge theory. For example,
Eq.(\ref{gauge-orig}) for the gauge field $B_{\mu}$ in the
fermionic vacuum reads now
\begin{equation}
\frac{1}{\sqrt{-\tilde{g}}}\partial_{\nu}
\left[\sqrt{-\tilde{g}}\tilde{g}^{\alpha\mu}\tilde{g}^{\beta\nu}B_{\alpha\beta}\right]
+ \tilde{g}^{\mu\nu}\frac{1}{8}g^{\prime 2}\upsilon^2B_{\nu}=0,
\label{gauge-Ein}
\end{equation}
and similar for $\vec{W}_{\mu}$ with the appropriate non-Abelian
structure.  It is straightforward now to construct linear
combinations of the gauge fields to produce the electromagnetic
field $A_{\mu}$ and  $W^{\pm}$ and $Z$ bosons.

 One can
show that not only in the case of the $SU(2)\times U(1)$ gauge
theory but  also in more general gauge theories, like GUT, gauge
fields equations of motion in the Einstein frame (in TMTF)
coincide
 with the appropriate equations of the standard field theory approach to
the gauge theory. Therefore after the Higgs field develops a non
zero vacuum expectation value (VEV), the Higgs phenomenon takes
place here exactly in the same manner as in the standard approach
to the unified gauge theories: fermions and part of the gauge
degrees of freedom become massive (see Eqs.(\ref{muferm1}),
(\ref{Lf-eff-Ein}) and (\ref{gauge-Ein})). Hence,  $SU(2)\times
U(1)$ gauge model serves an illustrative example  that in spite of
the very specific general structure of the TMT action
(\ref{totaction}), the scalar field $H$ (or $\upsilon$ in the
unitary gauge) indeed plays the role of the Higgs field. However
the detailed mechanism by means of which the symmetry breaking is
implemented in TMT may be very much different from how it is done
in the standard gauge theories. In particular, we are going to
show that it can be done without a tachyonic mass term in the
action.

Applying constraint (\ref{constraint2}) to Eq.(\ref{phief+ferm1})
one can reduce the latter to the form
\begin{equation}
\frac{1}{\sqrt{-\tilde{g}}}\partial_{\mu}\left[\frac{\zeta
+b_{\phi}}{\zeta
+b_{g}}\sqrt{-\tilde{g}}\tilde{g}^{\mu\nu}\partial_{\nu}\phi\right]-\frac{2\alpha\zeta}{(\zeta
+b_{g})^{2}M_{p}}M^{4}e^{-2\alpha\phi/M_{p}} =0,
\label{phi-after-con}
\end{equation}
where $\zeta$  is a solution of the constraint
(\ref{constraint2}).

Due to the constraint (\ref{constraint2}) which determines the
scalar field $\zeta$ as a function of scalar and fermion fields,
generically fermions in TMT are very much different from what one
is used to in normal field theory. For example the fermion mass
can depend upon the fermion density. In Appendix D we show that if
the local energy density of the fermion is many orders of
magnitude larger than the vacuum energy density in the space-time
region occupied by the fermion then the fermion can have a
constant mass. However this is exactly the case of atomic, nuclear
and particle physics, including accelerator physics and high
density objects of astrophysics. This is why to such "high
density" (in comparison with the vacuum energy density) phenomena
we  refer as "{\it normal particle physics conditions}" and the
appropriate fermion states in TMT we call "{\it regular
fermions}". For generic fermion states in TMT we  use the term
"{\it primordial fermions}" in order to distinguish them from
regular fermions.

\section{Scalar sector}
\subsection{Equations for General Case Including $k$-essence}

When neglecting  the fermions and gauge fields, the origin of the
gravity is the scalar sector of the matter fields which consists
of the dilaton $\phi$ and Higgs $\upsilon$ fields. The
gravitational, dilaton and Higgs equations of motion  in the
Einstein frame follow immediately from
Eqs.(\ref{gef})-(\ref{Veff1}), (\ref{phief+ferm1}) and
(\ref{chief+ferm}) where one should ignore all fermionic and gauge
fields terms; the scalar $\zeta$ is a solution of the constraint
(\ref{constraint2}) which for finite $\zeta$ has now the following
form
\begin{equation}
(b_{g}-\zeta)\left[M^{4}e^{-2\alpha\phi/M_{p}}+
V_{1}(\upsilon)\right]-2V_{2}(\upsilon)-\delta b_{g}(\zeta
+b_{g})X=0,
 \label{constr-vac}
 \end{equation}
where
\begin{equation}
X\equiv\frac{1}{2}\tilde{g}^{\alpha\beta}\phi_{,\alpha}\phi_{,\beta}.
\label{X}
 \end{equation}

In the absence of fermions case, the scalar sector can be
described as a perfect fluid with the following  energy and
pressure densities resulting from Eqs.(\ref{Tmn}) and
(\ref{Veff1}) after inserting the solution $\zeta
=\zeta(\phi,X,\upsilon)$ of the linear in $\zeta$
Eq.(\ref{constr-vac})
\begin{equation}
\rho =X+Y+
\frac{1}{4[b_{g}(M^{4}e^{-2\alpha\phi/M_{p}}+V_{1})-V_{2}]}
\left[(M^{4}e^{-2\alpha\phi/M_{p}}+V_{1})^{2}- 2\delta
b_{g}(M^{4}e^{-2\alpha\phi/M_{p}}+V_{1})X -3\delta^{2}
b_{g}^{2}X^2\right], \label{rho1}
\end{equation}
\begin{equation}
p =X+Y-
\frac{1}{4[b_{g}(M^{4}e^{-2\alpha\phi/M_{p}}+V_{1})-V_{2}]}
\left[(M^{4}e^{-2\alpha\phi/M_{p}}+V_{1})^{2}+ 2\delta
b_{g}(M^{4}e^{-2\alpha\phi/M_{p}}+V_{1})X +\delta^{2}
b_{g}^{2}X^2\right], \label{p1}
\end{equation}
where
$Y\equiv\frac{1}{2}\tilde{g}^{\alpha\beta}\upsilon_{,\alpha}\upsilon_{,\beta}$.

In a spatially flat FRW universe with the metric
$\tilde{g}_{\mu\nu}=diag(1,-a^2,-a^2,-a^2)$ filled with the
homogeneous scalar sector fields $\phi$ and $\upsilon$, the
dilaton $\phi$ and Higgs $\upsilon$ field  equations of motion
take the form
\begin{equation}
Q_{1}\ddot{\phi}+ 3HQ_{2}\dot{\phi}- \frac{\alpha}{M_{p}}Q_{3}
M^{4}e^{-2\alpha\phi/M_{p}}=0 \label{phi1}
\end{equation}
\begin{equation}
\ddot{\upsilon}+3H\dot{\upsilon}+Q_{\upsilon}(\phi,X,\upsilon)=0
\label{H1}
\end{equation}
 where $H$ is the Hubble parameter and we have used the following notations
\begin{equation}
\dot{\phi}\equiv \frac{d\phi}{dt}, \qquad \dot{\upsilon}\equiv
\frac{d\upsilon}{dt}, \label{phidot-vdot}
\end{equation}
\begin{equation}
Q_1=(b_{g}+b_{\phi})(M^{4}e^{-2\alpha\phi/M_{p}}+V_{1})-
2V_{2}-3\delta^{2}b_{g}^{2}X \label{Q1}
\end{equation}
\begin{equation}
Q_2=(b_{g}+b_{\phi})(M^{4}e^{-2\alpha\phi/M_{p}}+V_{1})-
2V_{2}-\delta^{2}b_{g}^{2}X\label{Q2}
\end{equation}
\begin{eqnarray}
Q_3&=&\frac{1}{[b_{g}(M^{4}e^{-2\alpha\phi/M_{p}}+V_{1})-V_{2}]}
\nonumber
\\
 &\times & \left[(M^{4}e^{-2\alpha\phi/M_{p}}+V_{1})
[b_{g}(M^{4}e^{-2\alpha\phi/M_{p}}+V_{1})-2V_{2}] +2\delta
b_{g}V_{2}X+3\delta^{2}b_{g}^{3}X^{2}\right] \label{Q3}
\end{eqnarray}
and
\begin{eqnarray}
Q_{\upsilon}(\phi,X,\upsilon)&=&\frac{M^{4}e^{-2\alpha\phi/M_{p}}+V_{1}
+\delta
b_{g}X}{4[b_{g}(M^{4}e^{-2\alpha\phi/M_{p}}+V_{1})-V_{2}]^{2}}
\nonumber
\\
&&\times\left[\left(b_{g}(M^{4}e^{-2\alpha\phi/M_{p}}+V_{1}-\delta
b_{g}X)
-2V_{2}\right)V^{\prime}_{1}+(M^{4}e^{-2\alpha\phi/M_{p}}+V_{1}+\delta
b_{g}X)V^{\prime}_{2}\right] \label{Q-H}
\end{eqnarray}

It is interesting that the non-linear $X$-dependence appears here
in the framework of the fundamental theory without exotic terms in
the original action (\ref{totaction}). This effect results just
from the fact that there are no reasons to choose the parameters
$b_{g}$ and $b_{\phi}$ in the action (\ref{totaction}) to be equal
in general. The above equations represent an explicit example of
$k$-essence\cite{k-essence}. In fact, one can check that the
system of equations (\ref{gef}), (\ref{rho1})-(\ref{H1})
(accompanied with the functions (\ref{Q1})-(\ref{Q-H}) and written
in the metric $\tilde{g}_{\mu\nu}=diag(1,-a^2,-a^2,-a^2)$) can be
obtained from the k-essence type effective action
\begin{equation}
S_{eff}=\int\sqrt{-\tilde{g}}d^{4}x\left[-\frac{1}{\kappa}R(\tilde{g})
+p\left(\phi,X,\upsilon,Y\right)\right] \label{k-eff},
\end{equation}
where $p(\phi,X,\upsilon,Y)$ is given by Eq.(\ref{p1}). In
contrast to the simplified models studied in
literature\cite{k-essence}, it is impossible here to represent the
Lagrangian density of the scalar sector in a factorizable form.
For example even in the case $\upsilon\equiv 0$,
 it is impossible to represent
$p(\phi,X)$, Eq.(\ref{p1}), in the form of the product
$K(\phi)\tilde{p}(X)$.

Recall that for the sake of simplicity we have chosen the
coefficient $b_{g}$ in front of $\sqrt{-g}$ in the first integral
of (\ref{totaction}) to be a common factor of the gravitational
term $-\frac{1}{\kappa}R(\omega ,e)$ and of the kinetic term for
the Higgs field $H$. It is evident that without such a fine tuned
choice we would obtain the k-essence type effective action
non-linear both in $X$ and $Y$.

\subsection{Equations for Simplified Dilaton - Gravity Models
in a Fine Tuned Case $\delta =0$ and No $k$-essence.}

In this section  we specialize the above model to  the
cosmological dynamics  in a simplified version of TMT where the
dilaton is the only matter field of the model.  The combined
effect of both dilaton and Higgs, leading to a new type of
cosmological mechanism for the gauge symmetry breaking will be
studied in Sec.VII.

However, even in the toy model without the Higgs field, the full
analysis of the problem is complicated because of the large space
of free parameters appearing in the action.  The qualitative
analysis of equations is significantly simplified if $\delta =0$.
This is what we will assume in this subsection. Although it looks
like a fine tuning of the parameters (i.e. $b_{g}=b_{\phi}$), it
allows us to understand qualitatively the basic features of the
model. In fact, only in the case $\delta =0$ the effective action
(\ref{k-eff}) takes the form of that of the  scalar field without
higher powers of derivatives. Role of $\delta\neq 0$ in a
possibility to produce an effect of superacceleration will be
studied in Sec.V.

So let us study the cosmology governed by the system of equations
\begin{equation}
\frac{\dot{a}^{2}}{a^{2}}=\frac{1}{3M_{p}^{2}}\rho
\label{cosm-phi}
\end{equation}
and (\ref{rho1})-(\ref{phi1}) where one should ignore all the
Higgs dynamics (therefore $V_{1}$ and $V_{2}$ are now constants
and $\dot{\upsilon}\equiv 0$) and set $\delta =0$.

In the described approximation  the constraint (\ref{constr-vac})
yields
\begin{equation}
\zeta =\zeta_{0}(\phi)\equiv b_{g}-\frac{2V_{2}}
{V_{1}+M^{4}e^{-2\alpha\phi/M_{p}}},
\label{zeta-without-ferm-delta=0}
 \end{equation}
 The energy density and pressure take then the canonical form,
\begin{equation}
\rho|_{\delta =0}=\frac{1}{2}\dot{\phi}^{2}+V_{eff}^{(0)}(\phi);
\qquad p|_{\delta
=0}=\frac{1}{2}\dot{\phi}^{2}-V_{eff}^{(0)}(\phi),
\label{rho-delta=0}
\end{equation}
where the effective potential of the scalar field $\phi$ results
from Eq.(\ref{Veff1})
\begin{equation}
V_{eff}^{(0)}(\phi)
=\frac{[V_{1}+M^{4}e^{-2\alpha\phi/M_{p}}]^{2}}
{4[b_{g}\left(V_{1}+M^{4}e^{-2\alpha\phi/M_{p}}\right)-V_{2}]}
\label{Veffvac-delta=0}
\end{equation}
and the $\phi$-equation (\ref{phi1}) is reduced to
\begin{equation}
\ddot{\phi}+3H\dot{\phi}+\frac{dV^{(0)}_{eff}}{d\phi}=0.
\label{eq-phief-without-ferm-delta=0}
\end{equation}

Notice that $V_{eff}^{(0)}(\phi)$ is non-negative for any $\phi$
provided
\begin{equation}
b_{g}V_{1}>V_{2} \label{bV1>V2},
\end{equation}
that we will assume in what follows. We assume also that
$b_{g}>0$.

\begin{figure}[htb]
\begin{center}
\includegraphics[width=18.0cm,height=12cm]{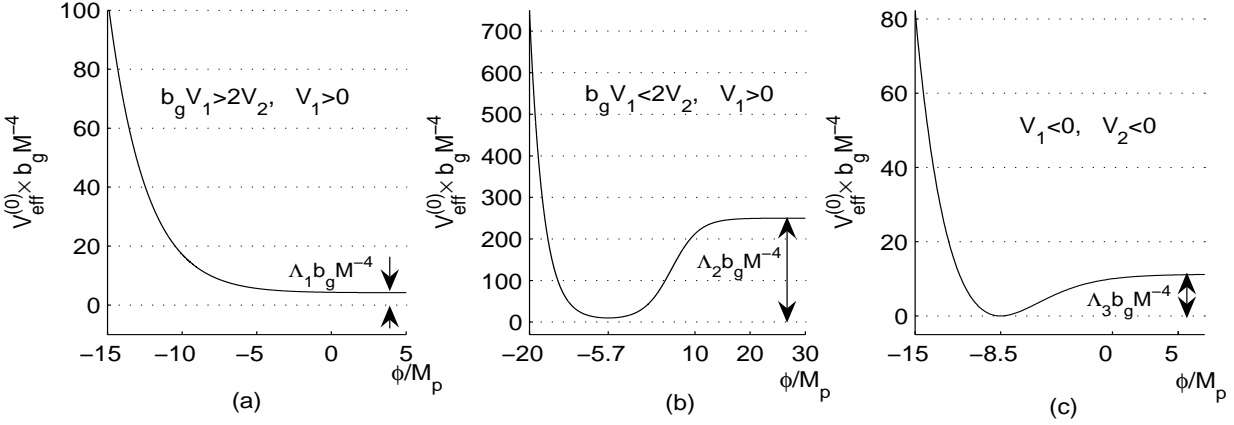}
\end{center}\caption{Three possible shapes of the effective potential
$V_{eff}^{(0)}(\phi)$
 in the models with  $b_{g}V_{1}>V_{2}$:
 \quad Fig.(a) \, $b_{g}V_{1}>2V_{2}$ (in the graph
 $V_{1}=10M^{4}$ and $V_{2}=4b_{g}M^{4}$); \quad Fig.(b) \, $b_{g}V_{1}<2V_{2}$
 (in the graph $V_{1}=10M^{4}$ and $V_{2}=9.9b_{g}M^{4}$).
 The value of $V_{eff}^{(0)}$ in the minimum $\phi_{min}=-5.7M_p$
 is larger than zero;
 \quad Fig.(c) \, $V_{1}<0$, $V_{2}<0$
 (in the graph $V_{1}=-30M^{4}$ and $V_{2}=-50b_{g}M^{4}$).
 $V_{eff}^{(0)}(\phi_{min})=0$ in the
 minimum $\phi_{min}=-8.5M_p$ .
In all the cases here as well as in all solutions presented in
this paper we choose $\alpha =0.2$.}\label{fig1}
\end{figure}

In the following three sections we consider three different
dilaton-gravity cosmological models determined by different choice
of the parameters $V_{1}$ and $V_{2}$ in the action
 with $\delta =0$: two models with $b_{g}V_{1}>V_{2}$ and $V_{1}>0$ and
one model with $V_{1}<0$. The appropriate three possible shapes of
the effective potential $V_{eff}^{(0)}(\phi)$ are presented in
Fig.1. A special case with the fine tuned condition
$b_{g}V_{1}=V_{2}$ is discussed in Appendix D where we show that
equality of the couplings to measures $\Phi$ and $\sqrt{-g}$ in
the action (equality $b_{g}V_{1}=V_{2}$ is one of the conditions
for this to happen) gives rise to a symmetric form of the
effective potential.

\section{Cosmological Dynamics in the Model With $b_{g}V_{1}>2V_{2}$ and $\delta =0$: \quad
Early Power Law Inflation Ending With Small $\Lambda$ Driven
Expansion}

 In this model
 the effective potential (\ref{Veffvac-delta=0}) is a
monotonically decreasing function of $\phi$ (see
Fig.{\ref{fig1}}a). As $M^{4}e^{-2\alpha\phi/M_{p}}\gg
Max\left(V_{1}, V_{2}/b_{g}\right)$, the effective potential
(\ref{Veffvac-delta=0}) behaves as the exponential potential
$V_{eff}^{(0)}\approx
\frac{1}{4b_{g}}M^{4}e^{-2\alpha\phi/M_{p}}$. So, as $\phi\ll
-M_{p}$  the model is able to describe a power law inflation of
the early universe\cite{power-law} if $\alpha < 1/\sqrt{2}$. The
latter condition will be assumed in all analytic solutions and
qualitative discussions throughout the paper.

 Applying this model to the
cosmology of the late time universe and assuming that the scalar
field $\phi\rightarrow\infty$ as $t\rightarrow\infty$, it is
convenient to represent the effective potential
(\ref{Veffvac-delta=0}) in the form
\begin{equation}
V_{eff}^{(0)}(\phi)=\Lambda_1+V_{q-l}(\phi).
\label{rho-without-ferm}
\end{equation}
where
\begin{equation}
\Lambda_1 =\frac{V_{1}^{2}} {4(b_{g}V_{1}-V_{2})}
\label{lambda-without-ferm-delta=0}
\end{equation}
is the positive cosmological constant and
\begin{equation}
V^{(0)}_{q-l}(\phi)
=\frac{(b_{g}V_{1}-2V_{2})V_{1}M^{4}e^{-2\alpha\phi/M_{p}}+
(b_{g}V_{1}-V_{2})M^{8}e^{-4\alpha\phi/M_{p}}}
{4(b_{g}V_{1}-V_{2})[b_{g}(V_{1}+
M^{4}e^{-2\alpha\phi/M_{p}})-V_{2}]}.
\label{V-quint-without-ferm-delta=0}
\end{equation}
We see that the evolution of the late time universe  is governed
both by the cosmological constant $\Lambda_1$ and by the
quintessence-like potential $V^{(0)}_{q-l}(\phi)$.

\begin{figure}[htb]
\begin{center}
\includegraphics[width=18.0cm,height=6.0cm]{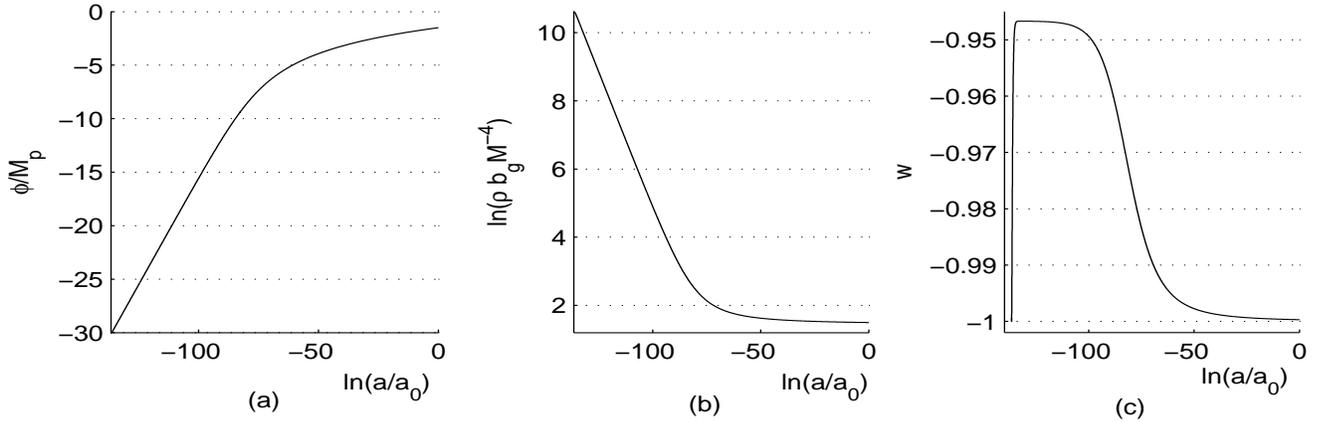}
\end{center}
\caption{Typical dependence of the field $\phi$ (fig. (a)), the
energy density $\rho$ (fig. (b)) and the equation-of-state $w$
(fig. (c)) upon $\ln(a/a_{0})$. Here and in all the graphs of this
paper describing scale factor $a$ dependences, $a(t)$ is
normalized such that at the end point of the described process
$a(t_{end})=a_{0}$. The values of $V_{1}$ and $V_{2}$ are as in
Fig.1a. The graphs correspond to the initial conditions
$\phi_{in}=-50M_{p}$, $\dot{\phi}_{in}=-5M^2b_{g}^{-1/2}$. The
early universe evolution is governed by the almost exponential
potential (see Fig.1a) providing the power low inflation
($w\approx -0.95$ interval in fig.(c)). After transition to the
late time universe the scalar $\phi$ increases with the rate
typical for a quintessence scenario. Later on  the cosmological
constant $\Lambda_{1}$ becomes a dominated component of the dark
energy that is displayed by the infinite region where $w\approx
-1$ in fig.(c).}\label{fig2}
\end{figure}

\begin{figure}[htb]
\begin{center}
\includegraphics[width=14.0cm,height=8.0cm]{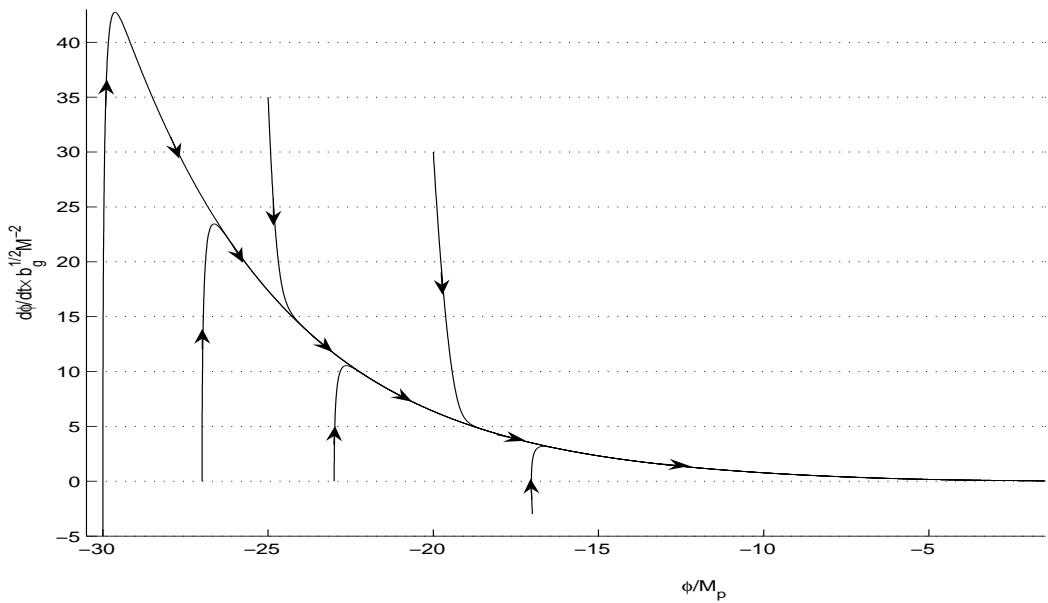}
\end{center}
\caption{Phase portrait (plot of $\frac{d\phi}{dt}$ versus $\phi$)
for the model with $b_{g}V_{1}>2V_{2}$.   All trajectories
 approach the attractor which in its turn
asymptotically (as $\phi\rightarrow\infty$) takes the form of the
straight line $\dot\phi =0$.}\label{fig3}
\end{figure}

The smallness of the observable cosmological constant, in this
model given by $\Lambda_1$, is known as the new cosmological
constant problem\cite{Weinberg2}. There are two ways to provide
the observable order of magnitude of the present day vacuum energy
density by an appropriate choice of the parameters of the theory.

(a) If $V_{2}<0$ then there is no need for $V_{1}$ and $V_{2}$ to
be small: it is enough that $b_{g}V_{1}<|V_{2}|$ and $V_{1}/
|V_{2}|\ll 1 $.  This possibility is a kind of  {\it seesaw}
 mechanism\cite{G1},\cite{seesaw}). For instance, if $V_{1}$ is
determined by the
 energy scale of electroweak symmetry breaking $V_{1}\sim
(10^{3}GeV)^{4}$ and $V_{2}$ is determined by the Planck scale
$V_{2} \sim (10^{18}GeV)^{4}$ then $\Lambda_{1}\sim
(10^{-3}eV)^{4}$. The range of the possible scale of the
dimensionless parameter $b_{g}$ remains very broad.

(b) If $V_2>0$ or alternatively $V_2<0$ and $b_gV_{1}>|V_{2}|$
then $\Lambda^{(0)}\approx \frac{V_1}{4b_{g}}$. Hence the second
possibility is to choose the {\it dimensionless} parameter
$b_{g}>0$ to be a huge number.   In this case the order of
magnitudes of $V_{1}$ and $V_{2}$ could be either as in the above
case (a) or to be  not too much different (or even of the same
order). For example, if $V_{1}\sim (10^{3}GeV)^{4}$ then for
getting $\Lambda_1\sim (10^{-3}eV)^{4}$ one should assume that
$b_{g}\sim 10^{60}$. Note that $b_{g}$ is the ratio of the
coupling constants of the scalar curvature to the measures
$\sqrt{-g}$ and $\Phi$ in the fundamental action of the theory
(\ref{totaction}). Taking into account our assumption that the
dimensionless parameters $b_g$, $b_{\phi}$, $k$ and $h_i$
$(i=N,E)$ are of the close order of magnitude, their huge values
can be treated as a sort of {\it a correspondence principle} in
the TMT. In fact, using the notations of the general form of the
TMT action (\ref{S}) in the case of the action (\ref{totaction}),
one can conclude that if these dimensionless parameters have the
order of magnitude $\sim 10^{60}$ then the relation between the
"usual" (i.e. entering in the action with the usual measure
$\sqrt{-g}$) Lagrangian $L_2$ and the new one $L_1$ (entering in
the action with the new measure $\Phi$) is  roughly speaking
$L_2\sim 10^{60}L_1$. It seems to be very interesting that such
{\it a correspondence principle may be responsible for the extreme
smallness of the cosmological constant}.

Summing the above analysis  we conclude that  the effective
potential (\ref{Veffvac-delta=0}) provides a possibility for a
cosmological scenario which starts with a power law inflation and
ends with a small cosmological constant $\Lambda_1$. It is very
important that the effective potential (\ref{Veffvac-delta=0})
appears here as the result (in a certain range of parameters) of
the TMT model undergoing the spontaneous breakdown of the global
scale symmetry\footnote{The particular case of this model with
$b_{g}=0$ and $V_{2}<0$ was studied in Ref.\cite{G1}. The
application of the TMT model with explicitly broken global scale
symmetry to the quintessential inflation scenario was discussed in
Ref\cite{K}.}.

Results of numerical solutions for such type of scenario are
presented in Figs.2 and 3 ($V_{1}=10M^{4}$, $V_{2}=4b_{g}M^{4}$)
The early universe evolution is governed by the almost exponential
potential (see Fig.1a) providing the power low inflation
($w\approx -0.95$ interval in fig.(c)) with the attractor behavior
of the solutions, see Ref.\cite{Halliwell}. After transition to
the late time universe the scalar $\phi$ increases with the rate
typical for a quintessence scenario. Later on  the cosmological
constant $\Lambda_{1}$ becomes a dominated component of the dark
energy that is displayed by the infinite region where $w\approx
-1$ in fig.(c). The phase portrait in Fig.3 shows that all the
trajectories started with $|\phi|\gg M_{p}$ quickly approach the
attractor which asymptotically (as $\phi\rightarrow\infty$) takes
the form of the straight line $\dot\phi =0$. Qualitatively similar
results are obtained also when $V_{1}$ is positive but $V_{2}$ is
negative.

\section{Dilaton - Gravity model with $V_{1}>0$ and $V_{2}<b_{g}V_{1}<2V_{2}$}
\subsection{ Cosmological Dynamics in the Case $\delta=0$}

In this case the effective potential (\ref{Veffvac-delta=0}) has
the minimum (see Fig.1b)
\begin{equation}
V^{(0)}_{eff}(\phi_{min})=\frac{V_{2}}{b_{g}^{2}} \qquad \text{at}
\qquad \phi =\phi_{min}=
-\frac{M_{p}}{2\alpha}\ln\left(\frac{2V_{2}-b_{g}V_{1}}{b_{g}}\right).
\label{minVeff}
\end{equation}

\begin{figure}[htb]
\begin{center}
\includegraphics[width=16.0cm,height=8.0cm]{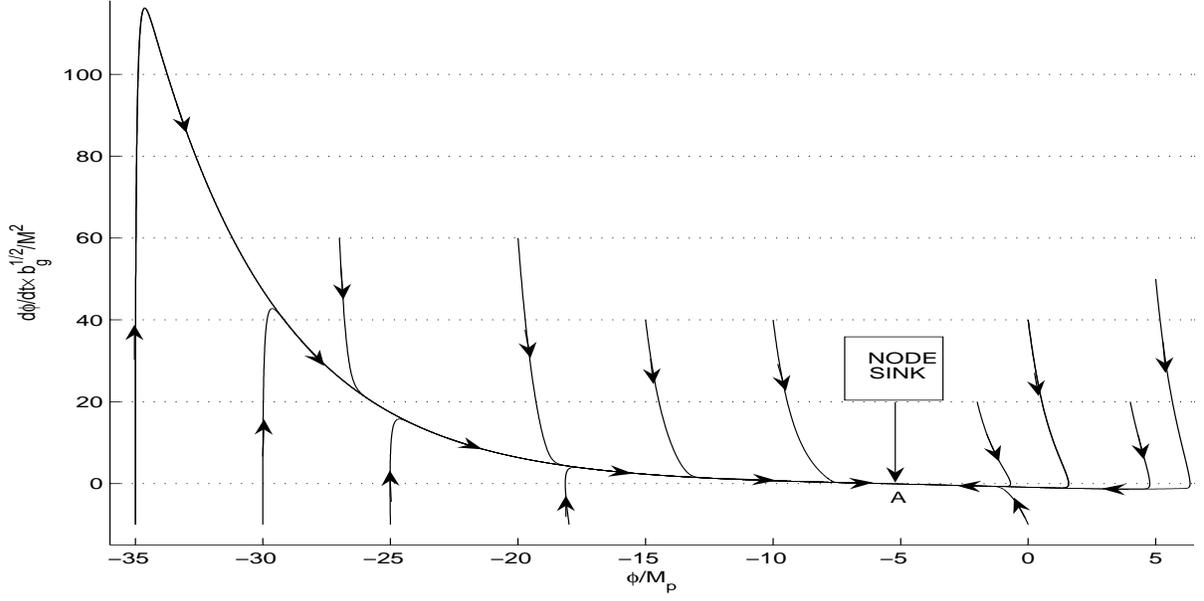}
\end{center}
\caption{Typical behavior of the phase trajectories in the plane
($\phi$,$\frac{d\phi}{dt}$) for the model with $b_{g}V_{1}<2V_{2}$
and $V_{1}>0$ (the parameters are chosen here as in Fig.1b ).
Trajectories started anywhere in the phase plane in a finite time
end up at  the same point $A(-5.7,0)$ which is a node sink.
However there exist two attractors ending up at $A$, one from the
left and other from the right in such a way that all phase
trajectories starting with $|\phi|\gg M_{p}$ quickly approach
these attractors.}\label{fig4}
\end{figure}

For the choice of the parameters as in Fig.1b, i.e.
$V_{1}=10M^{4}$ and $V_{2}=9.9b_{g}M^{4}$, the minimum is located
at $\phi_{min}=-5.7M_p$. The character of the phase portrait one
can see in Fig.\ref{fig4}.

\begin{figure}[htb]
\begin{center}
\includegraphics[width=18.0cm,height=6.0cm]{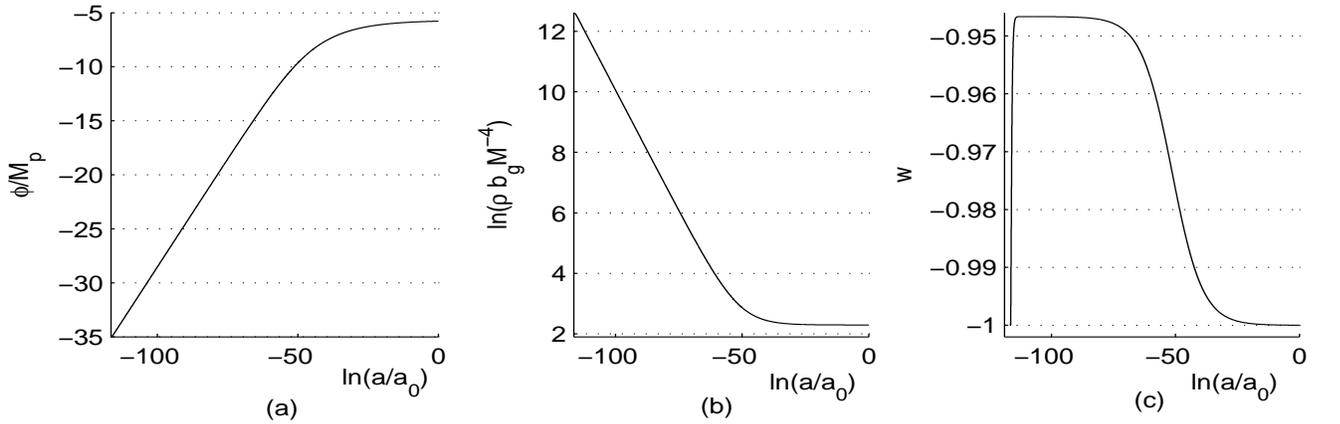}
\end{center}
\caption{Cosmological dynamics in the model with
$b_{g}V_{1}<2V_{2}$ and $V_{1}>0$: typical dependence of $\phi$
(Fig.(a)), the energy density $\rho$ (Fig.(b)) and the
equation-of-state $w$ (Fig.(c)) upon $\ln(a/a_{0})$ where the
scale factor $a(t)$  normalized as in Fig.2. The graphs correspond
to the initial conditions $\phi_{in}=-35M_{p}$,
$\dot{\phi}_{in}=-10b_{g}^{-1/2}M^{2}$. The early universe
evolution is governed by an almost exponential potential (see
Fig.1b) providing the power low inflation ($w\approx -0.95$
interval in Fig.(c)). After arriving the minimum of the potential
at  $\phi_{min}= -5.7M_p$ (see Fig.1b and
 the point $A(-5.7M_p,0)$ of the phase plane in Fig.4) the scalar $\phi$
remains constant. At this stage the dynamics of the universe is
governed by the  constant energy density $\rho
=V^{(0)}_{eff}(\phi_{min})$  (see the appropriate intervals $\rho
=const$ in Fig.(b) and  $w= -1$ in Fig.(c)).}\label{fig5}
\end{figure}

For the early universe as $\phi\ll -M_{p}$, similar to what we
have seen in the case of the monotonically decreasing potential in
Sec.IV, the model implies the power low inflation. However, the
phase portrait Fig.\ref{fig4} shows that now all solutions end up
without oscillations at the minimum $\phi_{min}=-5.7M_p$ with
$\frac{d\phi}{dt}=0$. In this final state of the scalar field
$\phi$, the evolution of the universe is governed by the
cosmological constant $V^{(0)}_{eff}(\phi_{min})$ determined by
Eq.(\ref{minVeff}). For some details of the cosmological dynamics
see Fig.\ref{fig5}. The desirable smallness of
$V^{(0)}_{eff}(\phi_{min})$ can be provided without fine tuning of
the dimensionfull parameters by the way similar to what was done
in item (b) of Sec.IV.  The absence of appreciable oscillations in
the minimum is explained by the following two reasons: a) the
non-zero friction at the minimum determined by the cosmological
constant $V^{(0)}_{eff}(\phi_{min})$; b) the shape of the
potential near to minimum is too flat.

The described properties of the model are evident enough after the
shape of the effective potential (\ref{Veff1}) in the Einstein
frame is obtained in TMT. Nevertheless we have presented them here
because this model is a particular (fine tuned) case of an
appropriate model with $\delta\neq 0$ studied in the next
subsection where we will demonstrate a possibility of states with
$w<-1$ without phantom in the original action.

\pagebreak
\subsection{ Dilaton - Gravity Cosmological Dynamics in the Case
$\delta\neq 0$. Super-acceleration}

We return now to the more general models of the scalar sector (see
Sec.III) where the parameter $\delta$, defined by
Eq.(\ref{delta}), is non zero. However we still ignore here the
Higgs dynamics. The appropriate simplified version of the
cosmological dynamics is of interest to us because  it allows,
without non relevant complications, to demonstrate the possibility
of solutions for the late time universe with equation-of-state
$w<-1$ (super-acceleration), which are favored by the present
data\cite{superaccel},\cite{Starobinsky}

\begin{figure}[htb]
\begin{center}
\includegraphics[width=17.0cm,height=12.0cm]{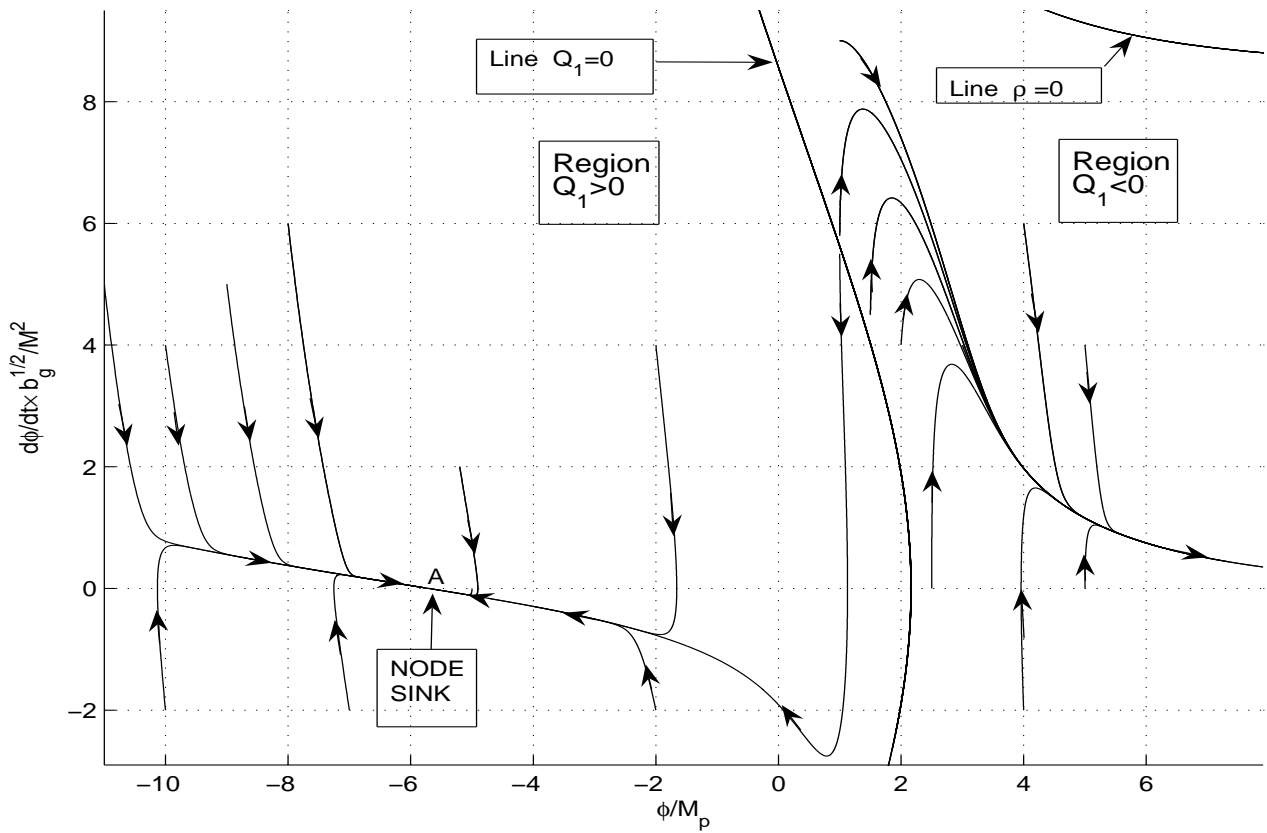}
\end{center}
\caption{The phase portrait for the model with $\alpha =0.2$,
$\delta =0.1$, $V_{1}=10M^{4}$ and $V_{2}=9.9b_{g}M^{4}$. The
phase plane is divided up into two dynamically disconnected
regions by the line $Q_{1}(\phi,\dot{\phi})=0$. To the left of
this line $Q_{1}>0$ and to the right $Q_{1}<0$. The phase portrait
in the left hand side, i.e. in the region $Q_{1}>0$, corresponds
to processes similar to those of Sec.V.A. Trajectories  in the
right hand side of phase portrait, i.e. in the region $Q_{1}<0$,
 correspond to processes with super-accelerating
expansion of the universe.}\label{fig6}
\end{figure}

\begin{figure}[htb]
\begin{center}
\includegraphics[width=16.0cm,height=5.0cm]{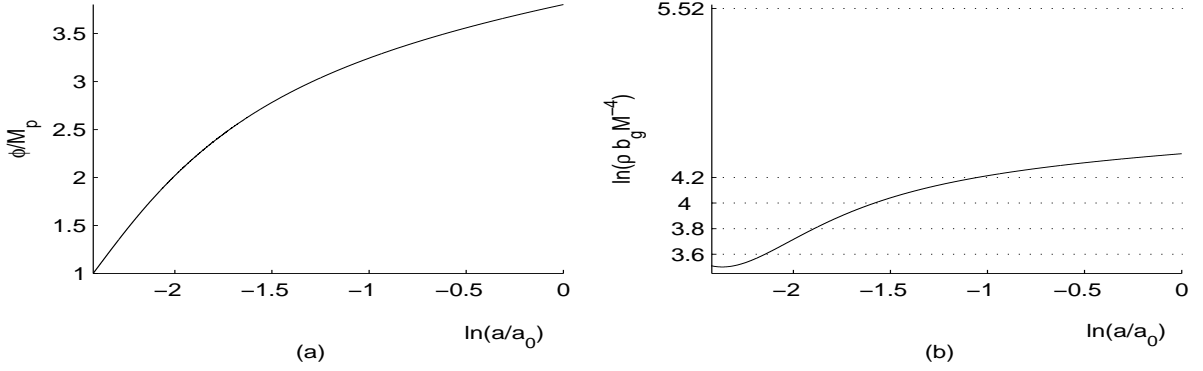}
\end{center}
\caption{ Scalar factor dependence of $\phi$  (Fig.(a)) and
 of the energy density $\rho$, defined by Eq.(\ref{rho1}),
(Fig(b)) in the regime of super-acceleration. Both graphs
correspond to the initial conditions $\phi_{in}=M_{p}$,
$\dot\phi_{in} =9M^4/\sqrt{b_g}$; $\rho$ increases approaching
asymptotically $\Lambda_{2}=
\frac{M^{4}}{b_{g}}e^{5.52}$.}\label{fig7}
\end{figure}
\begin{figure}[htb]
\begin{center}
\includegraphics[width=8.0cm,height=5.0cm]{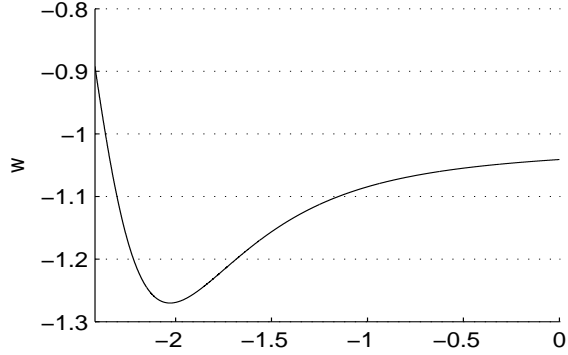}
\end{center}
\caption{The scale factor dependence of the equation-of-state $w$
for the initial conditions $\phi_{in}=M_{p}$, $\dot\phi_{in}
=9M^4/\sqrt{b_g}$.} \label{fig8}
\end{figure}

So, we consider now the dynamics of the FRW cosmology described by
Eqs.(\ref{cosm-phi}), (\ref{rho1}), (\ref{phi1}),
(\ref{Q1})-(\ref{Q3}) where $V_{1}$ and $V_{2}$ are constants and
$\upsilon\equiv 0$. Adding the non zero $\delta$ to the parameter
space enlarges significantly the number of classes of
qualitatively different models. Among the new possibilities the
most attractive one is the class of models giving rise to
solutions for the late time universe with equation-of-state $w<-1$
{\it without resorting to negative kinetic term in the fundamental
action} as in the phantom field models\cite{phantom}.

Before choosing the appropriate parameters for numerical studies,
let us start from the analysis of Eq.(\ref{phi1}). The interesting
feature of this equation is that for certain range of the
parameters, each of the factors $Q_{i}(\phi,X)$ \quad ($i=1,2,3$)
can get to zero. Equation $Q_{i}(\phi,X)=0$ determines a line in
the phase plane $(\phi,\dot\phi)$. In terms of a mechanical
interpretation of Eq.(\ref{phi1}), the change of the sign of
$Q_{1}$ can be treated as the change of the mass of "the
particle". Therefore one can think of situation where "the
particle" climbs up in the potential with acceleration. It turns
out that when the scalar field is behaving in this way, the flat
FRW universe undergoes a super-acceleration.

There are a lot of  sets of parameters providing this effect. For
example we are demonstrating here this effect with the following
set of the parameters of the original action(\ref{totaction}):
$\alpha =0.2$, $V_{1}=10M^{4}$ and $V_{2}=9.9b_{g}M^{4}$ used in
Sec.IV but now we choose $\delta =0.1$. The results of the
numerical solution are presented in Figs.6-8.

The phase plane, Fig.6, is divided into two dynamically
disconnected regions by the line $Q_{1}(\phi,X)=0$. To the left of
this line $Q_{1}>0$ and to the right  $Q_{1}<0$. Comparing
carefully the phase portrait in the region $Q_{1}>0$ with that in
Fig.\ref{fig4} of the previous subsection, one can see an effect
of $\delta\neq 0$ on the shape of phase trajectories. However the
general structure of these two phase portraits is very similar. In
particular, they have the same node sink $A(-5.7M_{p},0)$. At this
point "the force" equals zero since $Q_{3}|_A=0$. The value $\phi
=-5.7M_{p}$ coincides with the position of the minimum of
$V_{eff}^{(0)}(\phi)$ because in the limit $\dot{\phi}\rightarrow
0$ the role of the terms proportional to $\delta$ is negligible.
Among trajectories converging to node $A$ there are also
trajectories corresponding to a power low inflation of the early
universe, which is just a generalization to the case $\delta\neq
0$ of the similar result discussed in the previous subsection.

On the right side of the phase plane Fig.6, i.e. in the region
$Q_{1}<0$, all trajectories approach the attractor which in its
turn asymptotically (as $\phi\rightarrow\infty$) takes the form of
the straight line $\dot\phi =0$.

For a particular choice of the initial data $\phi_{in}=M_{p}$,
$\dot\phi_{in} =9M^4/\sqrt{b_g}$, the features of the solution of
the equations of motion are presented in Figs.7 and 8. The main
features of the solution as we observe from the figures are the
following: 1) $\phi$ slowly increases in time; 2) the energy
density $\rho$ slowly increases approaching the constant $\Lambda
=\Lambda_{2}$  defined by the same formula as in
Eq.(\ref{lambda-without-ferm-delta=0}), see also Fig.1b; for the
chosen parameters $\Lambda_{2}\approx\frac{M^{4}}{b_{g}}e^{5.52}$.
\quad
 3) $w\equiv p/\rho$ is
less than $-1$ and asymptotically approaches $-1$ from below.

Qualitative understanding of the fact that the energy density
$\rho$ approaches $\Lambda_{2}$  during the super-accelerated
expansion of the universe is based on the shape of the effective
potential, Fig.1b, that would be in the model with $\delta =0$.
However, the possibility of climbing up in the potential with
acceleration can be understood only due to the effect of changing
sign of $Q_{1}$, Eq.(\ref{Q1}), which becomes possible in the
model with $\delta\neq 0$. Due to such mechanism of the
super-acceleration it becomes clear why qualitatively the same
behavior one observes for all initial conditions
$(\phi_{in},\dot{\phi}_{in})$ disposed in the region $Q_{1}<0$.

\begin{figure}[htb]
\begin{center}
\includegraphics[width=15.0cm,height=8.0cm]{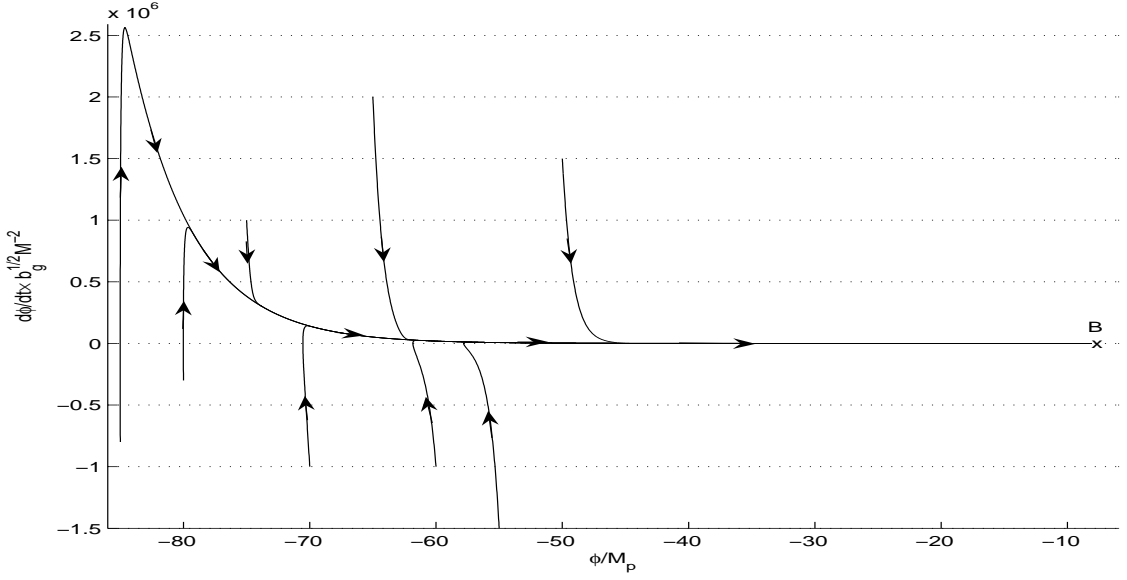}
\end{center}
\caption{Phase portrait (plot of $\frac{d\phi}{dt}$ versus $\phi$)
for the model with $V_{1}<0$ and $V_{2}<0$.   All trajectories
started with $|\phi|\gg M_{p}$ quickly approach the attractor long
before entering the oscillatory regime. The region appropriate to
the oscillatory regime is marked by point $B$. The oscillation
spiral is not visible in Fig.11 because of the choice of the scale
along the axis Y.}\label{fig9}
\end{figure}

\begin{figure}[htb]
\begin{center}
\includegraphics[width=16.0cm,height=7.0cm]{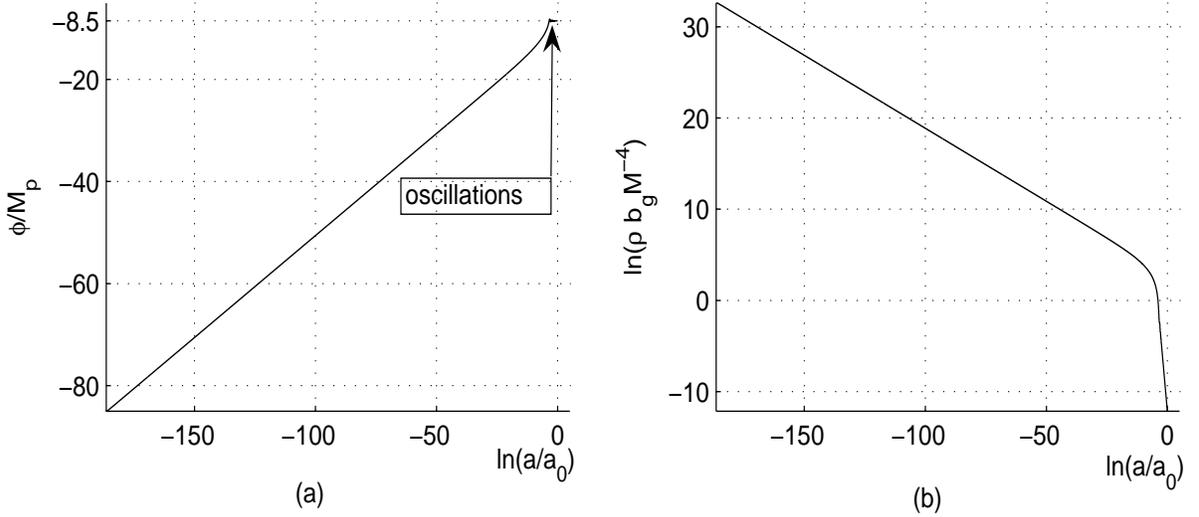}
\end{center}
\caption{(a) In the model with $V_{1}<0$ the power law inflation
ends with damped oscillations of $\phi$ around $\phi_{0}$
determined by Eq.(\ref{Veff=0}). For the choice $V_{1}=-30M^{4}$
Eq.(\ref{Veff=0}) gives $\phi_{0}=-8.5M_p$ . (b) The exit from the
early inflation is accompanied with approaching zero of the energy
density $\rho$. The graphs correspond to the evolution which
starts from the initial
  values  $\phi_{in} =-85M_{p}$, $\dot{\phi}_{in}=
  -8\cdot 10^5M^2/\sqrt{b_g}$. }\label{fig10}
\end{figure}

\begin{figure}[htb]
\includegraphics[width=8.5cm,height=8.0cm]{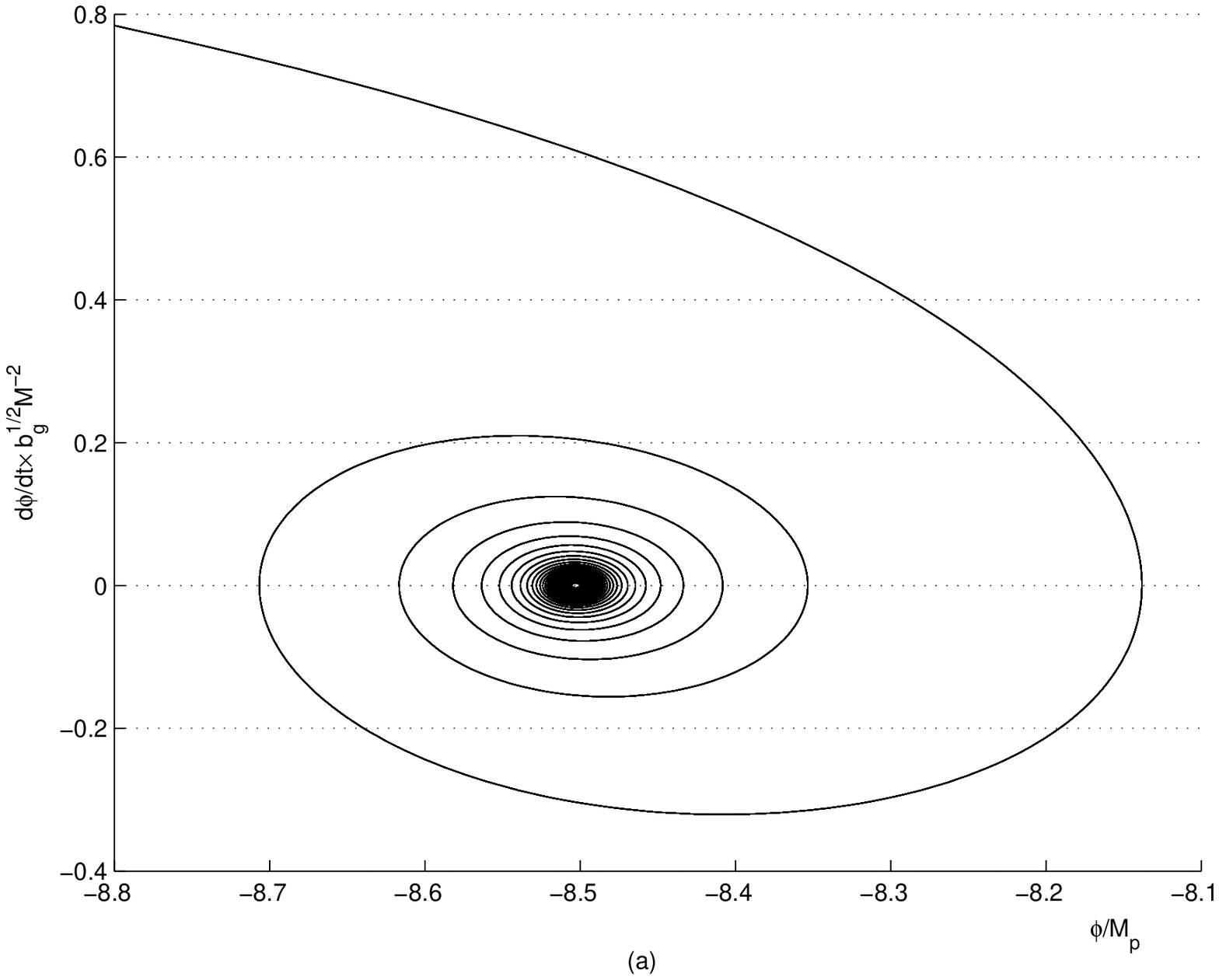}
\includegraphics[width=8.5cm,height=8.0cm]{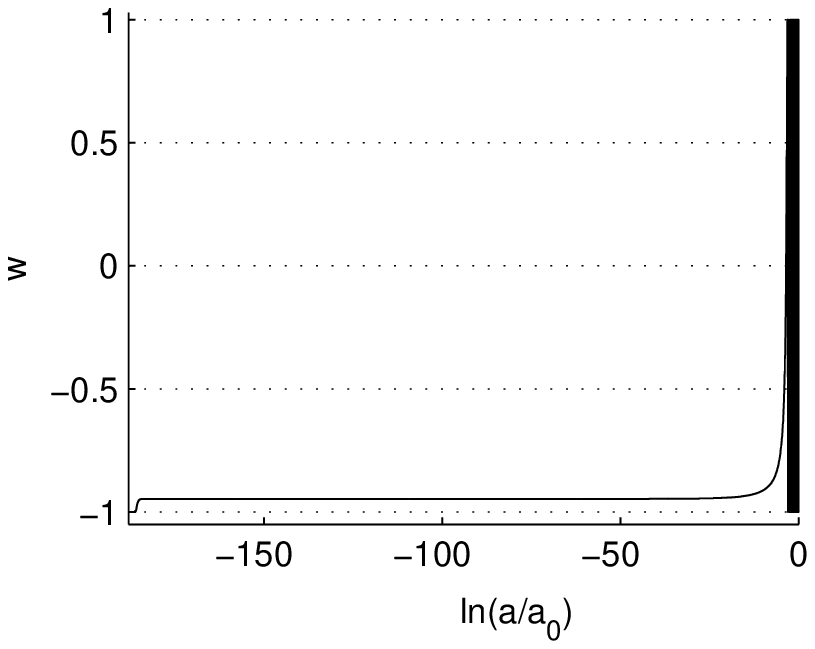}
\caption{Fig.(a) zoom in on the oscillatory regime which marked by
point B in Fig.9. \, (b) Equation-of-state $w=p/\rho$ as function
of the scale factor for the parameters and initial conditions as
in Fig.10. Most of the time the expansion of the universe is a
power law inflation with almost constant $w\approx -0.95$; $w$
oscillates between $-1$ and $1$ at the exit from inflation stage,
i.e. as $\phi\rightarrow\phi_{0}$ and $\rho\rightarrow
0$}\label{fig11}
\end{figure}

\section{Dilaton - Gravity Cosmological Dynamics in the Model with $V_{1}<0$ and
$V_{2}<0$.
 The Old Cosmological Constant Problem and
Non Applicability of the Weinberg Theorem}

The most remarkable features of the effective potential
(\ref{Veffvac-delta=0}) is that it is proportional to the square
of $V_{1}+ M^{4}e^{-2\alpha\phi/M_{p}}$. Due to this, as $V_{1}<0$
and $V_{2}<0$, {\it the effective potential has a minimum where it
equals zero automatically}, without any further tuning of the
parameters $V_{1}$ and $V_{2}$ (see also Fig.1c). This occurs in
the process of evolution of the field $\phi$ at the value of $\phi
=\phi_{0}$ where
\begin{equation}
V_{1}+ M^{4}e^{-2\alpha\phi_{0}/M_{p}}=0 \label{Veff=0}.
\end{equation}
This means that the universe evolves into the state with zero
cosmological constant without tuning  parameters of the model.

 If such type of the structure for the scalar field potential in a
usual (non TMT) model would be chosen "by hand" it would be a sort
of fine tuning. But in our TMT model it is not the starting point,
{\it it is rather a result} obtained in the Einstein frame of TMT
models with spontaneously broken global scale symmetry including
the shift symmetry $\phi\rightarrow \phi +const$. Later on we will
see the same effect in more general models including also the
Higgs field as well as in models with $\delta\neq 0$. Note that
the assumption of scale invariance is not necessary for the effect
of appearance of  the perfect square in the effective potential in
the Einstein frame and therefore for the described mechanism of
disappearance of the cosmological constant, see
Refs.\cite{GK2}-\cite{G1} and Appendix E.

On the first glance this effect contradicts the no-go Weinberg
theorem\cite{Weinberg1} which states that there cannot exist a
field theory model where the cosmological constant is zero without
fine tuning. Recall that one of the basic assumptions of this
no-go theorem is that all fields in the vacuum must be constant.
However, this is not the case in TMT.   In fact, in the vacuum
determined by Eq.(\ref{Veff=0}) the scalar field
$\zeta\equiv\frac{\Phi}{\sqrt{-g}}$ is non zero, see
Eq.(\ref{zeta-without-ferm-delta=0}). The latter is possible only
if all the $\varphi_{a}$ ($a=1,2,3,4$) fields (in the definition
of $\Phi$ by means of Eq.(\ref{Phi})) or the 3-index potential
$A_{\alpha\beta\gamma}$ (when using the definition of $\Phi$ by
means of Eq.(\ref{Aabg})) have non vanishing space-time gradients.
Moreover, exactly in the vacuum $\phi =\phi_{0}$ the scalar field
$\zeta$ has a singularity. However, in the conformal Einstein
frame all physical quantities are well defined and this
singularity manifests itself only in the vanishing of the vacuum
energy density. We conclude therefore that the Weinberg
theorem\cite{Weinberg1} is not applicable in the context of the
TMT models studied here. In fact, the possibility of such type of
situation was suspected by S. Weinberg in the footnote 8 of his
review\cite{Weinberg1} where he pointed out that when using a
3-index potential with non constant vacuum expectation value, his
theorem does not apply.

The results of numerical solutions are evident enough, but we want
to present them here because they will be useful for comparison
with the results of the next section.  For the potential of Fig.1c
(where we have chosen $V_{1}=-30M^{4}$ and $V_{2}=-50b_{g}M^{4}$)
the results of numerical solutions are presented in figures
9,10,11.

\section{Higgs-Inflaton Symbiosis: Inflation, Exit From Inflation
And New Cosmological Mechanism for gauge symmetry breaking}

\subsection{General Discussion}

Turning now to the general case of the scalar sector, Sec.III,
where both dilaton and Higgs dynamics are taken into account, we
have to make some nontrivial choices for the Higgs pre-potentials
$V_{1}(\upsilon)$ and $V_{2}(\upsilon)$. We will see that for
achieving a gauge symmetry breakdown there is no need for
tachyonic mass terms in $V_{1}(\upsilon)$ and $V_{2}(\upsilon)$
and  even for selfinteraction terms. We make the following
simplest  choice
\begin{equation}
V_{1}=V_{1}^{(0)}+\mu_{1}^{2}\upsilon^{2}; \qquad
V_{2}=V_{2}^{(0)}+\mu_{2}^{2}\upsilon^{2} \label{V1V2Higgs}
\end{equation}
assuming
\begin{equation}
 \mu_{1}^{2}>0, \qquad \mu_{2}^{2}>0 \qquad \text{and}
 \qquad V_{1}^{(0)}<0, \qquad V_{2}^{(0)}<0
\label{m2>0}
\end{equation}

In order to understand qualitatively what is the mechanism of the
gauge symmetry breaking it is useful to start from the model with
$\delta =0$. Then from Eqs.(\ref{Veff1}) and (\ref{constr-vac}) we
get for $\delta =0$
\begin{equation}
V_{eff}(\phi,\upsilon)
=\frac{[V_{1}^{(0)}+\mu_{1}^{2}\upsilon^{2}+M^{4}e^{-2\alpha\phi/M_{p}}]^{2}}
{4\left[\left(b_{g}V_{1}^{(0)}-V_{2}^{(0)}\right)+
\left(b_{g}\mu_{1}^{2}-\mu_{2}^{2}\right)\upsilon^{2}+b_{g}M^{4}e^{-2\alpha\phi/M_{p}}
\right]}\label{Veff+higgs}
\end{equation}

Even in this simple model the effective potential of the scalar
sector contains a very non-trivial coupling between dilaton,
playing the role of inflaton, and Higgs fields. It is easy to see
that this coupling disappears in the limit $\phi\ll -M_{p}$
corresponding to the very early universe:
\begin{equation}
V_{eff}(\phi,\upsilon)|_{early\,universe}\approx
\frac{1}{4b_{g}}M^{4}e^{-2\alpha\phi/M_{p}}+
\frac{1}{4b_{g}^{2}}(b_{g}\mu_{1}^{2}+\mu_{2}^{2})\upsilon^{2}+const.
\label{Veff+higgs-early}
\end{equation}

Therefore, in the very early universe, the minimization of
$V_{eff}$ is achieved at $\upsilon =0$ which means that the Higgs
field is in an unbroken symmetry phase without resorting to high
temperature effects. So, in the very early cosmological epoch, we
have $V_{1}=V_{1}^{(0)}<0$, \, $V_{2}=V_{2}^{(0)}<0$ and
$V_{eff}\approx \frac{1}{4b_{g}}M^{4}e^{-2\alpha\phi/M_{p}}$ and
therefore one can use the results of Sec.VI in what it concerns to
the early inflationary epoch. It turns out, however, that this
preliminary analysis is unable to give a complete scenario of the
early inflationary epoch, see the next two subsections.

If we want $V_{eff}(\phi,\upsilon)$ to be positive definite and
the symmetry broken state to be the absolute minimum we have to
assume in addition that
\begin{equation}
b_{g}V_{1}^{(0)}\geq V_{2}^{(0)}, \qquad
b_{g}\mu_{1}^{2}\geq\mu_{2}^{2}\label{bmu1>bmu2}
\end{equation}
This will be our choice.

 It is interesting that in the
particular case $b_{g}\mu_{1}^{2}=\mu_{2}^{2}$ the effective
potential (\ref{Veff+higgs}) can be written in  the form of the
Ginzburg-Landau type potential for the Higgs field $\upsilon$
\begin{equation}
V_{eff}(\phi,\upsilon)|_{b_{g}\mu_{1}^{2}=\mu_{2}^{2}}=
\lambda(\phi)[\upsilon^{2}-\sigma^{2}(\phi)]^{2},
\label{V-bmu1=bmu2}
\end{equation}
 where the coupling "constant" $\lambda$ and the mass parameter
 $\sigma$ are the following functions of $\phi$:
\begin{equation}
\lambda(\phi)=\mu_{1}^{4}\left[4\left(b_{g}V_{1}^{(0)}-V_{2}^{(0)}
+b_{g}M^{4}e^{-2\alpha\phi/M_{p}}\right)\right]^{-1}; \qquad
\sigma^{2}(\phi)=\mu_{1}^{-2}\left(|V_{1}^{(0)}|-M^{4}e^{-2\alpha\phi/M_{p}}\right).
\label{lambda-bmu1=bmu2}
\end{equation}

If the Higgs field were to remain in the unbroken phase $\upsilon
=0$, then according to Sec.VI, the power law inflation would be
ended with oscillations of the $\phi$-field accompanied with
approaching zero of the energy density. However decreasing
$M^{4}e^{-2\alpha\phi/M_{p}}$ causes that
$\partial^{2}V_{eff}/\partial\upsilon^{2}$ gets to be zero. The
continuation of this process changes the shape of the Higgs
dependence of the effective scalar sector potential such that
$\upsilon =0$ turns into local maximum and a nonzero $\upsilon
=<\upsilon>\neq 0$ appears as the {\it true minimum} of the
effective scalar field potential (\ref{Veff+higgs}) {\it in the}
$\upsilon$ {\it direction}. This happens as $<\upsilon>^{2}$ and
$\phi =\phi_{0}$ satisfy the equation of the following curve
\begin{equation}
V_{1}^{(0)}+\mu_{1}^{2}<\upsilon>^{2}+M^{4}e^{-2\alpha\phi_{0}/M_{p}}=0,
\label{min-v}
\end{equation}
which at the same time, {\it is also the condition for the minimum
in the} $\phi$ {\it direction} (see also
Eqs.(\ref{phi1})-(\ref{Q-H}) as $\delta =0$). The remarkable
feature of the effective potential (\ref{Veff+higgs}) is that in
this minimum it {\it equals zero without fine tuning of the
parameters of the action}. This vacuum with zero cosmological
constant is degenerate. Notice however that this degeneracy has no
relation to any symmetry of the action.

This degeneracy is a particular manifestation of the well known
more general feature of the systems with two interacting scalar
fields: if both of them are dynamically important then as noted in
Ref.\cite{Liddle}, there is no attractor behavior giving a unique
route into the potential minimum, as it happens in the single
field case. However, numerical and analytic solutions  show that
in spite of this general statement, for the system under
consideration, in a broad enough range of the parameters $\alpha$,
$\delta$, $V_{1}^{(0)}<0$, \, $V_{2}^{(0)}<0$, \, $\mu_{1}^{2}>0$
and $\mu_{2}^{2}>0$, both fields, i.e $\phi$ and $\upsilon$, are
dynamically important but there exists {\it an attractor
behavior}. As a result of this there is only one very short
segment  in the line (\ref{min-v}) where all the phase
trajectories end in the process of the cosmological evolution. In
other words, the magnitudes of $<\upsilon >^{2}$ and $\phi_{0}$
  are determined by the values of the parameters of the model but
dependence upon the initial values  $\phi_{in}$, \,
$\dot{\phi}_{in}$, \, $\upsilon_{in}$ and $\dot{\upsilon}_{in}$ is
extremely weak.

For the study of numerical and analytical solutions as well as for
detailed qualitative analysis of all the stages of the scalar
sector evolution during the cosmological expansion it will be
sometimes convenient to work with the equations of motion
(\ref{phi1})-(\ref{Q-H}) written in terms of dimensionless
parameters and variables. Then Eqs.(\ref{phi1}) and (\ref{H1})
take the following form (here we restrict ourselves with the
choice\footnote{One can show that in models with $\delta\neq 0$
the results are very similar: this is because the transition to
$<\upsilon >$ phase is accompanied with $\dot{\phi}\rightarrow
0$.} $\delta =0$)
\begin{equation}
\frac{d^2\varphi}{d\tau^2}+\sqrt{3\epsilon}\frac{d\varphi}{d\tau}
-\alpha e^{-2\alpha\varphi}
\frac{\left(\tilde{V}_1(\tilde{\upsilon})+e^{-2\alpha\varphi}\right)
\left(\tilde{V}_1(\tilde{\upsilon})-2\tilde{V}_2(\tilde{\upsilon})+e^{-2\alpha\varphi}\right)}
{2\left(\tilde{V}_1(\tilde{\upsilon})-\tilde{V}_2(\tilde{\upsilon})+e^{-2\alpha\varphi}\right)^2}=0
\label{phi-dimensionless}
\end{equation}
\begin{eqnarray}
&&\frac{d^2\tilde{\upsilon}}{d\tau^2}+\sqrt{3\epsilon}\frac{d\tilde{\upsilon}}{d\tau}
\nonumber\\
&+&\left(\frac{M_p}{M}\right)^2\frac{\tilde{V}_1(\tilde{\upsilon})+
e^{-2\alpha\varphi}}{2\left(\tilde{V}_1(\tilde{\upsilon})-\tilde{V}_2(\tilde{\upsilon})+e^{-2\alpha\varphi}\right)^2}
\left[\tilde{\mu}^2_1\left(\tilde{V}_1(\tilde{\upsilon})-2\tilde{V}_2(\tilde{\upsilon})+e^{-2\alpha\varphi}\right)
+\tilde{\mu}^2_2\left(\tilde{V}_1(\tilde{\upsilon})+e^{-2\alpha\varphi}\right)\right]\tilde{\upsilon}=0
\label{H-dimensionless}
\end{eqnarray}
where Eq.(\ref{cosm-phi}) and the following dimesionless
parameters and variables have been used
\begin{eqnarray}
&&\tilde{V}_1^{(0)}=\frac{V_1^{(0)}}{M^4}; \quad
\tilde{V}_2^{(0)}=\frac{V_2^{(0)}}{b_gM^4}; \quad
\tilde{\mu}_1^2=\frac{\mu_1^2}{M^2}; \quad
\tilde{\mu}_2^2=\frac{\mu_2^2}{b_gM^2}; \quad
\nonumber\\
&&\tau =\frac{M^2}{M_p\sqrt{b_g}}t; \qquad \varphi
=\frac{\phi}{M_p}; \quad \tilde{\upsilon}=\frac{\upsilon}{M};\quad
\tilde{V}_i(\tilde{\upsilon})=\tilde{V}_i^{(0)}+\tilde{\mu}_i^2\tilde{\upsilon}^2,
\quad i=1,2 \label{notations-dimensionless}
\end{eqnarray}
and
\begin{equation}
\epsilon =\frac{b_g}{M^4}\rho
=\tilde{X}+\frac{1}{2}\left(\frac{M}{M_p}\right)^2\left(\frac{d\tilde{\upsilon}}{d\tau}\right)^2
+V_{eff}^{(0)}(\varphi,\tilde{\upsilon}),
\label{rho-dimensionless}
\end{equation}
where
\begin{equation}
V_{eff}(\varphi,\tilde{\upsilon})=\frac{\left(\tilde{V}_1^{(0)}+
\tilde{\mu}_1^2\tilde{\upsilon}^2+
e^{-2\alpha\varphi}\right)^2}{4\left[\tilde{V}_1^{(0)}-\tilde{V}_2^{(0)}
+\left(\tilde{\mu}_1^2-\tilde{\mu}_2^2\right)\tilde{\upsilon}^2+e^{-2\alpha\varphi}\right]},
\label{Veff-dimensionless}
\end{equation}
\begin{equation}
\tilde{X}=\frac{1}{2}\left(\frac{d\varphi}{d\tau}\right)^2=\frac{b_g}{M^4}X;
\quad
\left(\frac{d\tilde{\upsilon}}{d\tau}\right)^2=\frac{b_gM_p^2}{M^6}\left(\frac{d\upsilon}{dt}\right)^2
\label{X-dimensionless}
\end{equation}
and $X$ is defined by Eq.(\ref{X}).

In terms of the dimensionless quantities, the equation for the
manifold of the true vacuum (\ref{min-v}) takes the form
\begin{equation}
e^{-2\alpha\varphi_{0}}=|\tilde{V}_1^{(0)}|-\tilde{\mu}_1^2<\tilde{\upsilon}>^2.
\label{vac-min-dimensionless}
\end{equation}
It follows from the positivity of the r.h.s. of
Eq.(\ref{vac-min-dimensionless}) that
\begin{equation}
<\tilde{\upsilon}>^2 \, < \,
\frac{|\tilde{V}_1^{(0)}|}{\tilde{\mu}_1^2}.
\label{upper-bound-on-v}
\end{equation}

Vacuum expectation value $<\tilde{\upsilon}>$ is indeed a minimum
of the effective potential (\ref{Veff-dimensionless}) in the
$\tilde{\upsilon}$ direction if the frequency squared of the
$\tilde{\upsilon}$ oscillations  around the vacuum manifold
(\ref{vac-min-dimensionless}) is positive:
\begin{equation}
\omega^2(<\tilde{\upsilon}>^2)\equiv \frac{\partial^2
V_{eff}}{\partial\tilde{\upsilon}^2}|_{vacuum \,
manifold}=\frac{2\tilde{\mu}_1^2\tilde{\mu}_2^2<\tilde{\upsilon}>^2}
{|\tilde{V}_2^{(0)}|-\tilde{\mu}_2^2<\tilde{\upsilon}>^2} \, > \,
0, \label{Omega-dimensionless-positive}
\end{equation}
We conclude from (\ref{upper-bound-on-v}) and
(\ref{Omega-dimensionless-positive}) that the vacuum expectation
value $<\tilde{\upsilon}>^2$ has to satisfy the upper bound
\begin{equation}
<\tilde{\upsilon}>^2 \quad < \quad
Min\left(\frac{|\tilde{V}_1^{(0)}|}{\tilde{\mu}_1^2}, \quad
\frac{|\tilde{V}_2^{(0)}|}{\tilde{\mu}_2^2}\right)
\label{upper-bound-on-v-full}
\end{equation}

\subsection{Numerical Solutions and Their Physical Meaning }

We are  presenting here the results of the numerical solutions for
equations of motion in the model with the following set of the
parameters:
\begin{equation}
 \alpha =0.2, \quad M=10^{-2}M_p\sim 10^{16}GeV, \quad
 V_{1}^{(0)}=-30M^{4},\,
V_{2}^{(0)}=-50b_{g}M^{4},\, \mu_{1}^{2}=20M^{2},\,
\mu_{2}^{2}=10M^{2}. \label{parameters}
\end{equation}
The choice of the GUT scale for the integration constant $M$ seems
to us to be the most natural. But restricting with a single Higgs
field we proceed actually in a toy model.

\begin{figure}[htb]
\begin{center}
\includegraphics[width=16.0cm,height=8.0cm]{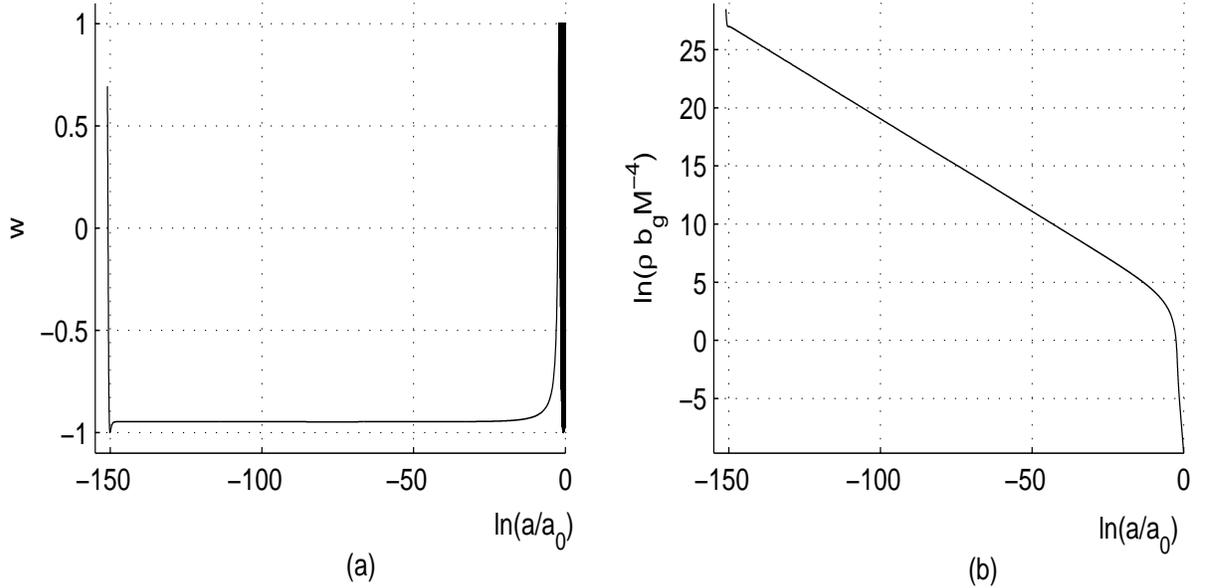}
\end{center}
\caption{Behavior of the equation-of-state $w$, Fig.(a), and the
energy density $\rho$, Fig.(b), as functions of $ln(a/a_0)$,
typical for all sets of the initial conditions,
Eqs(\ref{5-in-cond}). (\ref{6-th-in-cond}). During a short time at
the beginning $w>0$ because of the big value of the inflaton
kinetic energy. The graph of $w$ exhibit that most of the time of
the evolution the equation-of-state is close to the constant
$w\approx -0.95$. This power low inflation ends (i.e. $w\approx
-\frac{1}{3}$. )
 as $\ln a/a_0 \approx -3$. After the end of inflation the energy
 density $\rho$ quickly approaches zero. The
equation-of-state at this period is not well defined due to
oscillations of the Higgs field.}\label{fig12}
\end{figure}

\begin{figure}[htb]
\begin{center}
\includegraphics[width=17.0cm,height=7.0cm]{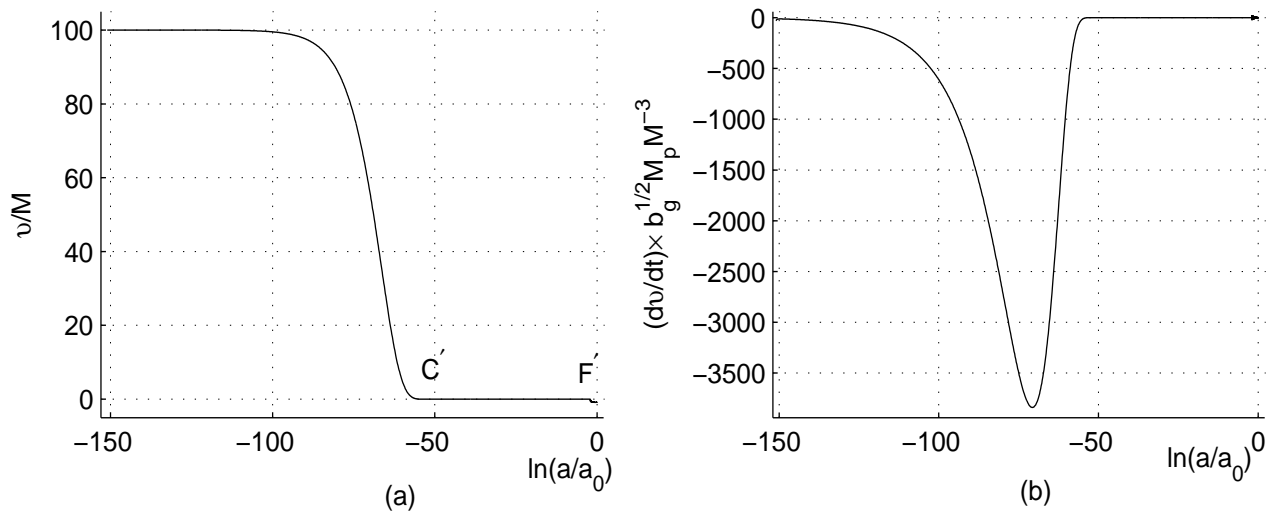}
\end{center}
\caption{The scale factor dependence of the Higgs field
$\upsilon$, Fig.(a), and its time derivative, Fig.(b), typical for
all the initial conditions, Eq.(\ref{5-in-cond}); here the
solution for the set (4) of the initial conditions is displayed.
During the first stage of the inflation (the GSB inflation, see
the main text) $\upsilon$ varies very slowly from its initial
value $\upsilon_{in}$. After the point $C'$, i.e. during the GSR
inflation, $\upsilon$ is very close to zero (oscillates around
$\upsilon =0$).  After the point $F'$ the Higgs-inflaton system
fulfils a transition to "zero cosmological constant and broken
gauge symmetry" phase where the Higgs field oscillates with
decaying amplitude around the vacuum manifold
(\ref{min-v}).}\label{fig13}
\end{figure}

\begin{figure}[htb]
\begin{center}
\includegraphics[width=17.0cm,height=7.0cm]{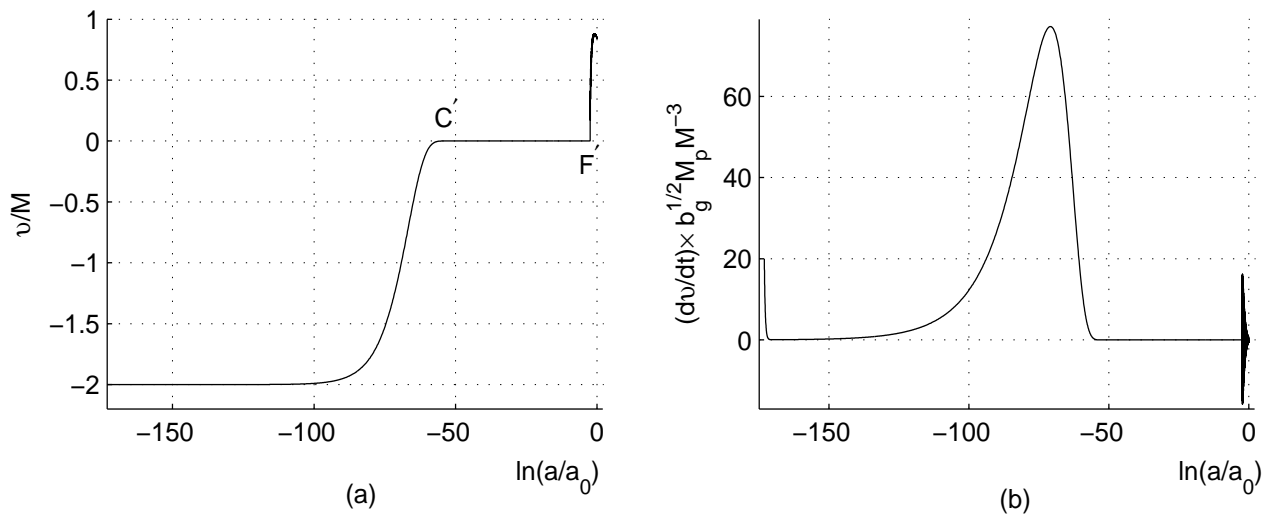}
\end{center}
\caption{The same as in Fig.13 for the set (2) of the initial
conditions.}\label{fig14}
\end{figure}

\begin{figure}[htb]
\begin{center}
\includegraphics[width=16.0cm,height=8.0cm]{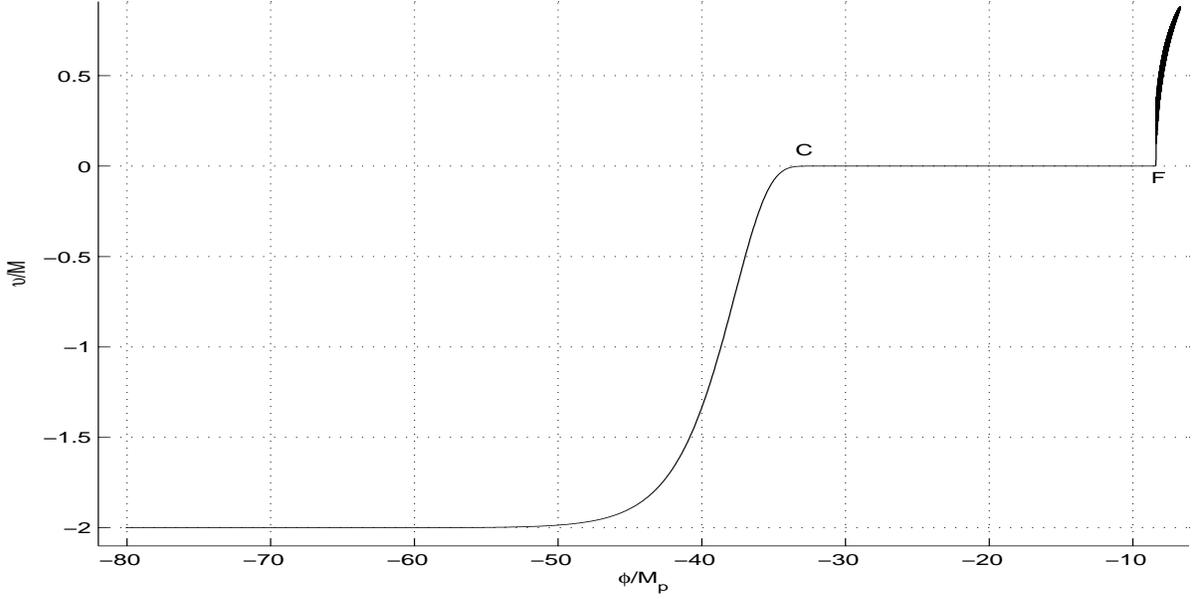}
\end{center}
\caption{Projection on the plane $(\phi,\upsilon)$ of the
4-dimensional phase trajectory corresponding to the solution for
the set (2) of the initial conditions. The function $\upsilon
(\phi)$ starts from $\upsilon =-2M$ and changes very slowly during
the GSB inflation. After transition to the GSR inflation (after
the point $C$) the phase trajectory  approaches the attractor
($\dot{\phi}=0$,  $\upsilon =0$, $\dot{\upsilon} =0$) via
oscillations of $\upsilon$ around $\upsilon =0$, see below. The
last phase transition during of which the Higgs-inflaton system
evolves (with damping oscillations of $\upsilon$) along  the
vacuum manifold (\ref{vac-min-dimensionless}) is presented by the
"horn" after the point $F$. The middle line of the horn is in fact
the vacuum manifold (\ref{vac-min-dimensionless}).}\label{fig15}
\end{figure}

\begin{figure}[htb]
\begin{center}
\includegraphics[width=17.0cm,height=7.0cm]{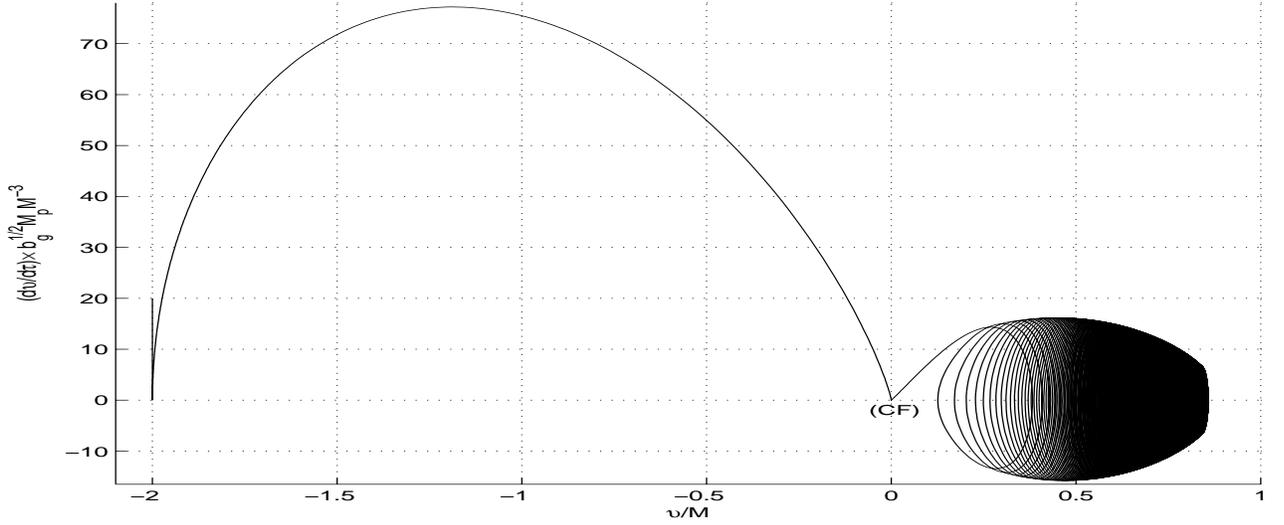}
\end{center}
\caption{Typical behavior of the projections of the 4-dimensional
phase trajectories on the plane $(\upsilon,\dot{\upsilon})$; here
the solution for the set (2) of the initial conditions has been
used. The vertical line $\upsilon\approx -2$ exhibits the stage of
the GSB inflation. The large curve corresponds to the transition
from the GSB inflation to the GSR inflation. The point labeled by
$(CF)$ is the projection of the interval $CF$ in Fig.15 on  the
$(\upsilon,\dot{\upsilon})$ plane. After $(CF)$ the Higgs field
$\upsilon$ evolves via oscillations from the practically zero
value to the gauge symmetry broken phase with the vacuum
expectation value $<\upsilon>$ which in the graph is about 0.84M.
}\label{fig16}
\end{figure}

\begin{figure}[htb]
\begin{center}
\includegraphics[width=17.0cm,height=7.0cm]{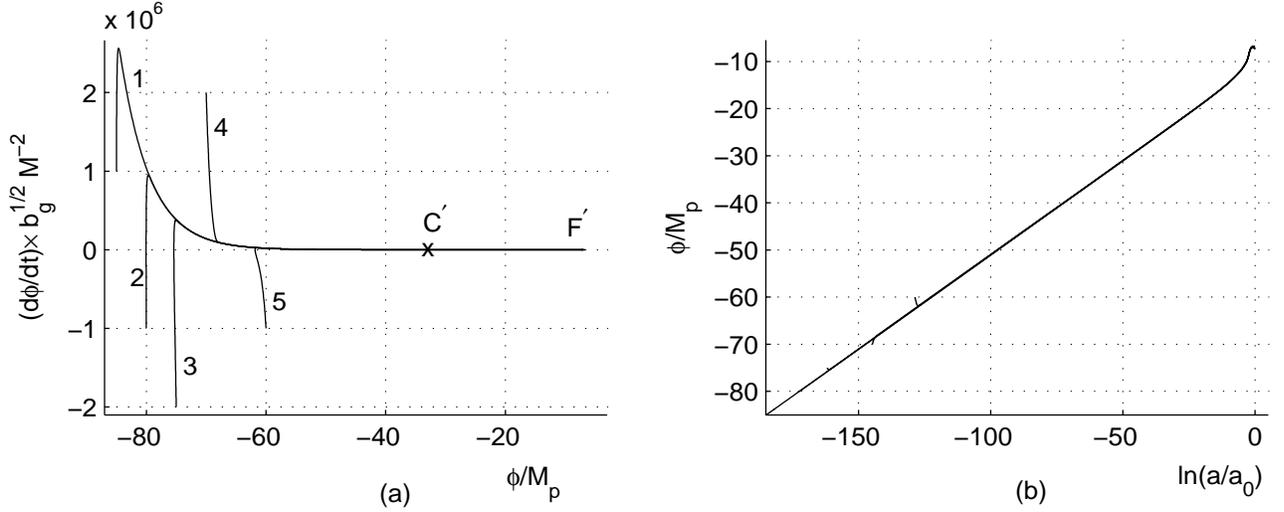}
\end{center}
\caption{Projections on the phase plane $(\phi,\dot{\phi})$ of
five 4-dimensional phase trajectories corresponding to five sets
of the initial conditions (\ref{5-in-cond}), Fig.(a). Similar to
Fig.11 in Sec.VI, the trajectories exhibit quick approach to the
attractor $\dot{\phi}=0$. Point $C'$ corresponds to the beginning
of the second stage of inflation (the GSR inflation) which
succeeds the first stage of inflation (the GSB inflation). The
phase trajectories become very close to the attractor
$\dot{\phi}=0$ long before the point $C'$. Point $F'$ corresponds
to the start of the phase transition to the "zero cosmological
constant and broken gauge symmetry" phase. Fig.(b) includes the
scale factor dependence of the inflaton for five sets of the
initial conditions (\ref{5-in-cond}).}\label{fig17}
\end{figure}

\begin{figure}[htb]
\begin{center}
\includegraphics[width=17.0cm,height=7.0cm]{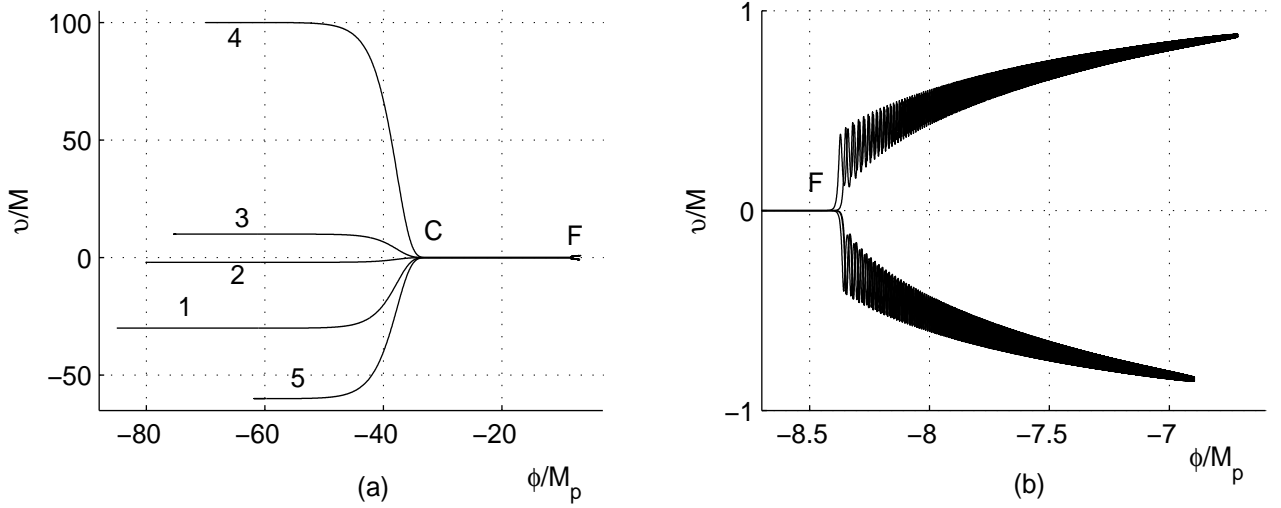}
\end{center}
\caption{Projections on the phase plane $(\phi,\upsilon)$ of five
4-dimensional phase trajectories corresponding to five sets of the
initial conditions (\ref{5-in-cond}). A surprising feature of the
dynamics consists in the observed fact that independently of the
initial conditions, all the phase trajectories achieve the line
$\upsilon =0$ at the same value of $\phi$, i.e. the point C is the
common point for all phase trajectories. Moreover, it turns out
that points F for different phase trajectories are also very close
to each other: see Fig.(b) which zooms in on the the behavior of
the five phase trajectories after the point F. Two horns show that
the five phase trajectories are divided into two directions giving
$<\upsilon>$ of both signs (recall that this sign has no physical
meaning). The final values of $|<\upsilon>|$ corresponding to
different conditions are very close (see however the appropriate
discussion in the main text).}\label{fig18}
\end{figure}

\begin{figure}[htb]
\begin{center}
\includegraphics[width=15.0cm,height=8.0cm]{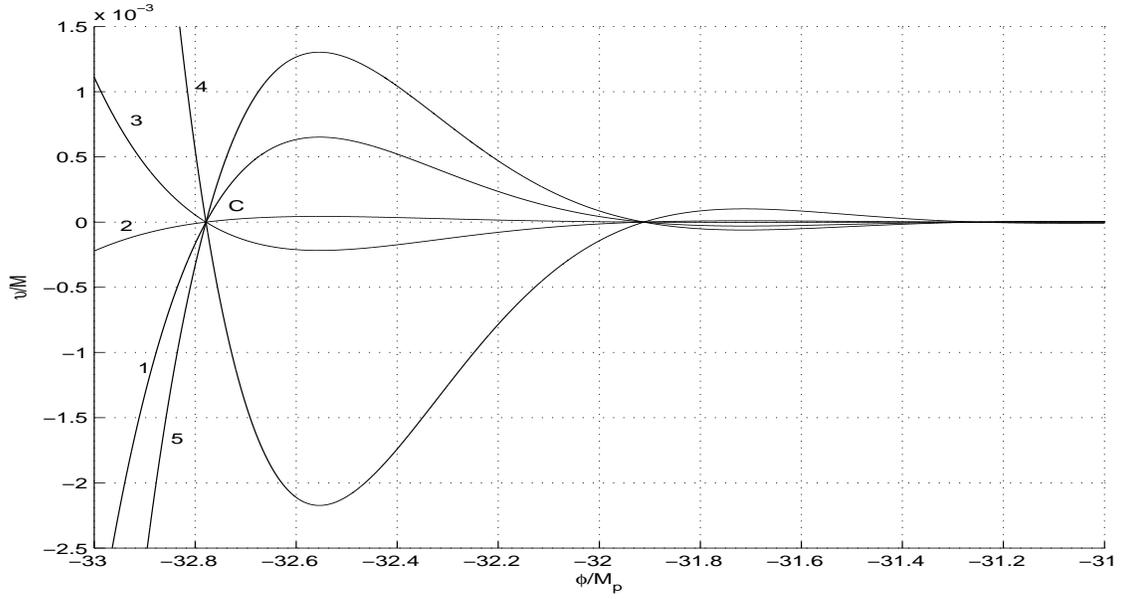}
\end{center}
\caption{This figure zooms in on the behavior of the  five phase
trajectories (corresponding to the five sets of the initial
conditions (\ref{5-in-cond})) near and after  the point $C$ in
Fig.18a. Solutions $\upsilon(\phi)$ exhibit the behavior typical
for Bessel functions. This feature as well as  the reason of an
observed synchronism will be explained in Sec.VIIC.}\label{fig19}
\end{figure}

\begin{figure}[htb]
\begin{center}
\includegraphics[width=15.0cm,height=8.0cm]{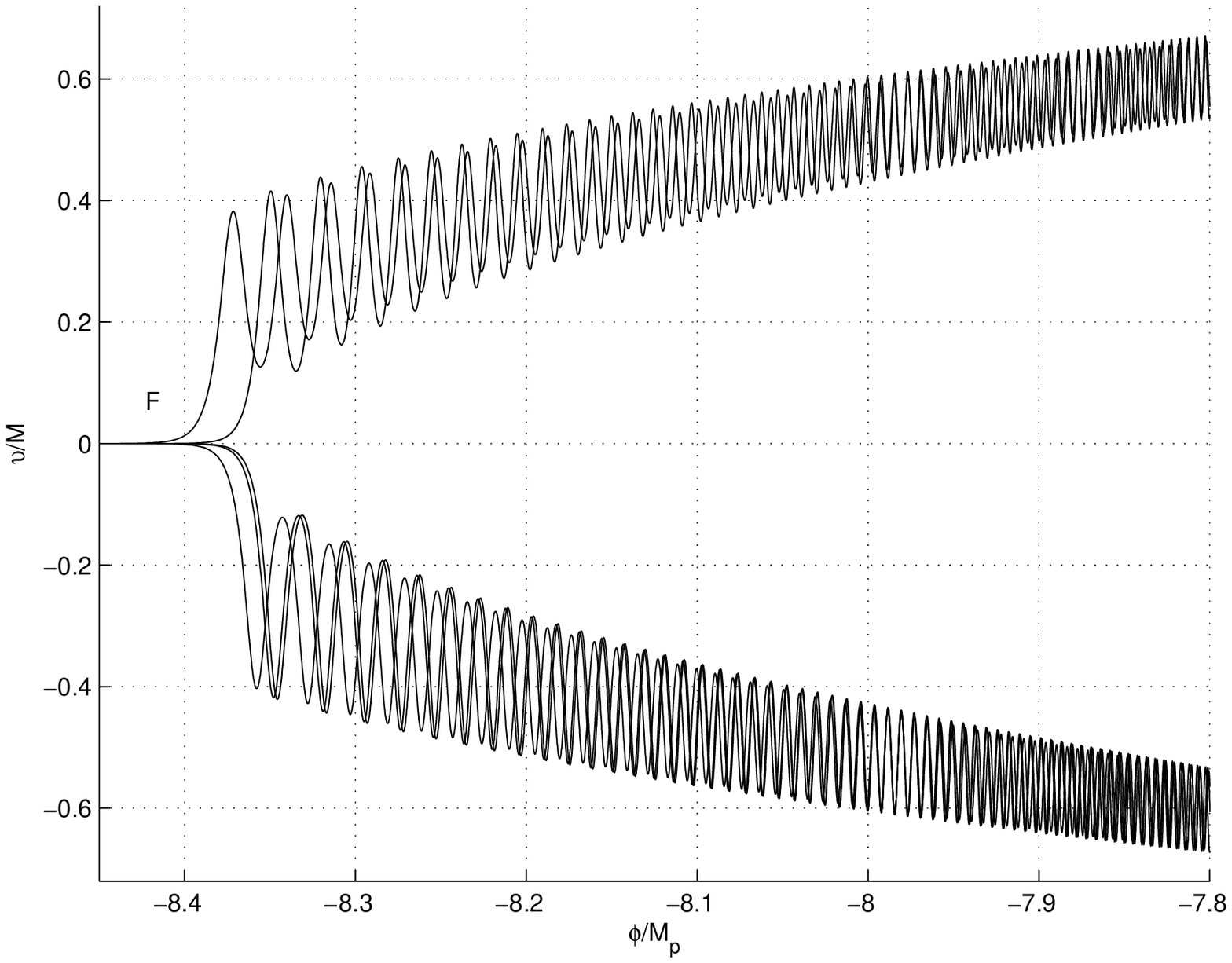}
\end{center}
\caption{This figure zooms in on two horns in Fig.18b. One can see
that amplitudes of the oscillations of $\upsilon$ around the line
of the vacuum manifold quickly become practically equal for
solutions with different initial conditions.  Frequency of
oscillations  increases with $\upsilon$, see
Eq.(\ref{Omega-dimensionless-positive}).}\label{fig20}
\end{figure}

In the main part of this subsection we choose $\delta =0$. But
afterwards, at the end of the subsection, we show that the main
results holds even if $\delta\neq 0$.

To explore the effect of the initial conditions ($\phi_{in},
\dot{\phi}_{in}, \upsilon_{in}, \dot{\upsilon}_{in})$ on the
evolution of the inflaton $\phi$ and the Higgs field $\upsilon$ we
have solved the equations for six sets of the initial conditions.
 Figures 12-20 demonstrate the results for following five sets of them:
\begin{eqnarray}
&(1)& \varphi_{in}=-85, \quad \tilde{\upsilon}_{in}=-30, \quad
\frac{d\varphi}{d\tau}|_{in}=10^6, \quad
\frac{d\tilde{\upsilon}}{d\tau}|_{in}=-20;
 \nonumber \\
&(2)& \varphi_{in}=-80, \quad \tilde{\upsilon}_{in}=-2, \quad
\frac{d\varphi}{d\tau}|_{in}=-10^6, \quad
\frac{d\tilde{\upsilon}}{d\tau}|_{in}=20;
\nonumber \\
&(3)& \varphi_{in}=-75, \quad \tilde{\upsilon}_{in}=10, \quad
\frac{d\varphi}{d\tau}|_{in}=-2\cdot 10^6,
 \quad \frac{d\tilde{\upsilon}}{d\tau}|_{in}=50;
 \nonumber \\
&(4)& \varphi_{in}=-70, \quad \tilde{\upsilon}_{in}=100, \quad
\frac{d\varphi}{d\tau}|_{in}=2\cdot 10^6,
 \quad \frac{d\tilde{\upsilon}}{d\tau}|_{in}=-30;
 \nonumber \\
&(5)& \varphi_{in}=-60, \quad \tilde{\upsilon}_{in}=-60, \quad
\frac{d\varphi}{d\tau}|_{in}=-10^6, \quad
\frac{d\tilde{\upsilon}}{d\tau}|_{in}=40. \label{5-in-cond}
\end{eqnarray}

 The results for the sixth set of the initial conditions
\begin{eqnarray}
&(6)& \varphi_{in}=-70, \quad \tilde{\upsilon}_{in}=-0.01, \quad
\frac{d\varphi}{d\tau}|_{in}=-2\cdot 10^6,
 \quad \frac{d\tilde{\upsilon}}{d\tau}|_{in}=-5
\label{6-th-in-cond}
\end{eqnarray}
 are shortly presented separately in Fig.21. This is done
just for technical reason because $\tilde{\upsilon}_{in}$ is too
small.

 So we
are exploring the numerical solutions in the broad enough range of
the initial values of the Higgs field $\upsilon$
\begin{equation}
10^{14}GeV\leq \upsilon_{in}\leq 10^{18}GeV \label{range-ups-in}
\end{equation}
while according to (\ref{upper-bound-on-v-full}) the true vacuum
expectation value $<\tilde{\upsilon}>$ is bounded by
\begin{equation}
|<\upsilon>| <  1.22\cdot 10^{16}GeV
 \label{upper-bound-on-v-numer}
\end{equation}

Fig.12 shows the behavior of the energy density and
equation-of-state typical for all the initial conditions we have
checked. For the set of the initial conditions (4),
Eq.(\ref{5-in-cond}), Fig.13 shows the evolution of $\upsilon$ and
$\dot{\upsilon}$ as functions of the scale factor. Using as a
pattern the set of the initial conditions (2),
Eq.(\ref{5-in-cond}), in Figs.14-16 we demonstrate the main
features of the cosmological evolution of the Higgs-inflaton
system. Figs.17-20 allow to show that dependence of these features
upon the initial conditions is very weak. In Fig.21 we show that
the interval of the initial conditions may be significantly
expanded without altering the latter conclusion. In Fig.22 we
present the scale factor dependence of the Higgs field for four
sets of initial conditions different from those in
Eqs.(\ref{5-in-cond}) and (\ref{6-th-in-cond}). Figs.23 and 24
show two additional facts: 1) the inflaton also oscillates during
the transition to the symmetry broken phase, although its
amplitudes are much less that those of the Higgs field; 2) exact
detection of the finishing point $(\phi_0,<\upsilon>)$ of the pure
classical transition to the symmetry broken phase is problematic.
Finally, in Figs.25-26 we show that the effect of the parameter
$\delta$ on the the main features of the cosmological evolution of
the Higgs-inflaton system, including the final gauge symmetry
broken phase, is also very weak.

\begin{figure}[htb]
\includegraphics[width=8.5cm,height=8.0cm]{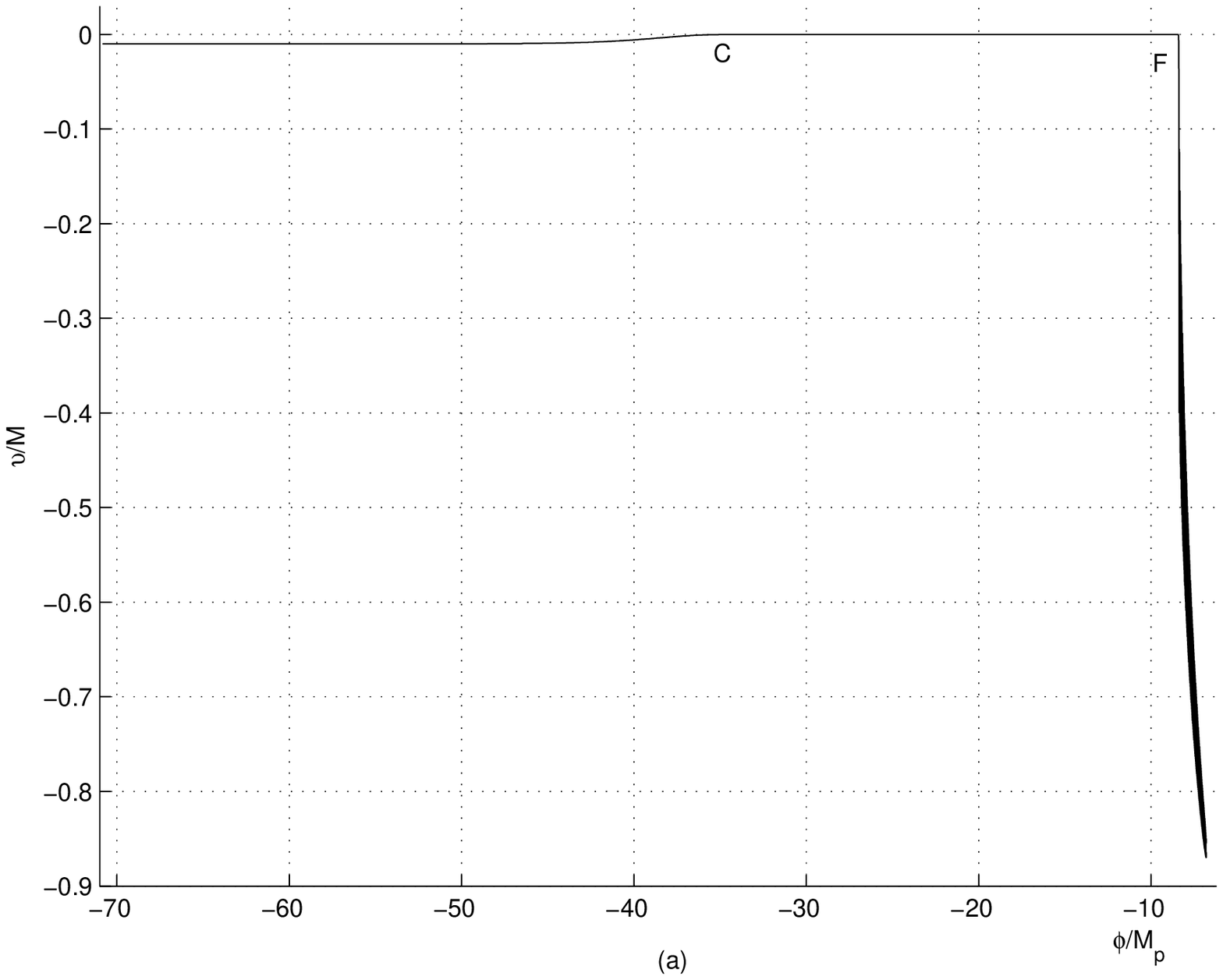}
\includegraphics[width=8.5cm,height=8.0cm]{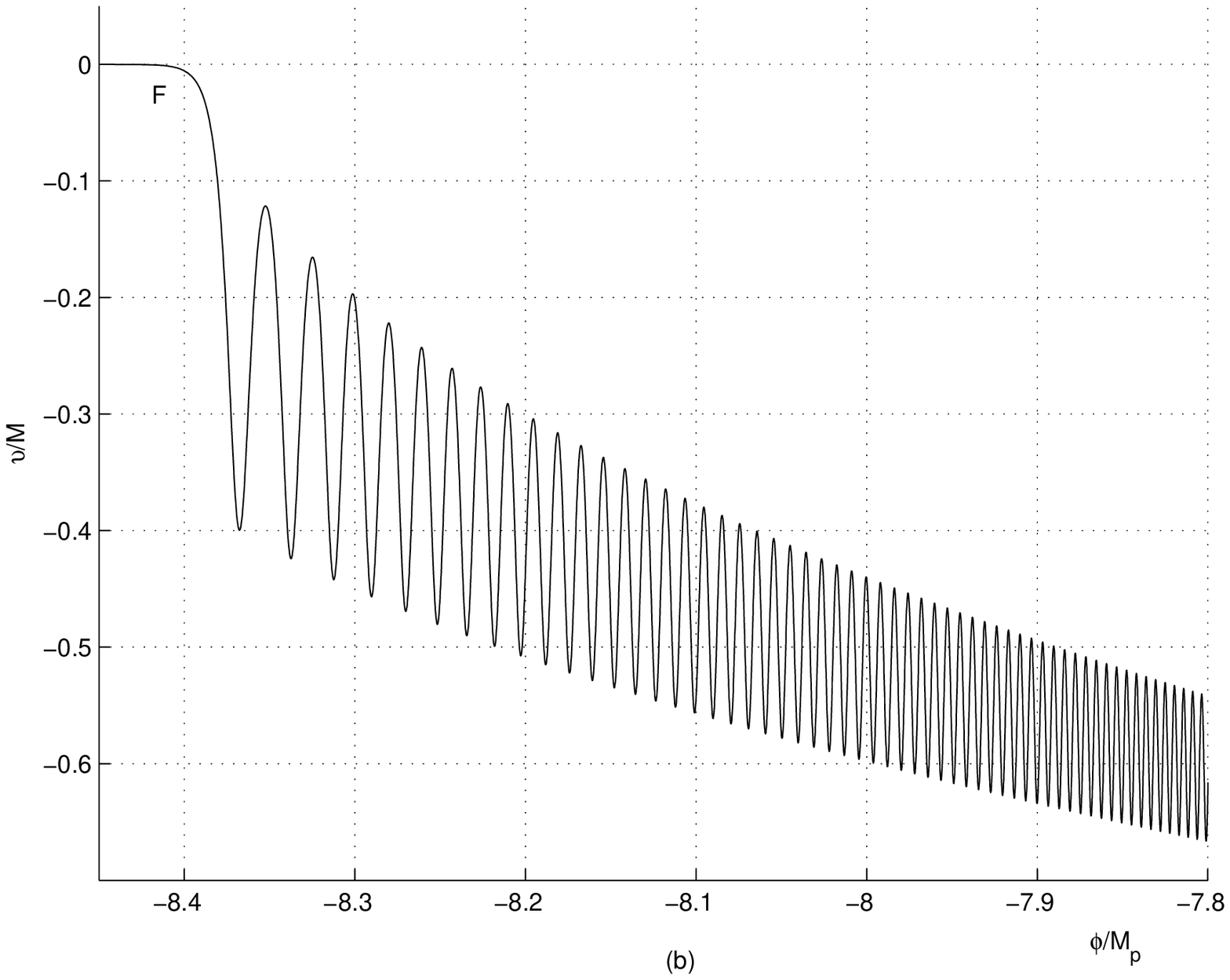}
\caption{The effects displayed for the sets of the initial
conditions (\ref{5-in-cond}) hold for larger range of the initial
values of $\upsilon$. Here for example we present the solution for
the set (\ref{6-th-in-cond}) of the initial conditions where
$\tilde{\upsilon}_{in}=-0.01$. The locations of points C and F are
the same as in Fig.18b; amplitudes of the oscillations of
$\upsilon$ around the line of the vacuum manifold are very close
to those in Fig.20. The final value of $|<\upsilon>|$ is very
close to those we have observed for the sets of the initial
conditions (\ref{5-in-cond}).}\label{fig21}
\end{figure}

\begin{figure}[htb]
\begin{center}
\includegraphics[width=18.0cm,height=8.0cm]{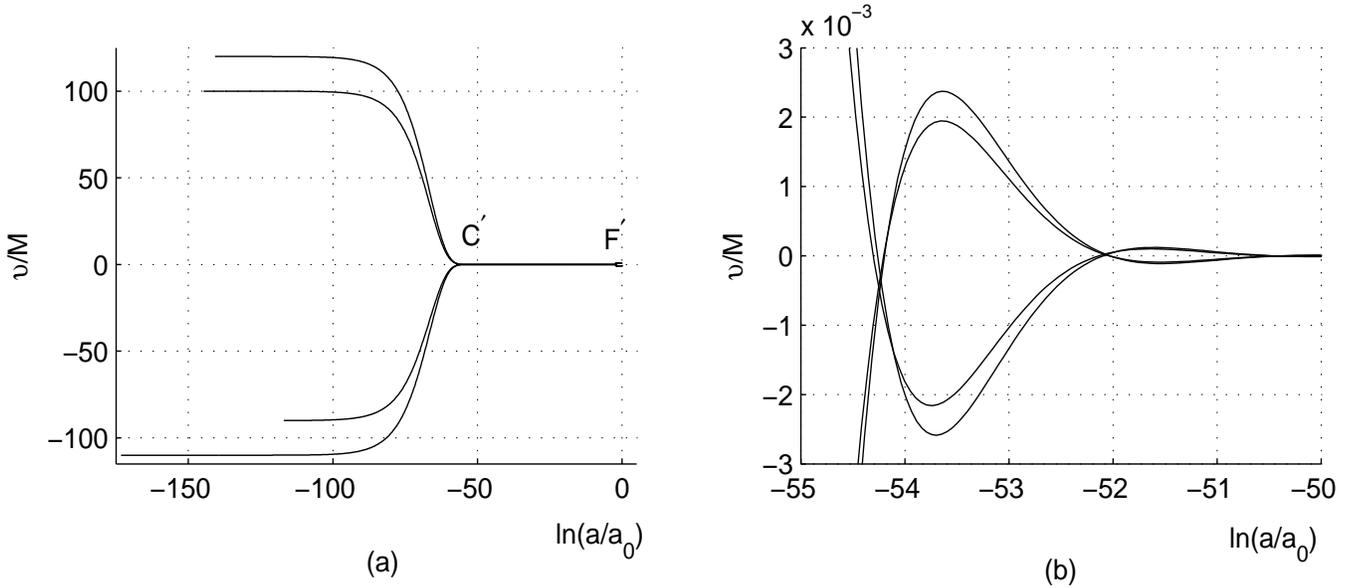}
\end{center}
\caption{Fig.(a) Scale factor dependence of the Higgs field for
four different initial conditions with $\tilde{\upsilon}_{in}\sim
10^2$ exhibits that the GSR inflation starts about 50 e-folding
before the end of inflation. Fig.(b) zooms in on the interval near
and after the point $C^{\prime}$.}\label{fig22}
\end{figure}

\begin{figure}[htb]
\begin{center}
\includegraphics[width=18.0cm,height=12.0cm]{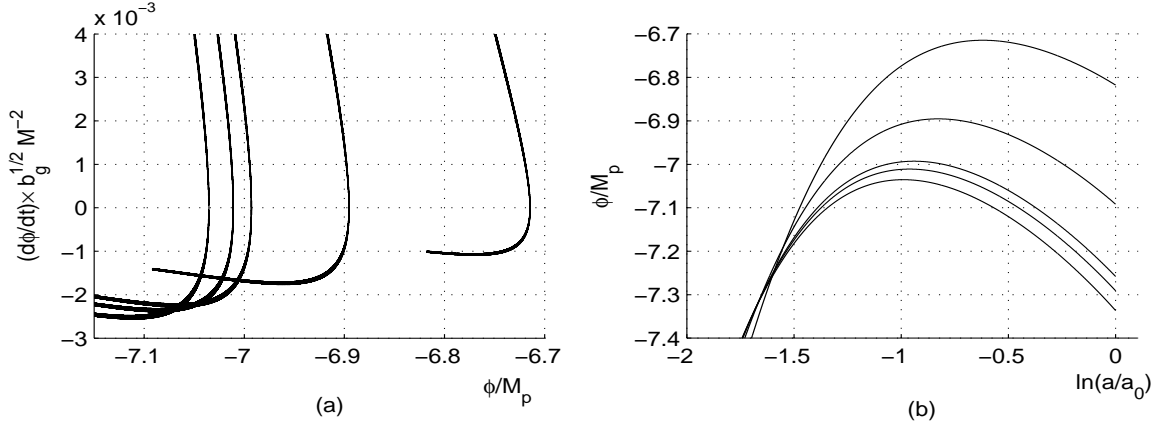}
\end{center}
\caption{Fig.(a) zooms in on the behavior of the five phase
trajectories in Fig.17a in the region after the point
$F^{\prime}$: evolving together with $\upsilon$ along the vacuum
manifold (\ref{vac-min-dimensionless}), $\phi$ performs a recoil
with change of the sign of $d\phi/dt$ from positive to negative as
the period average of $d\phi/dt$ gets to zero. Fig.(b) shows
details of the scalar factor dependence of $\phi$ at the end of
the process displayed in Fig.17b (i.e. during the transition to
the gauge symmetry broken phase) for the same five sets of the
initial conditions (\ref{5-in-cond}).} \label{fig23}
\end{figure}

\begin{figure}[htb]
\begin{center}
\includegraphics[width=16.0cm,height=8.0cm]{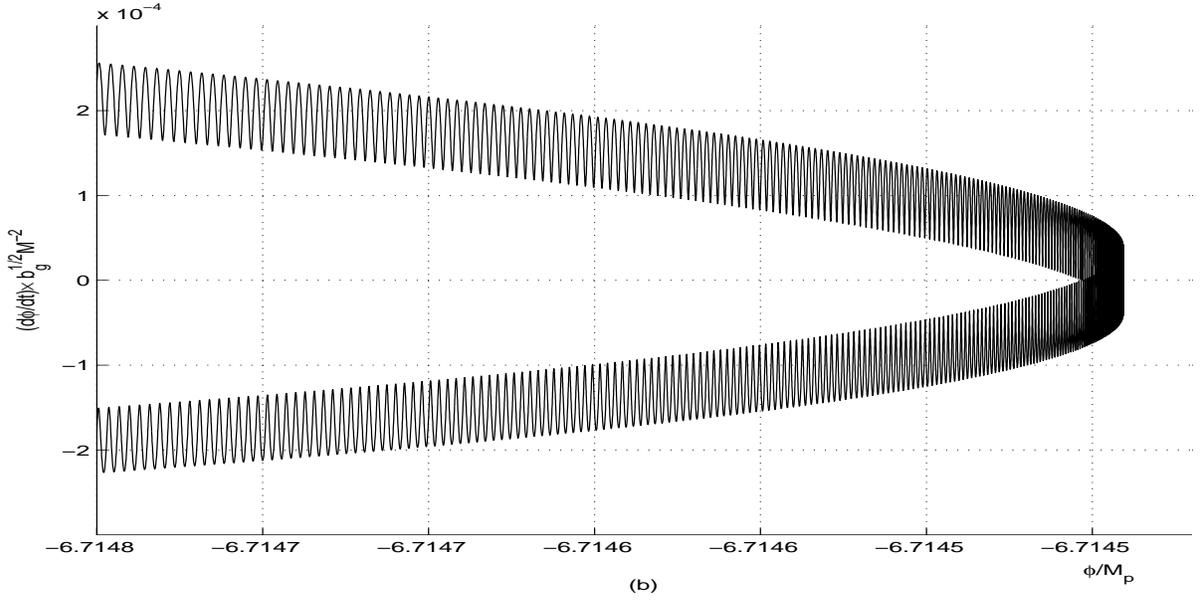}
\end{center}
\caption{Figure zooms in on one of the trajectories in Fig.23a
(namely for the trajectory with the set of the initial conditions
(2) in (\ref{5-in-cond})) near to the region where the period
average of $d\phi/dt$ gets to zero.}\label{fig24}
\end{figure}

\begin{figure}[htb]
\begin{center}
\includegraphics[width=18.0cm,height=8.0cm]{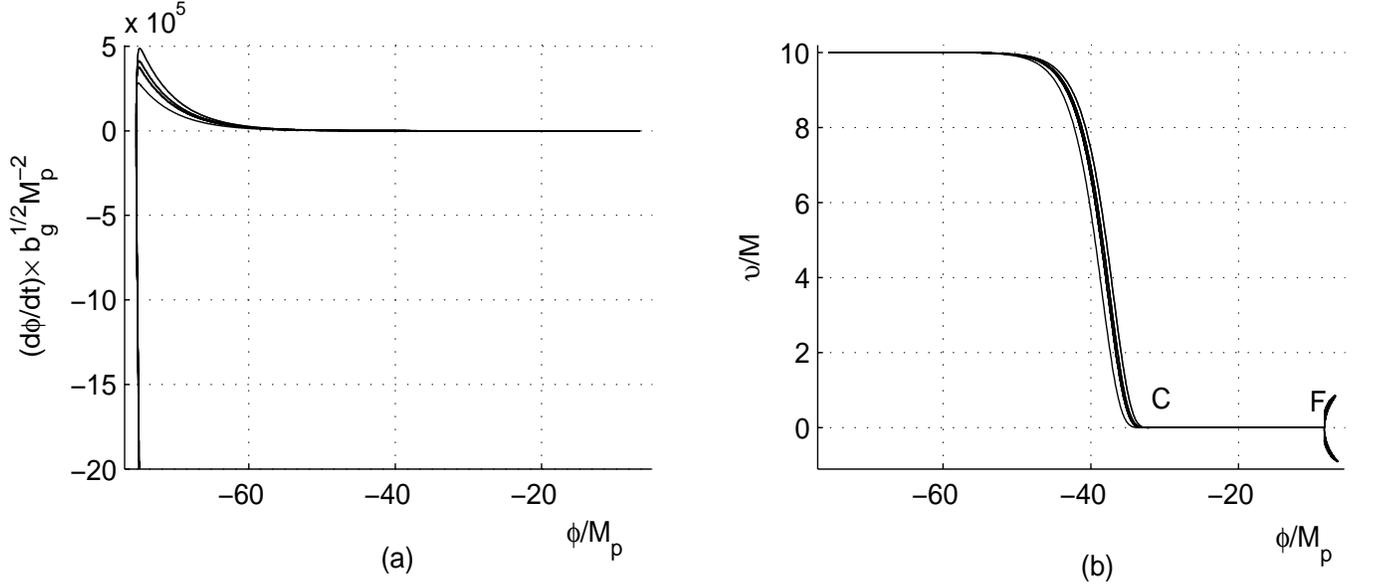}
\end{center}
\caption{The effect of the parameter $\delta$, Eq.(\ref{delta}),
on the dynamics of the Higgs-inflaton symbiosis. Four graphs in
Figs. (a) and (b) correspond to the solutions with the fixed
initial conditions (3) in Eq.(\ref{5-in-cond}) while for  the
parameter $\delta$ we have chosen the values $\delta =0$, \,
$\delta =0.2$, \, $\delta =0.5$, \, $\delta =0.7$. }\label{fig25}
\end{figure}

\begin{figure}[htb]
\includegraphics[width=8cm,height=8.0cm]{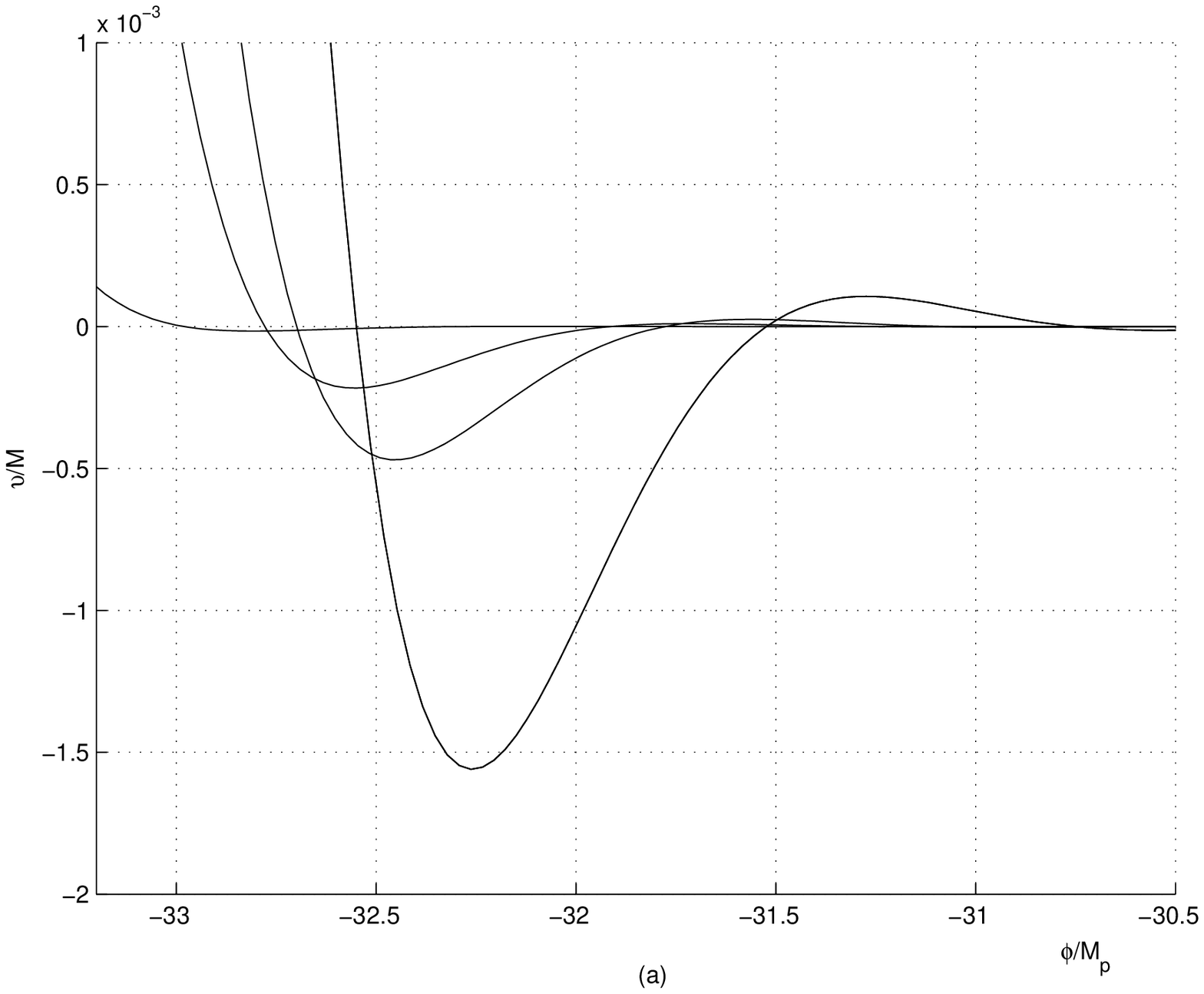}
\includegraphics[width=9cm,height=8.0cm]{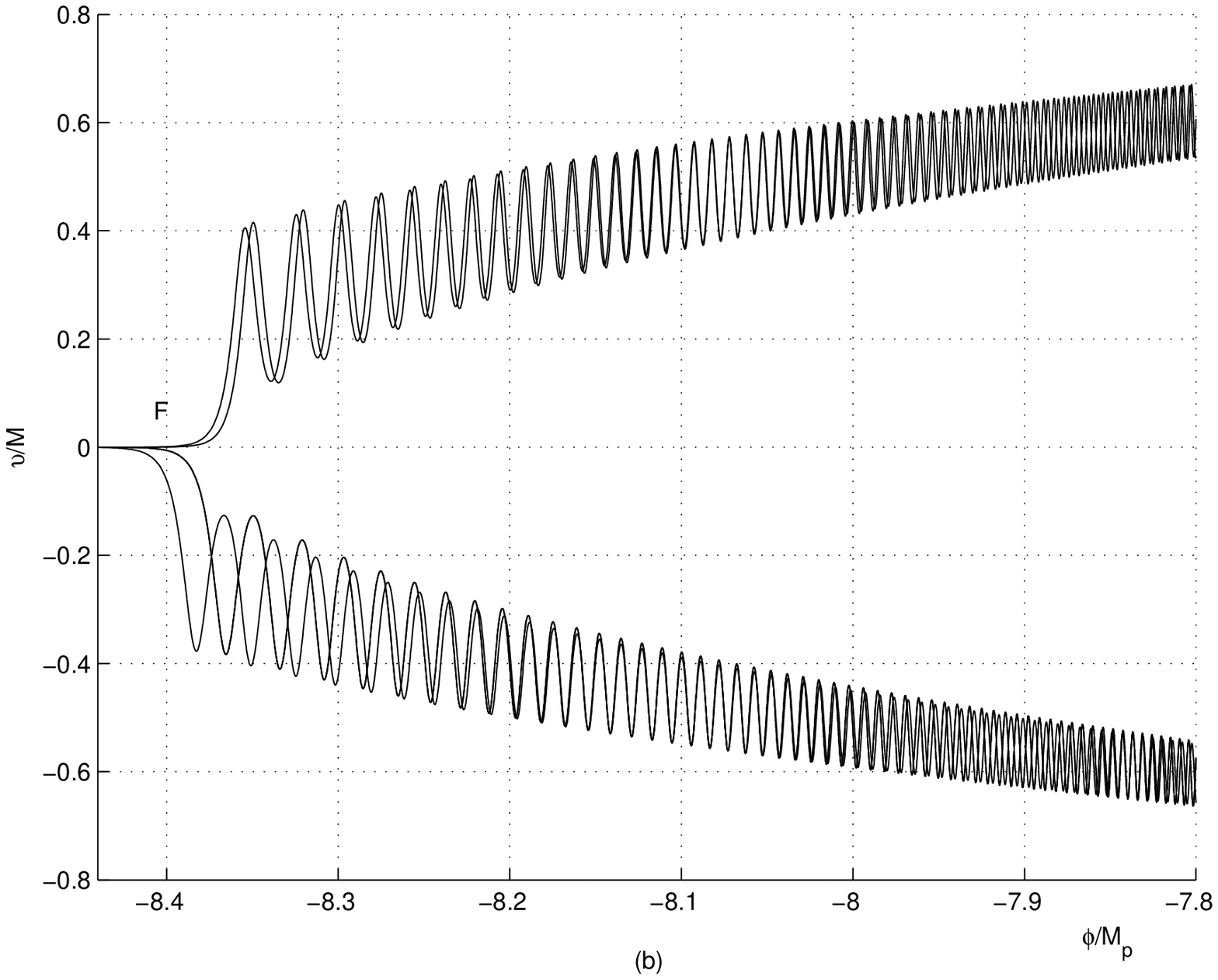}
\caption{Figures zoom in on the graphs in Fig.25b near and after
C, and near and after F. Although location of the point C is
$\delta$-dependent (as we see in Fig.(a)), the oscillatory regime
of the approach of the Higgs field to the final value of
$<\upsilon>$ is practically independent of $\delta$.}
\label{fig26}
\end{figure}

Fig.12b demonstrates the fact discussed in detail  after
Eq.(\ref{min-v})), that transition to the gauge symmetry broken
phase  at the same time is the transition to the state with zero
vacuum energy density.

 Fig.12a
demonstrates that most of the time of the evolution the
equation-of-state is close to the constant $w\approx -0.945$. This
corresponds to a  power low inflation which ends (i.e. $w\approx
-\frac{1}{3}$)
 as $\ln a/a_0 \approx -3$ and $\phi/M_p\approx -10$
 (the latter one can see analyzing Figs.12b and 17b).
  The qualitative explanation of this effect is evident enough.
 In fact, with our choice of the order of magnitudes of the
 parameters and initial conditions, contributions of
 $e^{-2\alpha\varphi}$ strongly dominate over all
 the terms both in numerator and denominator of the potential (\ref{Veff-dimensionless})
 as $\varphi\leq -15$, i.e in terms of the dimensionless
 quantities(\ref{notations-dimensionless}):
\begin{equation}
e^{-2\alpha\varphi}\gg Max\left[\left(\frac{M}{M_p}\cdot
\frac{d\tilde{\upsilon}}{d\tau}\right)^2, \tilde{V}_1^{(0)},
\tilde{V}_2^{(0)}, \tilde{\mu}_1^2\tilde{\upsilon}^2,
\tilde{\mu}_2^2\tilde{\upsilon}^2\right] \label{gg}
\end{equation}
Therefore for $\varphi\leq -15$ the back reaction of the Higgs
field on the inflaton dynamics is negligible. Therefore ignoring
all what is related to the Higgs field,
  the scalar sector potential $V_{eff}$,
Eq.(\ref{Veff-dimensionless}), acts as an exponential one
depending only on the field $\varphi$ (see also
Eq.(\ref{Veff+higgs-early})), and
 our model coincides with the well studied power law inflation
 model\cite{power-law}. With our choice of $\alpha =0.2$ the
 slow-roll parameter\cite{Liddle} is a constant
 \begin{equation}
\epsilon(\phi)=
\frac{1}{2}M_{p}^2\left(\partial_{\phi}V_{eff}/V_{eff}\right)^2
=2\alpha^2 =0.08. \label{eps-slow}
\end{equation}

Here are some other important features of the Higgs-inflaton
cosmological  evolution we observe in the numerical solution:

1. In spite of a practically  constant  equation-of state $w$,
Fig.12a, it is convenient to distinguish between two stages of the
inflation according to the value of the Higgs field. Starting from
a non zero initial value, $\upsilon$ varies very slowly during an
initial stage of the inflation for two reasons:  (a) a huge
friction (the Hubble parameter $\sim\sqrt{\epsilon}$, see the
second term in Eq.(\ref{H-dimensionless})); (b) the third term in
Eq.(\ref{H-dimensionless}) remains constant with very high
accuracy as $\varphi\ll -1$. We will refer to this initial stage
of the inflation   as {\it the gauge symmetry broken (GSB)
inflation}. When the Hubble parameter decreases significantly, the
Higgs field falls to its minimum $\upsilon =0$ or, more exactly it
performs a transition to the phase where it oscillates with
decaying amplitude around $\upsilon =0$.  Some features of  this
second stage of the inflation one can see in Figs.13, 14, 15, 17,
18a, 19, 21 and 22. We will refer to this stage of inflation as
{\it the gauge symmetry restored (GSR) inflation}. It lasts during
the time corresponding to the interval which starts from the point
$C$ and ends a little bit before the point $F$.

2. Projections of  all phase trajectories on the plane
$(\phi,\dot{\phi})$ (see Fig.17a) approach very closely the
attractor  long before the end of inflation and even before
transition to the GSR inflation. It becomes clear now that the
effect of the rapid approach to the attractor, we have observed in
the simpler model of Sec.VI, has {\it a key role for the extremely
weak dependence of the Higgs field vacuum expectation value
$<\upsilon>$ upon the initial values of $\phi_{in}$ and
$\dot{\phi}_{in}$}.

3.  In the process of the transition from the GSB inflation to the
GSR inflation, the Higgs field gets to its first zero {\it at the
same value of} $\phi$ {\it for all the phase trajectories}
corresponding to different initial conditions, see point $C$ in
Figs.18a and 19. The explanation of this surprising effect will be
given in the next subsection on the the basis of an analytic
solution. After point $C$ {\it all the phase trajectories quickly
approach the attractor} ($\dot{\phi}=0, \upsilon =0,
\dot{\upsilon}=0)$. This approach occurs through the damped
oscillations of $\upsilon$ around $\upsilon =0$.

4.  By means of graphs in Figs.12a and 17b one can check that with
our choice of the parameters the inflation ends
 as $\ln a/a_0 \approx -3$ and $\phi\approx -10M_p$. However
 it follows from Figs.13a, 14a, 15, 17b, 18, 21 and 22a that the last
 phase transition starts as
$\ln a/a_0 \approx -2$ and $\phi\approx -8.5M_p$. Therefore the
{\it transition to the "zero cosmological constant and broken
gauge symmetry" phase starts after the end of inflation}. This
result is also independent of the initial conditions.

5. {\it The initial stage of the exit from inflation} happens as
$e^{-2\alpha\varphi}\rightarrow |\tilde{V}_{1}^{0}|$ and, as a
result, the scalar sector energy density $\epsilon$,
Eqs.(\ref{rho-dimensionless}) and (\ref{Veff-dimensionless}),
starts to go to zero (see Fig.12b). The difference from the simple
model of Sec.VI here is that in the continuation of the exit from
inflation, instead of entering the regime of oscillations of the
inflaton, the modulus of the Higgs field starts to increase  in
such a way that the Higgs-inflaton system  very fast approaches
the vacuum manifold (\ref{min-v}), i.e.
 the Higgs field $\upsilon$ appears in the new phase where
 $\upsilon \neq 0$. Afterwards the
 Higgs-inflaton evolution proceeds along
 the vacuum manifold (see the horn in Fig.15)
via oscillations of both the Higgs field $\upsilon$ and the
inflaton $\phi$. However there is an essential difference between
the character of their oscillations: $d\upsilon/dt$ oscillates
around zero while $d\phi/dt$ generically oscillates without
changing its sign. Besides, the amplitude of the
$\upsilon$-oscillations  is a few orders of magnitude larger than
that of the inflaton. The character of the damped oscillations of
$\upsilon$ one can see in Figures.16, 18b, 20, 21. Some details
concerning  the $\phi$ oscillations are presented in Fig.24.

6. The point $(\phi_0,<\upsilon> )$ of the vacuum manifold
(\ref{min-v}) where the Higgs-inflaton system stops its classical
evolution depends generically on the initial conditions. Due to
the attractor behavior of the phase trajectories discussed in
items 2 and 3, the initial conditions can affect the values of
$<\upsilon>$ and $\phi_0$ only through the following two ways: a)
the amplitudes of the residual oscillations of $\upsilon$ around
$\upsilon =0$ (see Fig.19) which are extremely small in the region
close to point $F$ in Fig.18a: up to $\phi/M_p< -15$ the
amplitudes of the classical oscillations decay about 30 orders of
magnitude according to the behavior of the Bessel function, see
the next subsection and Fig.28 therein; b) the residual velocity
of the inflaton $\dot{\phi}$ which is also very small in the
region close to point $F$. These two circumstances explain why for
different initial conditions $\phi_{in}$, $\dot{\phi}_{in}$,
$\upsilon_{in}$, $\dot{\upsilon}_{in}$ there is only one  short
segment in the line (\ref{min-v}) where all the phase trajectories
end in the process of the cosmological evolution. We have found
that for our choice of the parameters (\ref{parameters}) the value
of $|<\tilde{\upsilon}>|$ approaches the interval $(0.83-0.88)$.

7. Analyzing the behavior of the numerical solution after the
point $F$ we conclude that most of the Higgs field kinetic energy
accumulated in the $\tilde{\upsilon}$ oscillations around the mean
value $<\tilde{\upsilon}>$ which monotonically evolves together
with $\varphi$. The kinetic energy of $\varphi$ consists of two
parts: the one due to the residual velocity of the monotonic
evolution of $\varphi$ and the other due to oscillations of
$\dot{\varphi}$, see Fig.24.  This circumstance causes a behavior
anomalous on the first glance: after the period average of
$\dot{\varphi}$ goes through zero, the Higgs and inflaton fields
start to go back,  that one can see in Figs.14a, 17b, 23 and 24.
This recoil with the change of the sign of $d\varphi/d\tau$ from
positive to negative can happen in the process of the
$\tilde{\upsilon}$-oscillations  at the moment when
$\tilde{V}_1(\tilde{\upsilon})+e^{-2\alpha\varphi}$ appears to be
negative (see Eq.(\ref{phi-dimensionless})). Once $\varphi$ starts
to evolve backwards, this causes the same for $<\upsilon>$. After
some time the period average of $\dot{\varphi}$ goes through zero
again and the recoil with the change of the sign of
$d\varphi/d\tau$ from   negative  to positive  happens now. This
process has apparently a tendency to repeat itself with decaying
amplitude of the period average of $\dot{\varphi}$, but  attempts
of more exact numerical solutions with the aim to obtain a
definite final value of $|<\tilde{\upsilon}>|$, where the
evolution of our system finishes, run against a computational
problem: enlarging running time of the process in our computations
it is impossible (or may be very hard) to observe a certain point
in the interval of vacuum manifold (although short enough) where
the trajectories stop. The main reason consists in the fact that
not only the elastic forces in Eqs.(\ref{phi-dimensionless}) and
(\ref{H-dimensionless}) approach zero (due to approaching zero the
amplitude of the oscillations around zero of the factor
$\tilde{V}_1(\tilde{\upsilon})+e^{-2\alpha\varphi}$) but also the
friction caused by the cosmological expansion
$\sim\sqrt{\epsilon}$ approaches zero. This makes the classical
mechanism of dissipation of the residual kinetic energies of
$\varphi$ and $\upsilon$ practically ineffective. However the
described problem  should disappear after taking into account
evident quantum effects, see Sec.VIIE.

8. We have chosen $a_0$ to be the value of the scale factor at the
end of the studied evolution process. In this section $a_0$ is the
value of the scale factor when the (last) transition to the "zero
cosmological constant and broken gauge symmetry" phase practically
ends. On the other hand, as we already discussed (see the item 3),
after the point $C$ in Fig.18a all the phase trajectories
corresponding to different initial conditions appear to be very
close to the attractor $(\dot{\phi}=0, \upsilon =0,
\dot{\upsilon}=0)$ and this fact holds true till the last phase
transition.  This explains why in the interval of $ln(a/a_0)$ from
its value corresponding to the fall to the point $C$ up to
$ln(a/a_0)=0$, we observe practically the same pictures for all
initial conditions (see for example Fig.22). In particular with
our choice of the parameters, we have found that the transition
from the GSB inflation to the GSR inflation ends about 50
e-folding before the end of inflation, and this result is
practically independent of the initial conditions.

9. We have checked by means of numerical solutions that in models
with $\delta\neq 0$ all the above conclusions are not affected,
see Figs.25 and 26. This is clear enough because contributions of
the terms with $\delta$ in equations of Sec.IIIA become negligible
as the inflaton kinetic term is very small. But this is exactly
what happens due to the attractor behavior of the phase
trajectories.

 \subsection{Analytic Solution for GSB and GSR Stages of Inflation}

  To understand the mechanism responsible for the  character of the attractor
  behavior of the phase trajectories (observed
  in the numerical solution) as they approach the point $C$ of the line $v(\phi)=0$
  in Figs.18a and 19,  one should to take into account in
  Eqs.(\ref{phi-dimensionless})-(\ref{X-dimensionless}) the
  inequalities (\ref{gg}) which hold for $\varphi\equiv\phi/M_{p}\leq 15$.
 When doing this in a consistent way one can
rewrite Eqs.(\ref{phi-dimensionless}) and (\ref{H-dimensionless}),
for $\varphi\leq 15$, with high enough accuracy in the following
simplified form
\begin{equation}
\frac{d^2\varphi}{d\tau^2}+\sqrt{3}\left[\frac{1}{2}\left(\frac{d\varphi}{d\tau}\right)^2+
\frac{1}{4}e^{-2\alpha\varphi}\right]^{1/2}\frac{d\varphi}{d\tau}
-\frac{\alpha}{2} e^{-2\alpha\varphi} =0,
\label{phi-dimensionless-simple}
\end{equation}
\begin{equation}
\frac{d^2\tilde{\upsilon}}{d\tau^2}+\sqrt{3}\left[\frac{1}{2}\left(\frac{d\varphi}{d\tau}\right)^2+
\frac{1}{4}e^{-2\alpha\varphi}\right]^{1/2}\frac{d\tilde{\upsilon}}{d\tau}
+\frac{1}{2}\left(\frac{M_p}{M}\right)^2(\tilde{\mu}_1^2+\tilde{\mu}_2^2)\tilde{\upsilon}=0,
\label{H-dimensionless-simple}
\end{equation}
while in the energy density $\epsilon$ in
Eq.(\ref{rho-dimensionless}) only the inflaton contribution has to
be taken into account.

It is well known that Eqs.(\ref{phi-dimensionless-simple}) and
(\ref{cosm-phi}) allow an exact solution corresponding to the
power low inflation\cite{power-law}
\begin{equation}
a(\tau)=a(\tau_{in})\left(\tau/\tau_{in}\right)^{1/2\alpha^{2}},
\qquad \varphi =\varphi_{in}
+\frac{1}{\alpha}\ln\frac{\tau}{\tau_{in}}, \label{p-l-solution}
\end{equation}
where $\tau_{in}$ and $\varphi_{in}$ satisfy the condition
\begin{equation}
l\equiv \tau_{in}
e^{-\alpha\varphi_{in}}=\sqrt{\frac{3}{\alpha^{4}}-\frac{2}{\alpha^{2}}}.
\label{l}
\end{equation}
Note that in addition to the above restriction $\varphi\leq 15$,
the solution (\ref{p-l-solution}) may be non applicable for a
relatively short period of time from the very beginning where
$\frac{d\varphi}{d\tau}$ may be non monotonic, see Fig.17a and
12a.

Using the solution (\ref{p-l-solution}) for $\varphi(\tau)$ one
can rewrite Eq.(\ref{H-dimensionless-simple}) in the form of the
equation for $\tilde{\upsilon}=\tilde{\upsilon}(\varphi)$:
\begin{equation}
\frac{d^2\tilde{\upsilon}}{d\varphi^2}+\left(\sqrt{\frac{9}{4\alpha^{2}}-1}-\alpha\right)
\frac{d\tilde{\upsilon}}{d\varphi}
+\frac{1}{2}\left(\frac{M_p}{M}\right)^2(\tilde{\mu}_1^2+\tilde{\mu}_2^2)\left(\frac{3}{\alpha^{2}}-2\right)
e^{2\alpha\varphi}\tilde{\upsilon}=0 \label{v(varphi)}
\end{equation}
To obtain the analytic solution of this equation we use the
following change of variable:
\begin{equation}
z=exp\left[\left(\alpha -\frac{3}{2\alpha}\right)\varphi\right].
\label{change}
\end{equation}
 Then
Eq.(\ref{v(varphi)}) takes the form
\begin{equation}
\frac{d^2\tilde{\upsilon}}{dz^2}
+\frac{2(\tilde{\mu}_1^2+\tilde{\mu}_2^2)}{3-2\alpha^2}\left(\frac{M_p}{M}\right)^2
z^{-3/(3-2\alpha^2)}\tilde{\upsilon}=0 \label{v(z)-eq}
\end{equation}
General solution of Eq.(\ref{v(z)-eq}) may be written in the form
\begin{equation}
\tilde{\upsilon}(z)=z^{1/2}\left[C_1J_{\nu}(\beta
z^{\gamma})+C_2Y_{\nu}(\beta z^{\gamma})\right] \label{v(z)-sol}
\end{equation}
where
\begin{equation}
\nu =\frac{1}{2}\left(\frac{3}{2\alpha^2}-1\right), \qquad \beta=
\frac{M_p}{M\alpha^2}\sqrt{\frac{1}{2}(\tilde{\mu}_1^2+\tilde{\mu}_2^2)(3-2\alpha^2)},
 \qquad \gamma =-\left(\frac{3}{2\alpha^2}-1\right)^{-1},
\label{beta}
\end{equation}
$J_{\nu}(\beta z^{\gamma})$ and $Y_{\nu}(\beta z^{\gamma})$ are
the Bessel function of the first kind and the second kind
respectively and $C_1$, $C_2$ are two arbitrary constants.
Returning to the original variable $\varphi$, we obtain the
general solution of Eq.(\ref{v(varphi)}):
\begin{equation}
\tilde{\upsilon}(\varphi)=\left[C_1J_{\nu}\left(\beta
e^{\alpha\varphi}\right)+C_2Y_{\nu}\left(\beta
e^{\alpha\varphi}\right)\right]\cdot
exp\left[-\left(\frac{3}{4\alpha}-\frac{\alpha}{2}\right)\varphi\right]
\label{v(phi)-sol-1}
\end{equation}

\begin{figure}[htb]
\includegraphics[width=8.0cm,height=7.0cm]{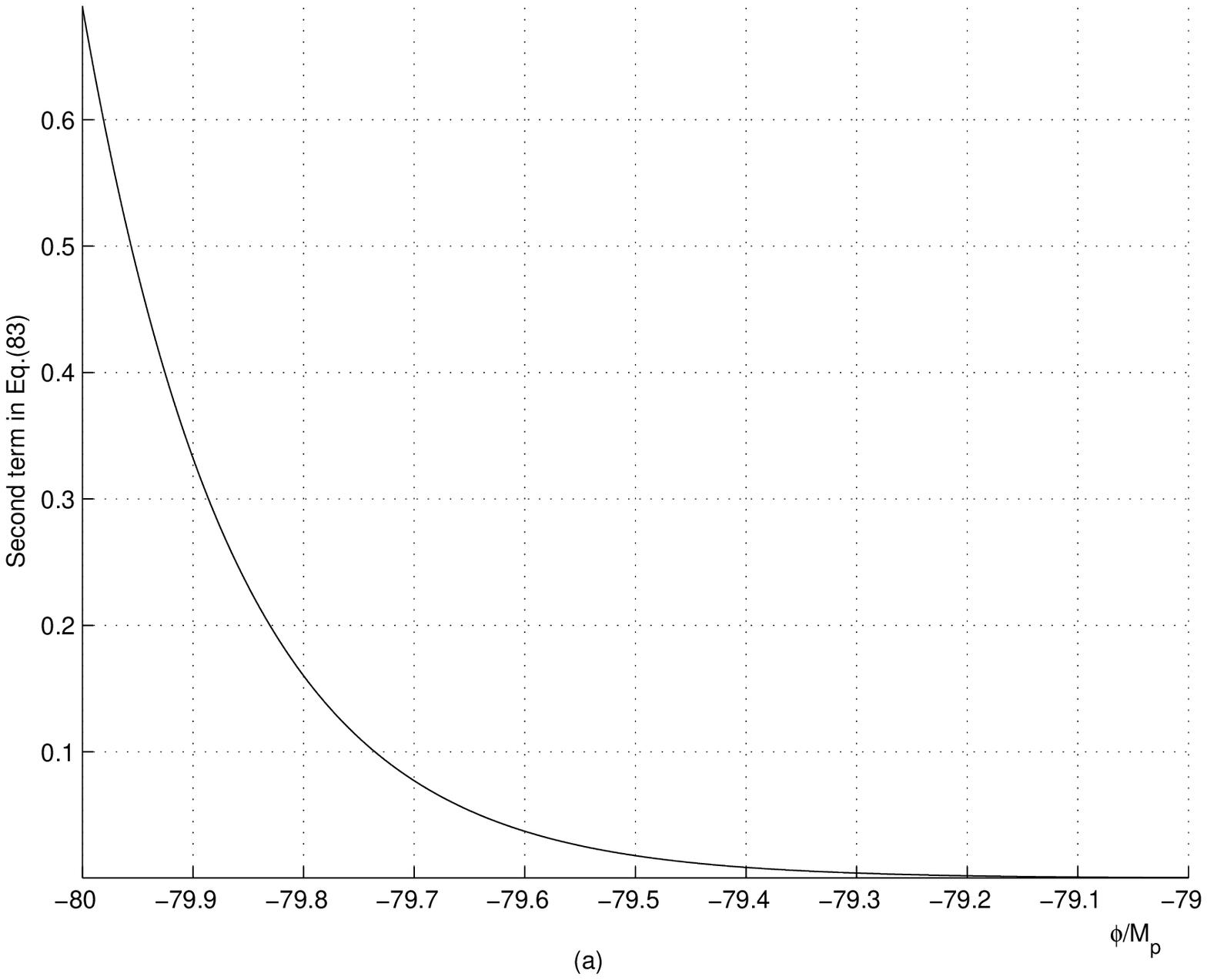}
\includegraphics[width=8.0cm,height=7.0cm]{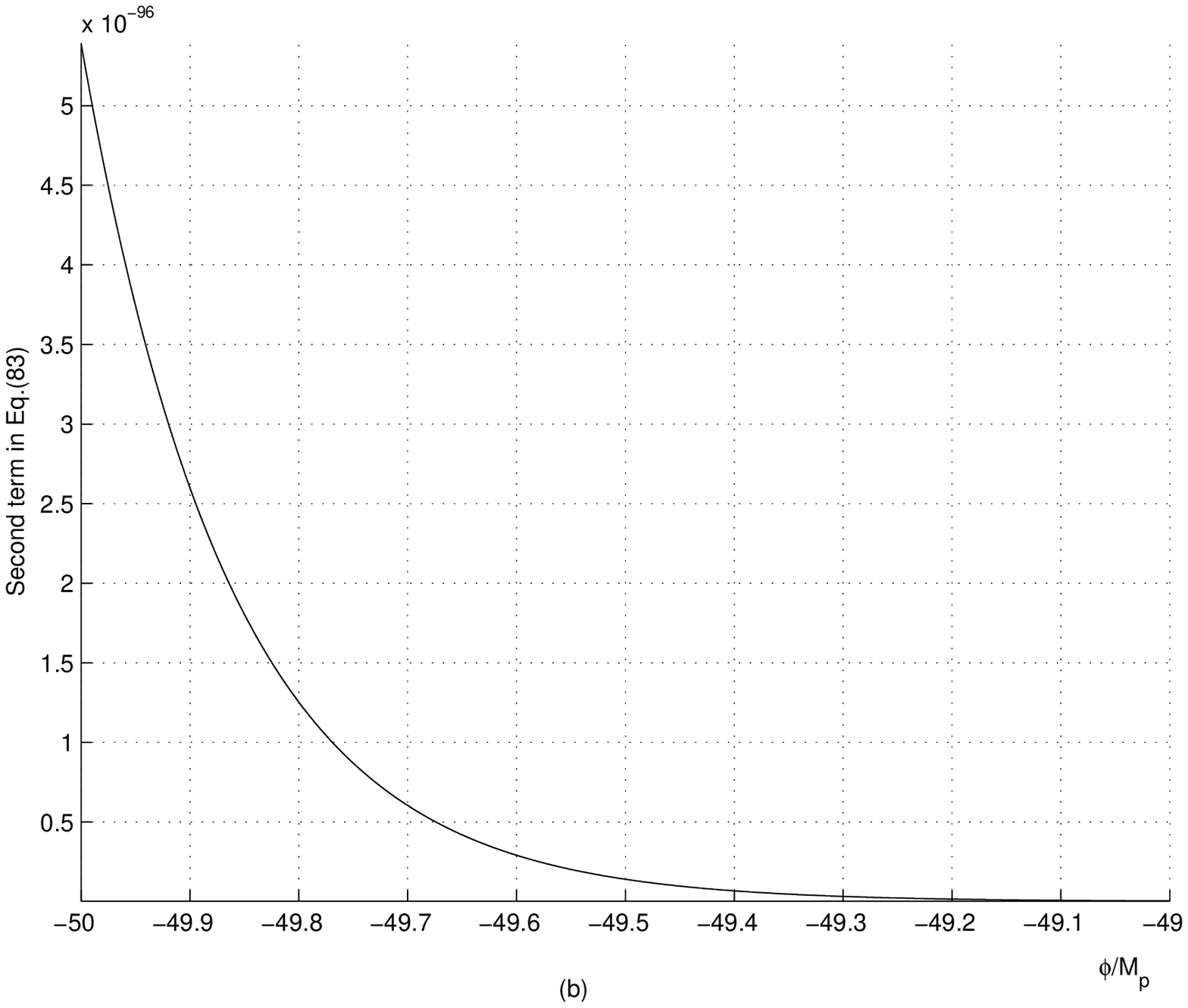}
\caption{Behavior of the second term $I\equiv
C_2Y_{\nu}\left(\beta e^{\alpha\varphi}\right)\cdot
exp\left[-\left(\frac{3}{4\alpha^2}-\frac{\alpha}{2}\right)\varphi\right]$
 in Eq.(\ref{v(phi)-sol-1}) is very
singular: to provide the initial value of $I$ to be of the order
of one as $\varphi_{in}= -80$ (Fig.(a)) we are forced to choose
the value of the constant $C_2\sim 10^{-197}$. Then for example
$I\sim 10^{-96}$ as $\varphi = -50$ (Fig.(b)). }\label{fig27}
\end{figure}

\begin{figure}[htb]
\begin{center}
\includegraphics[width=13.0cm,height=7.0cm]{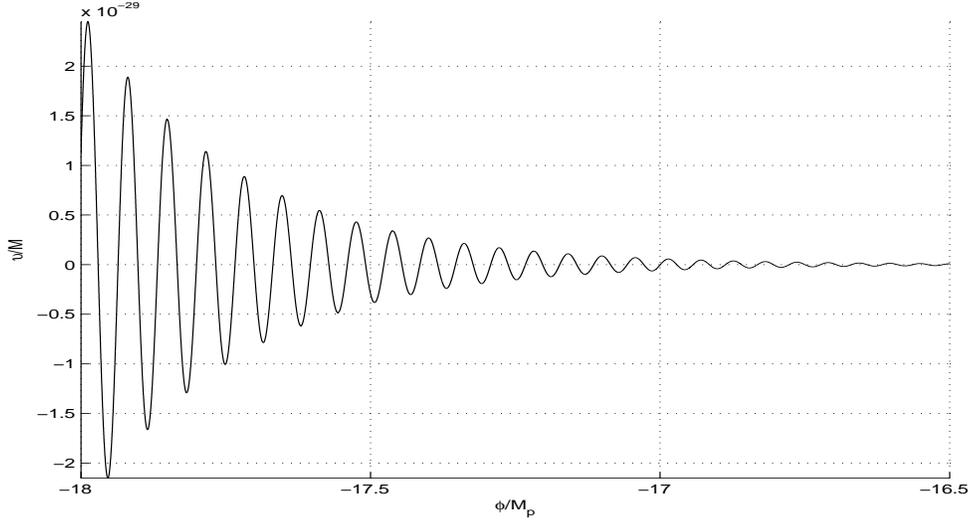}
\end{center}
\caption{Here  the analytic solution (\ref{v(phi)-sol}) has been
used in order to zoom in on  one of the five patterns of the phase
trajectories shown in Figs.18a and 19. This graph illustrates a
degree of damping of the Higgs field oscillations as $\varphi$ is
close to -15 (i.e. near to the end of the interval where the
solution (\ref{v(phi)-sol}) is applicable).}\label{fig28}
\end{figure}

To compare this analytic solution with the results of the numeric
solutions of subsection B we have used the same set of the values
of the parameters (\ref{parameters}) as in Sec.B. One can show
that the graph of the first term proportional to
$J_{\nu}\left(\beta e^{\alpha\varphi}\right)$ has the shape which
for five sets of the initial conditions (\ref{5-in-cond}) results
in the graphs which coincide with those  in Fig.19. But the graph
of the second term proportional to $Y_{\nu}\left(\beta
e^{\alpha\varphi}\right)$ exhibits a very singular behavior, see
Fig.27. Therefore when trying to satisfy the initial conditions
with natural, i.e. non anomalously large values of
$\tilde{\upsilon}(\varphi_{in})$ and
$\left(d\tilde{\upsilon}/d\varphi\right)|_{\varphi
=\varphi_{in}}$, we obtain that  $C_{2}$ is very close to zero.
Because of the extremely rapid decay of the second term, its
contribution into the solution (\ref{v(phi)-sol-1}) quickly
becomes negligible with growth of $\varphi$. Hence with very high
accuracy one can proceed with the following analytic solution of
Eq.(\ref{v(varphi)}):
\begin{equation}
\tilde{\upsilon}(\varphi)= C_1J_{\nu}\left(\beta
e^{\alpha\varphi}\right)\cdot
exp\left[-\left(\frac{3}{4\alpha^2}-\frac{\alpha}{2}\right)\varphi\right]
\label{v(phi)-sol}
\end{equation}

  Choosing now the values of $C_1$ so that $\tilde{\upsilon}_{in}=\tilde{\upsilon}(\varphi_{in})$
equals to the initial values considered in the numerical solutions
in Sec.VIIB, we obtain the same five graphs as in Figs.18a and 19
(but here only in the interval $\varphi\leq 15$). The curves
$\tilde{\upsilon}(\varphi)$, which are the projections of the
4-dimensional phase trajectories on the $(\varphi,
\tilde{\upsilon})$-plane, have the following characteristic
features:
\begin{itemize}
\item
From the beginning the solution $\tilde{\upsilon}(\phi)$ remains
almost constant in a long enough $\phi$-interval, that corresponds
to the GSB stage of the inflation The analytic solution
(\ref{v(phi)-sol}) provides  such a regime: the Bessel function of
the first kind $J_{\nu}\left(\beta e^{\alpha\varphi}\right)$
starts from very small values and increases exponentially; it
turns out that the decaying exponential factor in
Eq.(\ref{v(phi)-sol}) compensates the growth of the Bessel
function $J_{\nu}\left(\beta e^{\alpha\varphi}\right)$ with very
high accuracy resulting in a very slow  change of
$\tilde{\upsilon}(\phi)$ during an initial long enough stage of
the power law inflation (see Fig.18a). Note that in contrast to
this the Bessel function of the second kind $J_{\nu}\left(\beta
e^{\alpha\varphi}\right)$ decreases exponentially and therefore
its product with the decaying exponential factor in
Eq.(\ref{v(phi)-sol-1}) makes its relative contribution to the
solution negligible very quickly.

\item
Analytic solutions $\tilde{\upsilon}(\phi)$ with different initial
conditions fall to the same point $C$ of the line
$\tilde{\upsilon}(\varphi)=0$, exactly as in the numerical
solutions in Sec.VIIB, Figs.18a and 19. But now we have the
mathematical explanation of this effect: the point $C$ is the
first zero of the Bessel function. Since the choice of the initial
value $\tilde{\upsilon}(\varphi_{in})$ determines only the
appropriate value of the factor $C_1$ in Eq.(\ref{v(phi)-sol}), it
becomes clear why independently of initial values
$\tilde{\upsilon}(\varphi_{in})$ all the phase trajectories fall
to the same point $C$.

\item
After the point $C$ (see Figs.18a, 19, 28) all the 4-dimensional
phase trajectories approach the attractor
$\left(\frac{d\varphi}{d\tau}=0, \, \tilde{\upsilon}=0, \,
\frac{d\tilde{\upsilon}}{d\tau}=0\right)$ whose projection on the
$(\varphi, \tilde{\upsilon})$-plane we observe in Figs.17a as the
straight interval  after the point $C(-32.8, 0)$. As we see in
Figs.19, 28, approaches to this attractor occur through the very
strong damping of the oscillations of the Higgs field around its
zero value. Note also that locations of all the zeros of the
Bessel function $J_{\nu}\left(\beta e^{\alpha\varphi}\right)$  are
determined only by the parameters of the model and are independent
of the initial conditions.
\end{itemize}

 After
the  point $(\varphi\approx -15, \tilde{\upsilon}=0)$, the
presented analytic solution is not applicable since the
inequality(\ref{gg}) does not hold for $\varphi > -15$ and
therefore ignoring the back reaction of the Higgs field on the
inflaton dynamics does not hold anymore.

The fall of all the phase trajectories (corresponding to different
initial conditions) to the same point $C$ of the attractor
$\tilde{\upsilon}=0$, along with the attractor behavior with
respect to $\frac{d\varphi}{d\tau}=0$, has a decisive role in the
observed effect: the magnitudes of $<\upsilon>^{2}$ and $\phi_{0}$
where $\upsilon$ and $\phi$ stop in the vacuum manifold
(\ref{min-v}) depend very weak of
 the initial conditions. In fact, if different phase
 trajectories would fall to different points of the attractor $\tilde{\upsilon}=0$
  then the degree of damping of their oscillations
 right before they enter into the regime of the last phase transition
 (region $F$ in Fig.18a) could be generically very different.
 Those of them whose deviations from the attractor were not small would
 have the limiting point $(\phi_{0}, <\upsilon>^{2})$ of the phase transition significantly
 different from others.

\subsection{Higgs-Inflaton Symbiosis: Comparison with Hybrid Inflation
 and Summary of the Main Features}

The cosmological scenario studied in this section  includes a
number of dynamical effects demonstrating a strong enough tendency
of the inflaton and Higgs fields to coexistence and interference.
This is why we will refer to the appropriate phenomena as
Higgs-Inflaton Symbiosis (HIS). The dynamics of the HIS model
resembles the hybrid inflation model\cite{Linde-hybrid} but there
are essential differences.

Hybrid inflation models\cite{Linde-hybrid} contain more than one
scalar fields one of which (the inflaton $\phi$) drives the early
(long) stage of inflation and the dynamics of the other(s)
determines the character of the final (very short) stage of
inflation. One usually assumes\cite{Linde-hybrid} that the extra
scalar $\sigma$ is the Higgs field. Due to a direct inflaton-Higgs
coupling the Higgs effective mass is $\phi$ dependent. During the
early stage of inflation the Higgs effective mass square is
positive and may be very large. This is the reason why the Higgs
field is in the false vacuum $\sigma=0$ during the early stage of
inflation. Afterwards, slow roll of the inflaton below a certain
value causes the change of the sign of the Higgs effective mass
square. In other words, the inflaton plays the role of a trigger
which causes transition of the Higgs field into the true minimum
of the scalar sector potential in the Higgs direction. This
transition starts {\it before} the end of inflation because
locations of the minima in the directions of the inflaton and the
Higgs field do not coincide. The end of inflation and spontaneous
breaking of the gauge symmetry appears to be intrinsically
correlated: the last stages of inflation are supported not by the
inflaton potential $V(\phi)$ but by the "noninflationary"
potential $V(\sigma)$. To provide a desirable scenario there is a
need in strong enough restrictions on the parameters of the model.
Note also that this type of models preserves the original feature
of the standard approach to gauge symmetry breaking - the presence
of a tachyonic mass term in the action.

Here we summarize the main features of the HIS model including a
number of differences from hybrid inflation models.
\begin{itemize}
\item

Incorporation of both the inflation and the gauge symmetry
breaking is achieved {\it without tachyonic mass} terms in the
original action (see Eqs.(\ref{totaction}) and
(\ref{V1V2Higgs}),(\ref{m2>0})).

\item
The inflaton-Higgs coupling in Eq.(\ref{Veff+higgs}) disappears in
the inflationary epoch, i.e as the inflaton $\phi\ll -M_{p}$, see
Eq.(\ref{Veff+higgs-early}). Therefore the {\it effective Higgs
mass is constant during the inflation}.

\item
 The positivity of the effective mass square of the
Higgs field in the inflationary epoch (see
Eq.(\ref{Veff+higgs-early})) results directly from the positivity
of the original mass square parameters $\mu_{1}^{2}$ and
$\mu_{2}^{2}$ in the pre-potentials,
Eqs.(\ref{V1V2Higgs}),(\ref{m2>0}).

\item
As a direct result of TMT, the effective scalar sector potential
(\ref{Veff+higgs}) has a structure which provides automatically a
{\it common location} of the minima in the  inflaton and the Higgs
field directions. These minima form the line (\ref{min-v}) in the
plane $(\phi,\upsilon)$. Each point of this line describes a gauge
symmetry broken state with zero vacuum energy density. Instead of
resorting to tachyonic mass terms, these minima are achieved
 by assigning negative values to the constant parts
$V_{1}^{(0)}$, $V_{2}^{(0)}$ of the pre-potentials,
Eqs.(\ref{V1V2Higgs}),(\ref{m2>0}).

\item
Constancy of the effective Higgs mass during a significant part of
the early stage of inflation allows for the Higgs field $\upsilon$
to remain practically equal to its initial value when the Hubble
parameter is very large (GSB inflation).

\item
The fall of the Higgs field to the state $\upsilon =0$ (GSR
inflation) occurs at a quite definite value of the e-folding
before the end of inflation. This value of the e-folding is
practically independent of the initial conditions.

\item
After falling to  $\upsilon =0$ the Higgs field performs quickly
damping oscillations around $\upsilon =0$.

\item
At the end of inflation (i.e. when $w=-1/3$) all the phase
trajectories corresponding to different initial conditions appear
to be very close to the attractor $\left(\frac{d\varphi}{d\tau}=0,
\, \tilde{\upsilon}=0, \,
\frac{d\tilde{\upsilon}}{d\tau}=0\right)$.

\item
The effective mass square of the Higgs field {\it becomes negative
after the end of inflation}. Exponential form of the  $\phi$
dependence in Eq.(\ref{Veff+higgs})  makes this trigger effect
very sharp. The demonstrated extreme closeness of all  phase
trajectories (corresponding to different initial conditions) to
the attractor provides that the phase transition to the gauge
symmetry broken phase  occurs into a very short interval of  the
values of $<\upsilon>$ (and appropriate values of $\phi_0$) in the
vacuum manifold(\ref{min-v}).

\item
The remarkable structure of the scalar sector effective potential
(\ref{Veff+higgs}) (generated in the Einstein frame in our TMT
model) automatically provides that the effective potential in the
vacuum manifold (\ref{min-v}) is equal to zero  without fine
tuning of the parameters of the action and the initial conditions.

\item
Note finally that in contrast with the hybrid inflation models
where strong enough restrictions on the parameters are needed, in
the HIS model qualitatively the same results are obtained in a
very broad range of the parameters satisfying the conditions
(\ref{m2>0}), (\ref{bmu1>bmu2}).

\end{itemize}

\subsection{Particle Creation and Its Effects on  Primordial Perturbations
and \\ the Phase Transition:
  Qualitative Discussion}

  In the previous sections we did not touch quantum effects and their possible role
   in  the studied processes. One of the most important
   quantum effects is the particle creation. Here we want to
   discuss on the qualitative level what kind of influence one can
   expect from the particle creation on the studied processes.
In our qualitative discussion we have to take into account that in
the HIS evolution there are {\it two very different stages where
oscillations of the Higgs field must cause matter creation}.

(1) {\it The first oscillatory stage} consists of the transition
from GSB to GSR inflation and the GSR inflation itself. A rapid
transition of the Higgs field from a nonzero value to the
$\upsilon =0$ phase has to be accompanied with particle creation.
The subsequent coherent damping  oscillations of the Higgs field
around $\upsilon =0$, studied in detail in Secs.VIIB and VIIC must
be responsible for creation of massless particles, like for
example {\it massless gauge bosons}. This dissipative process
results in an additional damping of the Higgs field oscillations.
Hence the quantum effect of particle creation acts here as an
additional factor which forces the phase trajectories
corresponding to different initial conditions to achieve closer
unification with the attractor $\left(\frac{d\varphi}{d\tau}=0, \,
\tilde{\upsilon}=0, \, \frac{d\tilde{\upsilon}}{d\tau}=0\right)$.
As a result of this the final values of $<\upsilon>$ and $\phi_0$
in the symmetry broken phase become more strongly independent of
the initial conditions than in the pure classical model discussed
in the preceding subsections.

   One should note that with the chosen parameters, the
reheating caused by the first stage of the particle creation can
not play an essential role in the thermal history of the universe
because it happens about 50 e-folding before the end of inflation.
However possible effect of the appropriate thermal fluctuations on
the primordial density perturbations perhaps deserves a special
study.

(2) {\it The second stage of the particle creation} is realized
after the end of inflation during the transition from
$\tilde{\upsilon}=0$ phase to the gauge symmetry broken phase.
Coherent oscillations  of the Higgs field around the vacuum
manifold discussed in detail in Sec. VIIB may serve as a very
effective mechanism of the preheating, for example via  gauge
bosons creation. This effect has to be studied in detail, but it
is beyond the goals of this paper. Nevertheless it is interesting
to point out that {\it the back reaction of the particle creation}
on the finish of the phase transition may be very important.
Indeed, as we discussed in the preceding subsections, the
characteristic feature of the final stage of the classical
evolution of the Higgs-inflaton system is that the scalar sector
energy density $\rho$ approaches zero. If there is no other
matter, the friction $\sim \sqrt{\rho}$ caused by the cosmological
expansion also approaches zero and therefore the finish of the
classical evolution of the Higgs-inflaton system becomes
problematic (see item 8 in Sec.VIIB). However, if one takes into
account the quantum effects of matter creation due to the Higgs
field oscillations around the vacuum manifold, this immediately
produces additional friction in the Higgs field equation non
related to the cosmological expansion.  Besides, the presence of
massive gauge bosons, photons and fermions in addition to the
scalar sector (the energy density of the latter is very close to
zero) means that the universe has entered into a radiation/matter
dominated era. Therefore the friction caused by the cosmological
expansion (both in the inflaton and in the Higgs field equations)
is now $\sim \sqrt{\rho_{tot}}$ where $\rho_{tot}$ includes also
the radiation/matter energy density. Due to these two back
reaction effects the Higgs and inflaton fields must quickly
dissipate their kinetic energy and the evolution of the
Higgs-inflaton system has to finish in a certain point of the
vacuum manifold which is practically independent of the initial
conditions.

One should finally point out that in this paper we have restricted
ourselves only with a homogeneous distribution of the initial
values of the Higgs field\footnote{Inhomogeneous distribution of
the inflaton can be incorporated in the framework of a chaotic
inflationary model}. However  the discussed practical independence
of  $<\upsilon>$ of the initial conditions in the framework of the
homogeneous problem may serve as an indication that in a more
realistic case of inhomogeneous distribution of the initial values
of the Higgs field, its value in the final gauge symmetry broken
phase will be practically the same everywhere in our universe.

\section{Discussion and Conclusion}

\subsection{Differences of TMT from the standard field theory in
curved space-time} The main  idea of TMT is that the general form
of the action $\int L\sqrt{-g}d^{4}x$ is not enough in order to
account for some of the fundamental problems of particle physics
and cosmology. The key difference of TMT from the standard field
theory in curved space-time consists in the
hypothesis\cite{GK1}-\cite{GK3} that in addition to the term in
the action with the volume element $\sqrt{-g}d^{4}x$ there should
be one more term where the volume element is metric independent
but rather it is determined either by four (in the 4-dimensional
space-time) scalar fields $\varphi_{a}$ or by a three index
potential $A_{\alpha\beta\gamma}$,  see
Eqs.(\ref{S})-(\ref{Aabg}). We would like to emphasize that
including in the action of TMT the coupling of the Lagrangian
density $L_{1}$ with the measure $\Phi$, we modify in general both
the gravitational and matter sectors as compared with the standard
field theory in curved space-time. Besides we made two more
assumptions: the measure fields ($\varphi_{a}$ or
$A_{\alpha\beta\gamma}$) appear only in the volume element; one
should proceed in the first order formalism. {\it These
assumptions constitute all the modifications of the general
structure of the theory we have made as compared with the standard
field theory in curved space-time}. In fact, the Lagrangian
densities $L_{1}$ and $L_{2}$ in the models we have studied in the
present paper, contain  all types of
 terms which should be present if the standard model would be
formulated in curved space-time. However we do not consider any
exotic terms and it turns out that there is {\it no need for any
exotic term} in order to achieve desirable results. In particular
there is no need for the phantom type terms in $L_{1}$ and $L_{2}$
in order to obtain super-acceleration at the late time universe;
there is no need for tachyonic mass terms in the Higgs
pre-potentials $V_{1}$ and $V_{2}$ in order to obtain spontaneous
breakdown of gauge symmetry.

After making use of the variational principle and formulating the
resulting equations in the Einstein frame, we have seen that in
the absence of fermions, the  dynamics of the scalar sector is
described by the effective action (\ref{k-eff}) which is a
concrete realization of the $k$-essence\cite{k-essence} obtained
from first principles.

\subsection{Short summary of results}

\subsubsection{The early universe inflation}

For pedagogical reasons in Secs.IV-VI we have studied first models
without Higgs field. As $\delta =0$, the dynamics of $\phi$ can be
analyzed by means of its effective potential
(\ref{Veffvac-delta=0}). As $\phi\ll -M_{p}$ the effective $\phi$
potential has the exponential form. We have seen that
independently of the values of the parameters $V_{1}$, $V_{2}$ and
under very general initial conditions, solutions rapidly approach
a regime characterized by a power law inflation. If $\delta\neq
0$, we deal with the  intrinsically $k$-essence dynamics. The
numerical solutions in this case have showed that there is no
qualitative difference of the  power law inflation from the case
with $\delta =0$.

For the model with the Higgs field and $\delta =0$, we have seen
that with the choice of positive mass squared parameters
$\mu_{1}^{2}$ and $\mu_{2}^{2}$, the effective  potential of the
scalar sector in the very early universe,
Eq.(\ref{Veff+higgs-early}), describes the Higgs field with
positive mass square\footnote{Note that the choice of the negative
parameters $\mu_{1}^{2}$, $\mu_{2}^{2}$ would result in the
tachyonic mass term in the effective potential
(\ref{Veff+higgs-early}) in the early universe; in such a case the
minimization of the effective potential in the early inflation
epoch could not be realized in the phase $\upsilon =0$.}, i.e. the
minimization of the effective potential (\ref{Veff+higgs-early})
in the Higgs direction in the inflationary epoch  is achieved if
the Higgs field is in the unbroken symmetry phase $\upsilon =0$.
The inflationary epoch with $\upsilon =0$ we called the GSR
inflation. However in contrast with the hybrid
inflation\cite{Linde-hybrid}, before the GSR inflation the
universe suffers a long lasting GSB inflation, where the Higgs
field is in a broken gauge symmetry phase with slow varying
$\upsilon\neq 0$. With our choice of the parameters, the
transition from the GSB inflation to the GSR inflation occurs
about 60 e-folding before the end of inflation, see Fig.22.

\subsubsection{End of inflation in models without the Higgs field}

 In these toy models there
are three regions of the parameters $V_{1}$ and $V_{2}$ and
appropriate three shapes of the effective potentials, Fig.1.
Therefore three different types of scenarios for exit from
inflation can be realized:

a) $V_{1}>0$ and $b_{g}V_{1}>2V_{2}$, \, Sec.IV. In this case the
power law inflation monotonically transforms to the late time
inflation asymptotically governed by the cosmological constant
$\Lambda_1$.

b) $V_{2}<b_{g}V_{1}<2V_{2}$, \, Sec.V. In this case the power law
inflation is ended without oscillations at the final value
$\phi_{min}$,
 corresponding to the (non zero) minimum of the effective
potential,.

c) $V_{1}<0$ and $V_{2}<0$, \,  Sec.VI. In this case the power law
inflation is ended with damped oscillations of $\phi$ approaching
the point of the phase plane ($\phi =\phi_0, \dot\phi =0$) where
the vacuum energy $V_{eff}^{(0)}(\phi_0)=0$. This occurs without
fine tuning  of the parameters $V_{1}$, $V_{2}$, $\delta$ and the
initial conditions.

\subsubsection{ New cosmological mechanism for gauge symmetry
breaking
\\
 and reheating}.

From the summary of the main features of the HIS (see Sec. VIID)
one can conclude that our TMT model implies  a {\it new
cosmological mechanism for a transition to a gauge symmetry broken
phase soon after the end of inflation}. From the field theory
point of view, one of the advantages of the model is that the
gauge symmetry breaking is obtained {\it without tachyonic mass
terms} in the original action. Another important advantage of the
model is that the inflaton driven transition to the the gauge
symmetry broken phase is automatically (i.e. without tuning)
accompanied with approaching zero of the scalar sector energy
density. From the point of view of the realistic cosmology, the
scenario based on the HIS model exhibits a new mechanism of
reheating which most likely may be very effective: coherent
oscillations of the Higgs field must result in intensive creation
of the gauge bosons. Besides, transition to the gauge symmetry
broken phase starts soon after the end of the GSR inflation where
$\upsilon=0$ and therefore masses of quantum of all the gauge
fields are very close to zero at the beginning of the reheating.

\subsubsection{Cosmological constant problems}

\vspace{0.5cm}

{\it The old cosmological constant problem}.
 In Secs.VI and VII,
we have seen in details that if $V_1<0$ then, for a broad range of
other parameters, the vacuum energy turns out to be zero without
fine tuning; this is possible both in the model with $\phi$ as the
scalar sector and in the model with $\phi$ and $\upsilon$. This
effect is a direct consequence of the result of the TMT: the
effective scalar sector potential generated in the Einstein frame
is proportional to {\it a perfect square}. If such type of the
structure for the scalar field potential in a usual (non TMT)
model would be chosen "by hand" it would be a sort of fine tuning.

There is  a need to explain how this result avoids the well known
no-go theorem by Weinberg\cite{Weinberg1} stating that generically
in field theory one cannot have zero value of the potential in the
minimum without fine tuning. The reason why this theorem is not
applicable here is that one of the basic assumptions of the
theorem does not hold. Indeed in the mentioned theorem, it is
assumed that all fields in the vacuum are constants. In our case,
the basic assumption that the measure of integration in the action
$\Phi\neq 0$ implies that  all the measure fields ($\varphi_{a}$
($a=1,2,3,4$) in the definition (\ref{Phi}) or
$A_{\alpha\beta\gamma}$ in the definition (\ref{Aabg})) have non
vanishing space-time gradients.

In this paper we have restricted ourselves with the simplest form
(\ref{V1V2Higgs}) of the Higgs field dependence of the
pre-potentials $V_{1}$ and $V_{2}$. If however we had studied
models with more complicated Higgs field dependence in $V_{1}$ and
$V_{2}$, including for example the quartic Higgs
self-interactions, then there could be  vacua with zero
cosmological constant disconnected from each other (again without
fine tuning). This is an explicit realization of the "Multiple
Point Principle" proposal\cite{Nielsen} which is based on the idea
that if there is a mechanism that sets a certain state to have a
zero cosmological constant  then the same mechanism  may act also
in  other  field configurations with the same result.

\vspace{0.5cm} {\it The new cosmological constant problem}.

Interesting result following from the general structure of the
scale invariant TMT model with $V_1>0$ is that the cosmological
constant $\Lambda_1$, Eq.(\ref{lambda-without-ferm-delta=0}), is a
ratio of quantities constructed from pre-potentials $V_{1}$,
$V_{2}$ and the dimensionless parameter $b_{g}$. Such structure of
$\Lambda_1$ allows to propose at least two independent ways (see
Sec.IV) for resolution of the problem of the smallness of
$\Lambda_1$ that should be $\Lambda_1\sim (10^{-3}eV)^{4}$.

The first way is a kind of a {\it seesaw} mechanism\cite{seesaw}.
For instance, if  $V_{1}\sim (10^{3}GeV)^{4}$ and $V_{2} \sim
(10^{18}GeV)^{4}$ then $\Lambda_1\sim (10^{-3}eV)^{4}$.

The second way is realized if the dimensionless parameters
$b_{g}$, $b_\phi$, etc. of the action (\ref{totaction}) are huge
numbers. For example, if $V_{1}\sim (10^{3}GeV)^{4}$ then for
getting $\Lambda_1\sim (10^{-3}eV)^{4}$ one should assume that
$b_{g}\sim 10^{60}$. It is interesting that this idea may be
treated as an indication that there is a connection between the
resolution of the new cosmological constant problem and a new type
of {\it the correspondence principle} in the action, see details
in item (b) of Sec.IV.

\subsubsection{Possibility of the late time  super-acceleration.
Speculations about realistic cosmological scenario. }

We have seen in this paper that even restricting ourselves with
only the scalar sector we discover a  number of interesting
effects which may be useful in the construction of a realistic
scenario of the evolution of the universe. Studying these effects
makes it difficult in the same paper to address also questions
like particle creation, perturbations, etc.. For this reason we
cannot yet present a more complete cosmological scenario. In
particular, we have seen in Sec.V that there is a region of the
parameters where the late time universe can evolve with
equation-of-state $w<-1$ without introducing explicit phantom
terms in the action. This occurs due to the very interesting
effect of the classical field  dynamics  described in Sec.V in
detail. However  it is impossible to obtain {\it a pure classical
solution} which connects the early universe inflation with the
late time super-acceleration. This problem  may be related with
the toy character of the scenario where the role of the matter has
been ignored.

In TMT, the effects related to  particle creation can change the
dynamics of the scalar sector. In fact, if the particle creation
results in the matter domination epoch, then in the presence of
fermions, the constraint (\ref{constraint2}) differs from the
constraint in the absence of fermions, Eq.(\ref{constr-vac}).
Therefore in the stage of a transition from the scalar sector
(dark energy) domination epoch to the matter domination epoch,
$\zeta$ must significantly vary that will lead to a change in the
dynamics of the scalar sector due to the $\zeta$-dependence of the
scalar sector potential (\ref{Veff1}). Similar conclusion is true
also for the dynamics of the scalar sector in the matter
domination epoch itself. In such a case the matter creation and
subsequent matter domination epoch might be "bridges" connecting
the early inflation with the possible late time
super-acceleration. This should include also a so far unclear
mechanism responsible for the freezing of  the Higgs  vacuum
expectation value $<\upsilon>$ to survive till the late time
universe. This kind of questions we are going to study in the near
future.

\subsection{What can we expect from quantization}

In this paper we have studied only classical two measures field
theory and its effects in the context of cosmology. However
quantization of TMT as well as influence of quantum effects on the
processes explored in this paper may have a crucial role. We
summarize here some ideas and speculations which gives us a hope
that quantum effects can keep and even strengthen the main results
of this paper.

1) Recall first two fundamental facts of TMT as a classical field
theory: (a) The measure degrees of freedom appear in the equations
of motion only via the scalar $\zeta$, Eq.(\ref{zeta}); (b) The
scalar $\zeta$ is determined by  the constraint which is nothing
but a consistency condition of the equations of motion (see
Eqs.(\ref{app1}), (\ref{app3}) and (\ref{app4}) in Appendix B).
Therefore the constraint plays a key role in TMT. Note however
that if we were ignore the gravity from the very beginning in the
action (\ref{totaction}) then instead of the constraint
(\ref{app4}) we would obtain Eq.(\ref{app1}) (where one has to put
zero the scalar curvature). In such a case we would  deal with a
different theory. This notion shows that the gravity and matter
intertwined in TMT in a much more complicated manner than in GR.
Hence introducing the new measure of integration $\Phi$ we have to
expect that the quantization of TMT may be a complicated enough
problem. Nevertheless we would like here to point out that in the
light of the recently proposed idea of Ref.\cite{Giddings}, the
incorporation of four scalar fields $\varphi_a$  together with the
scalar density $\Phi$, Eq.(\ref{Phi}), (which in our case are the
measure fields and the new measure of integration respectively),
is  a possible way to define local observables in the local
quantum field theory approach to quantum gravity. We regard this
result as an indication that the effective gravity $+$ matter
field theory has to contain the new measure of integration $\Phi$
as it is in TMT.

2)  The assumption formulated in item 2 in Sec.IIA, that the
measure fields $\varphi_a$ (or $A_{\alpha\beta\gamma}$) appear in
the action (\ref{S}) {\it only} via the measure of integration
$\Phi$, has a key role in the TMT results and in particular for
the resolution of the old cosmological constant problem, see
Sec.VI. In principle one can think of breakdown of such a
structure by quantum corrections. However, fortunately there
exists an infinite dimensional symmetry mentioned in item 2 of
Sec.IIA which, as we hope, is able to protect  the postulated
structure of the action from a deformation caused by quantum
corrections or at least to suppress such a quantum anomaly in
significant degree. Therefore one can hope that the proposed
resolution of the old cosmological constant problem holds in the
quantized TMT as well.

3) As we have noticed in Sec.VIIA, the vacuum manifold
(\ref{min-v}) of the HIS is degenerate: the effective potential of
the scalar sector remains equal to zero when shifting along this
line in the $(\phi,\upsilon)$ plane. However this degeneracy has
no relation to any symmetry of the action, and in particular the
kinetic energy of the scalar sector does not respect this shift as
a symmetry. Therefore there are no reasons for quantum corrections
not to break this shift symmetry of the vacuum manifold.
Disappearance of the degeneracy of the vacuum manifold provides us
with two important conclusions: a) our model is out of danger to
run against the problem of the appearance of a pseudo-Goldstone
boson; b) transition to the gauge symmetry broken phase will end
in a certain point $(\phi_0,<\upsilon>^2)$ instead of a short
interval of the line (\ref{min-v}) as it was in the pure classical
problem studied in Sec.VII.

4) Recall finally that such quantum effects as the first and the
second stages of particle creation  discussed in details in
Sec.VIIE, supplement and strengthen the obtained classical
results: the first  stage of particle creation allows to hope that
VEV of the Higgs field is independent of the initial conditions;
the second stage of particle creation should work as a mechanism
of dissipation responsible for realization of the final state of
the HIS in the gauge symmetry broken phase.

Exploration of the quantization problems of TMT will be the
subject of forthcoming research.

\section{Acknowledgements}

We thank  L. Amendola, S. Ansoldi, R. Barbieri, J. Bekenstein, A.
Buchel, A. Dolgov,  S.B. Giddings, P. Gondolo, J.B. Hartle, B-L.
Hu, P.Q. Hung, D. Kazanas, D. Marolf, D.G. McKeon, J. Morris, V.
Miransky,
 H. Nielsen, Y.Jack Ng, H. Nishino, E. Nissimov,  S. Pacheva, L.
Parker, R. Peccei, M. Pietroni, S. Rajpoot, R. Rosenfeld,  V.
Rubakov, E. Spallucci, A. Vilenkin, S. Wetterich and  A. Zee for
helpful conversations on different stages of this research. Our
special gratitude to M. Gedalin for his advice concerning some
questions of dynamical systems and to L. Prigozhin for great help
in implementation of numerical solutions.

\appendix

\section{Equations of motion  in
the original frame}

Variation of the measure fields $\varphi_{a}$ with the condition
$\Phi\neq 0$ leads, as we have already seen in Sec.II, to the
equation $ L_{1}=sM^{4}$ where $L_{1}$ is now defined, according
to  Eq. (\ref{S}), as the part of the integrand of the action
(\ref{totaction}) coupled to the measure $\Phi$.  Equation
(\ref{varphi}) in the context of the model (\ref{totaction}) reads
(with the choice $s=1$):
\begin{eqnarray}
&&\left[-\frac{1}{\kappa}R(\omega, e)+
\frac{1}{2}g^{\mu\nu}\phi_{,\mu}\phi_{,\nu}
+\frac{1}{2}g^{\mu\nu}(D_{\mu}H)^{\dag}D_{\nu}H
\right]e^{\alpha\phi /M_{p}} -V_{1}e^{2\alpha\phi /M_{p}} +L_{fk}
e^{\alpha\phi /M_{p}}
\nonumber\\
&&-\frac{\upsilon}{\sqrt{2}}\left[
f_{N}\overline{N}N+f_{E}\overline{E}E\right] =M^{4}, \label{app1}
\end{eqnarray}
where the representation of the Higgs field $H$ in the unitary
gauge
\begin{equation}
H=
\begin{pmatrix} 0 \\
 2^{-1/2}\upsilon (x)
 \end{pmatrix}
\end{equation}
has been used in the last term of the l.h.s. of Eq.(\ref{app1});
to simplify the coming calculations it is convenient in the
meanwhile not to do this explicitly with $D_{\mu}H$ which includes
interactions to gauge fields.

It can be noticed that the appearance of a nonzero integration
constant $M^{4}$ spontaneously breaks the scale invariance
(\ref{stferm}).

Variation of the action (\ref{totaction}) with respect to
$e^{a\mu}$ yields
\begin{eqnarray}
&&(\zeta +b_{g})\left[-\frac{2}{\kappa}R_{a,\mu}(\omega ,e) +
e^{\beta}_{a}(D_{\mu}H)^{\dag}D_{\beta}H \right] +(\zeta
+b_{\phi})e^{\beta}_{a}\phi_{,\mu}\phi_{,\beta}
+\frac{1}{\sqrt{-g}}\frac{\partial(\sqrt{-g}L_{gauge}) }{\partial
e^{a,\mu}} e^{-\alpha\phi /M_{p}}
\nonumber\\
&+& g_{\mu\beta}e^{\beta}_{a}\left[\frac{b_{g}}{\kappa}R(\omega
,e) -\frac{b_{\phi}}{2}g^{\mu\nu}\phi_{,\mu}\phi_{,\nu}
-\frac{b_{g}}{2}g^{\mu\nu}(D_{\mu}H)^{\dag}D_{\nu}H
+V_{2}e^{\alpha\phi /M_{p}}\right]
\nonumber\\
&+& g_{\mu\beta}e^{\beta}_{a}\left[-kL_{fk}
+\frac{\upsilon}{\sqrt{2}}\left(h_{N}f_{N}\overline{N}N
+h_{E}f_{E}\overline{E}E\right)e^{\frac{1}{2}\alpha\phi /M_{p}}
\right]
\nonumber\\
&+& (\zeta +b)\frac{i}{2} \left[\overline{L}_{L}
\left(\gamma^{a}\overrightarrow{D}_{\mu}
-\overleftarrow{D}_{\mu}\gamma^{a}\right)L_{L} +\overline{E}_{R}
\left(\gamma^{a}\overrightarrow{D}_{\mu}
-\overleftarrow{D}_{\mu}\gamma^{a}\right)E_{R} +\overline{N}_{R}
\left(\gamma^{a}\overrightarrow{D}_{\mu}
-\overleftarrow{D}_{\mu}\gamma^{a}\right)N_{R} \right] =0.
\label{app2}
\end{eqnarray}

Contraction of Eq.(\ref{app2}) with $e^{a,\mu}$ gives
\begin{eqnarray}
&&(\zeta -b_{g})\left[-\frac{1}{\kappa}R(\omega, e)+
+\frac{1}{2}g^{\mu\nu}(D_{\mu}H)^{\dag}D_{\nu}H \right] +(\zeta
-b_{\phi})\frac{1}{2}g^{\mu\nu}\phi_{,\mu}\phi_{,\nu}
+2V_{2}e^{\alpha\phi /M_{p}} +\frac{1}{2}(\zeta -3k)L_{fk}
\nonumber\\
&+&\frac{2\upsilon}{\sqrt{2}}\left(h_{N}f_{N}\overline{N}N
+h_{E}f_{E}\overline{E}E\right) e^{\frac{1}{2}\alpha\phi /M_{p}}
=0, \label{app3}
\end{eqnarray}
where the identity $e^{a\mu}\left(\partial\sqrt{-g}L_{gauge}/
\partial e^{a\mu}\right)\equiv 0$ has been used.

Excluding $R(\omega, e)$ from Eqs.(\ref{app1}) and (\ref{app3}) we
obtain the {\it consistency condition} of these two equations:
\begin{eqnarray}
&& (\zeta -b_{g})\left(M^{4}e^{-\alpha\phi
/M_{p}}+V_{1}e^{\alpha\phi /M_{p}}\right)+2V_{2}e^{\alpha\phi
/M_{p}}+(b_{g}-b_{\phi})\frac{1}{2}g^{\mu\nu}\phi_{,\mu}\phi_{,\nu}
 -\frac{1}{2}(\zeta-2b_{g}+3k)L_{fk}
\nonumber\\
&+&
\frac{\upsilon}{\sqrt{2}}\left[(\zeta-b_{g}+2h_{N})f_{N}\overline{N}N
+(\zeta-b_{g}+2h_{E})f_{E}\overline{E}E)\right]e^{\frac{1}{2}\alpha\phi
/M_{p}} =0, \label{app4}
\end{eqnarray}

It is well known that  the Lagrangian for a single fermion field
$\Psi$ equals zero on the mass-shell. This results from the
linearity of the Lagrangian both in $\Psi$ and in
$\overline{\Psi}$. The same is true for a system of fermion fields
in our case because in spite of the presence of interactions, the
fermion sector in (\ref{totaction}) is linear in all fermion
degrees of freedom. Therefore we have
\begin{equation}
(\zeta +k)L_{fk}= \frac{\upsilon}{\sqrt{2}} \left[(\zeta
+h_{N})f_{N}\overline{N}N +(\zeta
+h_{E})f_{E}\overline{E}E\right]e^{\frac{1}{2}\alpha\phi /M_{p}}
\label{app5}
\end{equation}

Inserting $L_{fk}$ into the consistency condition,
Eq.(\ref{app4}), we get {\em the constraint (in the original
frame)}
\begin{eqnarray}
&&(\zeta -b_{g})\left(M^{4}e^{-2\alpha\phi /M_{p}}+V_{1}\right)
+2V_{2}+(b_{g}-b_{\phi})e^{-\alpha\phi
/M_{p}}\frac{1}{2}g^{\mu\nu}\phi_{,\mu}\phi_{,\nu} \nonumber \\
&+&\frac{\upsilon}{2\sqrt{2}(\zeta +k)}e^{-\frac{1}{2}\alpha\phi
/M_{p}}\sum_{i=N,E}(\zeta-\zeta_1^{(i)})(\zeta-\zeta_1^{(2)})\overline{\Psi}_{i}\Psi_{i}
=0. \label{app10}
\end{eqnarray}
where $\Psi_{i}$ ($i=N,E$) is the general notation for the
primordial fermion fields $N$ and $E$ and
\begin{equation}
\zeta_{1,2}^{(i)}=\frac{1}{2}\left[k-3h_{i}\pm\sqrt{(k-3h_{i})^{2}+
8b_g(k-h_{i}) -4kh_{i}}\,\right], \qquad i=N,E. \label{zetapm}
\end{equation}

Contracting Eq.(\ref{app2}) with factor $e^{a}_{\nu}$ and using
Eqs.(\ref{app3}), (\ref{app5}) we get
\begin{eqnarray}
 \frac{2}{\kappa}R_{\mu\nu}(\omega, e)&=&
 \frac{\zeta +b_{\phi}}{\zeta +b_g}\phi_{,\mu}\phi_{,\nu}
+(D_{\mu}H)^{\dag}D_{\nu}H
+\frac{2}{\sqrt{-g}}\frac{\partial\sqrt{-g}L_{gauge}}{\partial
g_{\mu\nu}} e^{-\alpha\phi /M_{p}}\nonumber\\
 &-&g_{\mu\nu}\frac{1}{\zeta
+b_g}\left[b_gM^{4}e^{-\alpha\phi /M_{p}}
      +(b_gV_{1}-V_{2})e^{\alpha\phi /M_{p}}
      -(b_g-b_{\phi})\frac{1}{2}g^{\alpha\beta}\phi_{,\alpha}\phi_{,\beta}\right]
\nonumber\\
&+&\frac{\zeta +k}{\zeta +b_g} \frac{i}{2}e_{a\nu}
\left[\overline{L}_{L} \left(\gamma^{a}\overrightarrow{D}_{\mu}
-\overleftarrow{D}_{\mu}\gamma^{a}\right)L_{L} +\overline{E}_{R}
\left(\gamma^{a}\overrightarrow{D}_{\mu}
-\overleftarrow{D}_{\mu}\gamma^{a}\right)E_{R} +\overline{N}_{R}
\left(\gamma^{a}\overrightarrow{D}_{\mu}
-\overleftarrow{D}_{\mu}\gamma^{a}\right)N_{R} \right]
\nonumber\\
&+&g_{\mu\nu}\frac{\upsilon}{\sqrt{2}(\zeta
+k)}e^{\frac{1}{2}\alpha\phi /M_{p}}
\left[(h_{N}-k)f_{N}\overline{N}N+(h_{E}-k)f_{E}\overline{E}E\right].
\label{app12}
\end{eqnarray}

The scalar field $\phi$ equation of motion in the original frame
can be written in the form
\begin{eqnarray}
&& \frac{1}{\sqrt{-g}}\partial_{\mu}\left[e^{\alpha\phi /M_{p}}
(\zeta +b_{\phi})\sqrt{-g}g^{\mu\nu}\partial_{\nu}\phi\right]
\nonumber\\
&-&\frac{\alpha}{M_{p}}e^{\alpha\phi /M_{p}}\left[(\zeta
+b_g)M^{4}e^{-\alpha\phi /M_{p}}+
[(b_g-\zeta)V_{1}-2V_{2}]e^{\alpha\phi /M_{p}}
+(b_g-b_{\phi})\frac{1}{2}g^{\alpha\beta}\phi_{,\alpha}\phi_{,\beta}\right]
\nonumber\\
&=&-\frac{\alpha}{M_{p}}e^{\frac{3}{2}\alpha\phi/M_{p}}
\frac{\upsilon}{2\sqrt{2}(\zeta +k)}
\sum_{i=N,E}(\zeta-\zeta_1^{(i)})(\zeta-\zeta_1^{(2)})\overline{\Psi}_{i}\Psi_{i},
\label{phi-orig}
\end{eqnarray}
where Eqs.(\ref{app1}) and (\ref{app5})  have been used.

In the vacuum of the gauge fields, the Higgs field equation in the
unitary gauge reads
\begin{equation}
\frac{1}{\sqrt{-g}}\partial_{\mu}\left[e^{\alpha\phi /M_{p}}
(\zeta +b_g)\sqrt{-g}g^{\mu\nu}\partial_{\nu}\upsilon\right] +
e^{2\alpha\phi /M_{p}}[\zeta V^{\prime}_1+V^{\prime}_2]=
-\frac{\upsilon}{\sqrt{2}} \, e^{\frac{3}{2}\alpha\phi/M_{p}}
\sum_{i=N,E}f_i(\zeta +h_i)\overline{\Psi}_{i}\Psi_{i},
\label{Higgs-orig}
\end{equation}
where $V_{i}^{\prime}\equiv dV_i/d\upsilon$,  $(i=1,2)$.

The gauge field $B_{\mu}$ equation in the unitary gauge and in the
fermion vacuum reads
\begin{equation}
\partial_{\nu}\left[\sqrt{-g}g^{\alpha\mu}g^{\beta\nu}B_{\alpha\beta}\right]
+ e^{\alpha\phi /M_{p}} (\zeta
+b_g)\sqrt{-g}g^{\mu\nu}\frac{1}{8}g^{\prime 2}\upsilon^2B_{\nu}=0
\label{gauge-orig}
\end{equation}
and similar for $\vec{W}_{\mu}$.

The fermion $\Psi_i$ equation in the unitary gauge and for
simplicity, in the gauge fields vacuum has the following form
\begin{eqnarray}
&&ie_a^{\mu}\left[\gamma^a\vec{\partial}_{\mu}
+\frac{1}{4}\omega_{\mu}^{cd}\left(\gamma^a\sigma_{cd}+\sigma_{cd}\gamma^a\right)
\right]\Psi_i + \frac{ie^{-\alpha\phi /M_{p}}}{2(\zeta
+k)\sqrt{-g}}\left[\partial_{\mu}\left((\zeta
+k)\sqrt{-g}e^{\alpha\phi
/M_{p}}e_a^{\mu}\right)\right]\gamma^a\Psi_i \nonumber\\
&-&\frac{\zeta +h_i}{\zeta
+k}e^{\frac{1}{2}\alpha\phi/M_{p}}\frac{f_i\upsilon}{\sqrt{2}}\Psi_i=0.
 \label{ferm-orig}\end{eqnarray}

\bigskip
\section{Connection in the original and Einstein frames}
\bigskip

We present here what is the dependence of the spin connection
$\omega_{\mu}^{ab}$ on $e^{a}_{\mu}$, $\zeta$, $\Psi$ and
$\overline{\Psi}$. Varying the action (\ref{totaction}) with
respect to $\omega_{\mu}^{ab}$ and making use that
\begin{equation}
R(V,\omega)\equiv
-\frac{1}{4\sqrt{-g}}\varepsilon^{\mu\nu\alpha\beta}\varepsilon_{abcd}
e^{c}_{\alpha}e^{d}_{\beta}R_{\mu\nu}^{ab}(\omega) \label{C1}
 \end{equation}
we obtain
\begin{equation}
\varepsilon^{\mu\nu\alpha\beta}\varepsilon_{abcd}\left[(\zeta
+b_{g}) e^{c}_{\alpha}D_{\nu}e^{d}_{\beta}
+\frac{1}{2}e^{c}_{\alpha}e^{d}_{\beta}\left(\zeta_{,\nu}+\frac{\alpha}{M_{p}}\phi_{,\nu}\right)\right]+
\frac{\kappa}{4}\sqrt{-g}(\zeta
+k)e^{c\mu}\varepsilon_{abcd}\overline{\Psi}
\gamma^{5}\gamma^{d}\Psi=0, \label{C2}
 \end{equation}
where
\begin{equation}
D_{\nu}e_{a\beta}\equiv\partial_{\nu}e_{a\beta} +\omega_{\nu
a}^{d}e_{d\beta} \label{C3}
 \end{equation}
The solution of Eq. (\ref{C2}) is represented in the form
\begin{equation}
\omega_{\mu}^{ab}=\omega_{\mu}^{ab}(e)   +
K_{\mu}^{ab}(e,\overline{\Psi},\Psi) + K_{\mu}^{ab}(\zeta, \phi)
\label{C4}
 \end{equation}
where
\begin{equation}
\omega_{\mu}^{ab}(e)=e_{\alpha}^{a}e^{b\nu}\{
^{\alpha}_{\mu\nu}\}- e^{b\nu}\partial_{\mu}e_{\nu}^{a} \label{C5}
 \end{equation}
is the Riemannian part of the spin-connection,
\begin{equation}
K_{\mu}^{ab}(e,\overline{\Psi},\Psi)= \frac{\kappa}{8}\frac{\zeta
+k}{\zeta +b_{g}}
\eta_{cn}e_{d\mu}\varepsilon^{abcd}\overline{\Psi}
\gamma^{5}\gamma^{n}\Psi \label{C7}
 \end{equation}
is the fermionic contribution that differs from the familiar one
\cite{..},\cite{Gasperini} by the factor $\frac{\zeta +k}{\zeta
+b_{g}}$ and
\begin{equation}
K_{\mu}^{ab}(\zeta, \phi)=\frac{1}{2(\zeta
+b_{g})}\left(\zeta_{,\alpha}+\frac{\alpha}{M_{p}}\phi_{,\alpha}\right)(e_{\mu}^{a}e^{b\alpha}-
e_{\mu}^{b}e^{a\alpha}) \label{C6}
 \end{equation}
is  the non-Riemannian part of the spin-connection originated by
specific features of TMT.

In the Einstein frame, i.e. in terms of variables defined by
Eq.(\ref{ctferm}), the spin-connection read
\begin{equation}
\omega^{\prime ab}_{\mu}=\omega^{ab}_{\mu}(\tilde{e}) +
\frac{\kappa}{8}
\eta_{cn}\tilde{e}_{d\mu}\varepsilon^{abcd}\overline{\Psi}^{\prime}
\gamma^{5}\gamma^{n}\Psi^{\prime} \label{C7}
 \end{equation}
 which is exactly the spin-connection of the Einstein-Cartan
 space-time\cite{Gasperini} with the vierbein $\tilde{e}^{a}_{\mu}$.

 \section{General Relativity and Regular Fermions}

\subsection{Reproducing Einstein Equations and Regular Fermions}

In Sec.III we have seen that in the absence of fermions case (and
if $\delta =0$) the gravitational equations (\ref{gef}) coincide
with the Einstein equations. Analyzing
Eqs.(\ref{gef})-(\ref{muferm1}) in more general cases it is easy
to see that Eqs.(\ref{gef}) and (\ref{Tmn}) are reduced to the
Einstein equations in the appropriate field theory model (i.e.
when the scalar field, electromagnetic field and massive fermions
are sources of gravity) if $\zeta$ is constant and
\begin{equation}
\Lambda_{dyn}^{ferm}=0 \qquad \text{or at least} \qquad
|T_{\mu\nu}^{(ferm,noncan)}|\ll |T_{\mu\nu}^{(ferm,can)}|.
\label{noncan-ll-can}
\end{equation}

According to Eqs.(\ref{Tmn-noncan})-(\ref{muferm1}), in the case
when a single massive fermion is a source of gravity, the
condition (\ref{noncan-ll-can}) is realized  if
\begin{equation}
Z_{i}(\zeta)\approx 0 \quad \Longrightarrow \quad
\zeta=\zeta_{1}^{i} \quad or \quad \zeta=\zeta_{2}^{i}, \quad
i=N^{\prime},E^{\prime}, \label{Z-0}
\end{equation}
where $\zeta_{1,2}^{i}$ are defined in Eqs.(\ref{Zeta}).

Recall that existence of a noncanonical contribution to the
energy-momentum tensor (\ref{Tmn-noncan}), along with the $\zeta$
dependence of the fermion mass (\ref{muferm1}) discovered in
Sec.IIC, displays  the fact that generically primordial fermion is
very much different from the regular one  (see definitions in item
(ii) of the Introduction section). To answer the question what are
the characteristic features of the regular massive fermion we have
to take into account the undisputed fact that the classical tests
of GR deal only with regular fermionic matter. Hence {\it we
should identify the regular fermions with states of the primordial
fermions satisfying the  condition} (\ref{Z-0}).

\subsection{Meaning of the Constraint and Regular Fermions}

We are going  now  to understand the meaning of the constraint
(\ref{constraint2}). We start from the detailed analysis of two
limiting cases: (a) in space-time regions without  fermions; \,
(b) in space-time regions occupied by regular fermions; \, and
afterwards we will be able to formulate the meaning of the
constraint in general case.

We will proceed choosing the parameters $V_1$, $V_2$ and/or $b_g$
 as in Sec.IV. It is convenient to divide the analysis into a few steps:

(1) It follows from the condition $b_gV_1>2V_2$ that
$\zeta_{0}(\phi)$ determined by the constraint in the absence of
fermions case,  Eq.(\ref{zeta-without-ferm-delta=0}), has the same
order of magnitude as the parameter $b_g$.

(2) Recall that $V_{eff}^{(0)}(\phi)$, Eq.(\ref{Veffvac-delta=0}),
having the order of magnitude typical for the dark energy density
(in the absence of fermions case) is obtained from
$V_{eff}(\phi;\zeta)$, Eq.(\ref{Veff1}), as $\zeta
=\zeta_{0}(\phi)$. Therefore with the help of the item (1) we
conclude that  each time when $\zeta$ has the order of magnitude
close to that of the parameter $b_g$ (and if no special tuning is
assumed) $V_{eff}(\phi;\zeta)$, Eq.(\ref{Veff1}), has the order of
magnitude close to that of the dark energy density (in the absence
of fermions case).

(3) It is easy to see that  {\it each time when} $\zeta$ {\it has
the order of magnitude close to that of the parameter} $b_g$ (if
no special tuning is assumed and in particular $\zeta\neq
\zeta_{0}(\phi)$), the left hand side (l.h.s.) of the constraint
(\ref{constraint2}) has the order of magnitude close to that of
$V_{eff}(\phi;\zeta)$, Eq.(\ref{Veff1}), i.e. {\it the l.h.s. of
the constraint has the order of magnitude close to that of the
dark energy density in the absence of fermions case}.

(4) Let us now turn  to the right hand side (r.h.s.) of the
constraint (\ref{constraint2}) in the presence of a single massive
primordial fermion. It contains factor
$m_{i}(\zeta)\overline{\Psi}^{\prime}_{i}\Psi^{\prime}_{i}$, \,
($i=N^{\prime},E^{\prime}$) which have typical order of magnitude
of the fermion canonical energy density $T_{00}^{(ferm,can)}$.
 If the
primordial fermion is in a state of a regular fermion then
according to the conclusion made at the end of the previous
subsection,  in the space-time region where the fermion is
localized, the scalar $\zeta$ must be $\zeta=\zeta_{1}^{i}$ (or
$\zeta=\zeta_{2}^{i}$).  Therefore in the space-time region
occupied by a single regular fermion, the r.h.s. of
(\ref{constraint2}) is
\begin{equation}
\Lambda_{dyn}^{(ferm)}|_{regular}\equiv
Z_{i}(\zeta_{1,2}^{i})m_{i}(\zeta_{1,2}^{i})
\left(\overline{\Psi^{\prime}_{i}}\Psi^{\prime}_{i}\right)_{regular}
\label{rhs-reg}
\end{equation}

Due to our assumption that the dimensionless parameters $b_g$,
$b_{\phi}$, $k$, $h_N$ and $h_E$ have close orders of magnitude
(see paragraph after Eq.(\ref{Lgauge})), it follows from the
definitions (\ref{Zeta}) that both $\zeta_{1}^{i}$ and
$\zeta_{2}^{i}$ have the order of magnitude close to that of
$b_g$. Hence in the space-time region occupied by a single regular
fermion, the l.h.s. of (\ref{constraint2}) has the order of
magnitude close to that of the dark energy density in the fermion
vacuum. It is evident that {\it in normal particle physics
conditions} (see the last paragraph of Sec.II), that is when the
energy density of a single fermion $\sim
m_{i}(\zeta)\overline{\Psi}^{\prime}_{i}\Psi^{\prime}_{i}$
 is tens of orders of magnitude larger than the
fermion vacuum energy density, the balance dictated by the
constraint is satisfied in the present day universe just due to
the condition (\ref{Z-0}).

5 In more general cases, i.e. when primordial fermion is in a
state different from the regular one, the meaning of the
constraint is similar: {\it the balance between the scalar dark
energy contribution  to the l.h.s. of the constraint and  the
 fermionic contribution to
the r.h.s. of the constraint} is realized due to the factors
$Z_{i}(\zeta)$. In other words, {\it the constraint describes the
local balance between the fermion energy density and the scalar
dark energy density} in the space-time region where the  wave
function of the primordial fermion is not equal to zero; {\it by
means of this balance the constraint determines the scalar}
$\zeta(x)$.  Note also that due to this balance, {\it the degree
of localization of the fermion and values of $\zeta(x)$ may be
strongly interconnected}.

One can suggest the following two alternative approaches to the
question of how a primordial fermion can be realized as a regular
one:

{\it The first approach} discussed in Refs.\cite{GK4}, \cite{GK5},
is based on the idea of the "maximal economy". We start from one
primordial fermion field for each type of fermions: one neutral
primordial lepton field $N$, one charged primordial lepton field
$E$ and similar for quarks. In other words we start from one
 generation of fermions.
Splitting of the primordial fermions into families occurs only in
normal particle physics conditions, i.e. when the fermion energy
density is huge in comparison with the vacuum energy density. One
of the possibilities for this to be realized is the above
mentioned condition $Z_{i}(\zeta)\approx 0$. The appropriate two
constant solutions for $\zeta$, i.e. $\zeta=\zeta_{1,2}^{i}$,
correspond to two different states of the primordial fermions with
{\it different constant masses} determined Eqs.(\ref{muferm1})
where we have to substitute $\zeta_{1,2}^{i}$ instead of $\zeta$.
So, in the normal particle physics conditions, the scalar $\zeta$
plays the role of an additional degree of freedom determining
different mass eigenstates  of the primordial fermions which we
want to identify with different fermion generations. Note that the
classical tests of GR deal in fact with matter built of the
fermions of the first generation (may be with a small touch of the
second generation). This is why one can identify the states of the
primordial fermions realized as $\zeta =\zeta_{1,2}^{i}$ with the
first two generations of the regular fermions. For example, if the
free primordial electron $E$ is in the state with $\zeta
=\zeta_{1}^{(E)}$ (or $\zeta =\zeta_{2}^{(E)}$), it is detected as
the regular electron $e$ (or muon $\mu$) and similar the
primordial neutrino $N$ splits into the regular electron and muon
neutrinos with masses respectively:
\begin{equation}
 m_{e(\mu)}=
\frac{\mu_{E}(\zeta_{1(2)}^{(E)} +h_{E})}{(\zeta_{1(2)}^{(E)}
+k)(\zeta_{1(2)}^{(E)} +b_g)^{1/2}};\quad m_{\nu_{e}(\nu_{\mu})}=
\frac{\mu_{N}(\zeta_{1(2)}^{(N)} +h_{N})}{(\zeta_{1(2)}^{(N)}
+k)(\zeta_{1(2)}^{(N)} +b_g)^{1/2}}\label{m-12}
\end{equation}
It turns out that  there is only one more additional possibility
to satisfy  the constraint (\ref{constraint2}) when primordial
fermion is in the normal particle physics conditions. This is the
solution $\zeta^{i}=\zeta_{3}^{i}\approx -b_g$ which one can
associate with the third generation of fermions (for details see
Refs.\cite{GK4}, \cite{GK5}). It is interesting that in contrast
to the first two generations, the third generation defined by this
way, may have gravitational interaction with unusual features
since the condition (\ref{noncan-ll-can}) may not hold. The
described splitting of the primordial fermions into three
generations in the normal particle physics conditions is the
family replication mechanism proposed in Refs.\cite{GK4},
\cite{GK5}.

{\it The second approach} is based on the idea that the three
families of fermions of the standard model exist from the
beginning in the original action, i.e. not to use the family
replication mechanism for explanation of the observed three
generations of fermions. In this case again, exactly as it was in
the first approach, the primordial fermions turn into the regular
fermions only in the normal particle physics conditions. Now
however if we interpret the state of the primordial fermions with,
for example, $\zeta^{(i)} =\zeta_{1}^{(i)}$ as the observable
regular fermions, then some role should be assigned to the states
with $\zeta^{(i)} =\zeta_{2}^{(i)}$ and $\zeta^{(i)}
=\zeta_{3}^{(i)}$. By means of a choice of the parameters one can
try for example to provide very large masses of the regular
fermions with $\zeta^{(i)} =\zeta_{2}^{(i)}$ and $\zeta^{(i)}
=\zeta_{3}^{(i)}$ that might explain the reason why they are
unobservable so far. However these questions are beyond of the
goals of this paper and  together with many other aspects of
fermions in TMT will be studied in a separate publication.

\subsection{Resolution of the 5-th Force Problem for Regular Fermions}

Reproducing Einstein equations when the primordial fermions are in
the states of the regular fermions is not enough in order to
assert that GR is reproduced. The reason is that at the late
universe, as $\phi\gg M_{p}$, the scalar field $\phi$ effective
potential is very flat and therefore due to the Yukawa-type
coupling of massive fermions to $\phi$, (the r.h.s. of
Eq.(\ref{phief+ferm1})),  the long range scalar force appears to
be possible in general. The Yukawa coupling "constant" is
$\alpha\frac{m_{i}(\zeta)}{M_{p}}Z_{i}(\zeta)$. Applying our
analysis of the meaning of the constraint
 in two previous subsections, it is easy to see that  for
regular fermions with $\zeta^{(i)} =\zeta_{1,2}^{(i)}$ the factor
$Z_{i}(\zeta)$ is of the order of the ratio of the vacuum energy
density to the regular fermion energy density. Thus we conclude
that the 5-th force is extremely suppressed for the fermionic
matter observable in classical tests of GR. It is very important
that this result is obtained automatically, without tuning of the
parameters and it takes place for both approaches to realization
of the regular fermions in TMT discussed in previous subsection.

\section{Asymmetry between early and late time dynamics of the
universe as result of asymmetry in the couplings to measures
$\Phi$ and $\sqrt{-g}$ in the action. }

The results of the previous three sections depend very much on the
choice of the parameters $V_{1}$, $V_{2}$ and $\delta$ in the
action (\ref{totaction}).  Let us recall  that the curvature term
in the action (\ref{totaction}) couples to the measure $\Phi
+b_{g}\sqrt{-g}$ while the $\phi$ kinetic term couples to the
measure $\Phi +b_{\phi}\sqrt{-g}$. This is the reason of
$\delta\neq 0$. If we were choose the fine tuned condition $\delta
=0$ then both the curvature term and the $\phi$ kinetic term would
be coupled to the same measure $\Phi +b_{g}\sqrt{-g}$. One can
also pay attention that depending on the choice of one of the
alternative  conditions $b_{g}V_{1}>2V_{2}$ or $b_{g}V_{1}>2V_{2}$
we realize different shapes of the effective potential if
$b_{g}V_{1}>V_{2}$ (see Fig.1). And again, if instead we were
choose the fine tuned condition $b_{g}V_{1}=V_{2}$ then the action
would contain only one prepotential coupled to the measure $\Phi
+b_{g}\sqrt{-g}$.

So, in order to avoid fine tunings we have  introduced asymmetries
in the couplings of the different terms in the Lagrangian
densities $L_{1}$ and $L_{2}$ to measures $\Phi$ and $\sqrt{-g}$.
In order to display the role of these asymmetries it is useful to
consider what happens if such asymmetries are absent in the action
at all. In other words we want to study the gravity+dilaton model
where both $\delta =0$ and $b_{g}V_{1}=V_{2}$. In such a case the
action contains only one Lagrangian density coupled to the measure
$\Phi +b_{g}\sqrt{-g}$:
\begin{equation}
S=\int (\Phi +b_{g}\sqrt{-g}) d^{4}x e^{\alpha\phi /M_{p}}
\left(-\frac{1}{\kappa}R +
\frac{1}{2}g^{\mu\nu}\phi_{,\mu}\phi_{,\nu}-Ve^{\alpha\phi /M_{p}}
\right), \label{S-symm}
\end{equation}
where $V=V_{1}=V_{2}/b_{g}$. An equivalent statement is that
$L_{1}=b_{g}L_{2}$; it is an example of the very special class of
the TMT models where $L_{1}$ is proportional to $L_{2}$.

To see the cosmological dynamics in this model one can use the
results of Sec.IIIB. If we assume in addition $b_{g}>0$ and
$V_{1}>0$, then after the shift $\phi\rightarrow \phi +\Delta\phi$
where $\Delta\phi =-{M_{p}}{2\alpha}\ln(V/M^{4})$ (which is not a
shift symmetry in this case), the effective potential
(\ref{Veffvac-delta=0}) takes the form
\begin{equation}
V_{eff}^{(symm)}(\phi)=\frac{V^{2}}{b_{g}M^{4}}\cosh^{2}(\alpha\phi/M_{p}).
\label{Veff-symm}
\end{equation}
In contrast to general cases ($b_{g}V_{1}\neq V_{2}$)  this
potential has no flat regions and it is symmetric around a certain
point in the $\phi$-axis. This form of the potential (with an
additional  constant) has been used in a model of the early
inflation\cite{Barrow}.

\section{Some remarks on the measure fields independence of
$L_{1}$ and $L_{2}$}

Although we have assumed in the main text that $L_{1}$ and $L_{2}$
are $\varphi_{a}$ independent, a contribution equivalent to the
term $\int f(\Phi/\sqrt{-g})\Phi d^{4}x$ can be effectively
reproduced in the action (\ref{S})  if a nondynamical field
(Lagrange multipliers) is allowed in the action. For this purpose
let us consider the contribution to the action of the form
\begin{equation}
S_{auxiliary}=\int[\sigma\Phi+l(\sigma)\sqrt{-g}]d^{4}x
\label{L-m-action}
\end{equation}
where $\sigma$ is an auxiliary nondynamical field and $l(\sigma)$
is an analytic function. Varying $\sigma$ we obtain
$dl/d\sigma\equiv l^{\prime}(\sigma)=-\Phi/\sqrt{-g}$ that can be
solved for $\sigma$: $\sigma =l^{\prime -1}(-\Phi/\sqrt{-g})$
where $l^{\prime -1}$ is the inverse function of $l^{\prime}$.
Inserting this solution for $\sigma$ back into the action
(\ref{L-m-action}) we obtain
\begin{equation}
S_{aux.integrated}=\int f(\Phi/\sqrt{-g})\Phi d^{4}x
\label{L-m-action-int}
\end{equation}
where the auxiliary field has disappeared and
\begin{equation}
f(\Phi/\sqrt{-g})\equiv l^{\prime -1}(-\Phi/\sqrt{-g})+l(l^{\prime
-1}(-\Phi/\sqrt{-g}))\frac{\sqrt{-g}}{\Phi}.\label{f-aux}
\end{equation}

To see the difference between effect of this type of auxiliary
fields as compared with a model where the $\sigma$ field is
equipped with a kinetic term, let us consider two toy models
including gravity and $\sigma$ field: one - without kinetic term
\begin{equation}
S_{toy}=\int\left[\left(-\frac{1}{\kappa}R+\sigma\right)\Phi
+b\sigma^{2}\sqrt{-g}\right]d^{4}x \label{toy-1}
\end{equation}
and the other - with a kinetic term
\begin{equation}
S_{toy,k}=\int\left[\left(-\frac{1}{\kappa}R+\sigma +
\frac{1}{2}g^{\alpha\beta}\frac{\partial_{\alpha}\sigma
\partial_{\beta}\sigma}{\sigma^{2}}\right)\Phi
+b\sigma^{2}\sqrt{-g}\right]d^{4}x \label{toy-2}
\end{equation}
where $b$ is a real constant. For both of them it is assumed the
use of the first order formalism. The first model is invariant
under local transformations $\Phi\rightarrow J\Phi$, \,
$g_{\mu\nu}\rightarrow Jg_{\mu\nu}$, \, $\sigma\rightarrow
J^{-1}\sigma$ where $J$ is an arbitrary space-time function while
in the second model the same symmetry transformations hold only if
 $J$ is  constant.

Variation of the measure fields $\varphi_{a}$ in the model
(\ref{toy-2}) leads (if $\Phi\neq 0$) to
\begin{equation}
-\frac{1}{\kappa}R+\sigma +
\frac{1}{2}g^{\alpha\beta}\frac{\partial_{\alpha}\sigma
\partial_{\beta}\sigma}{\sigma^{2}}=M^{4},\label{toy-1-phi}
\end{equation}
where $M^{4}$ is the integration constant. On the other hand
varying the action (\ref{toy-2}) with respect to $g^{\mu\nu}$
gives
\begin{equation}
\chi\left(-\frac{1}{\kappa}R_{\mu\nu} +
\frac{1}{2}\frac{\partial_{\mu}\sigma
\partial_{\nu}\sigma}{\sigma^{2}}\right)-
\frac{1}{2}b\sigma^{2}g_{\mu\nu}=0,\label{toy-1-gmunu}
\end{equation}
where $\chi\equiv\frac{\Phi}{\sqrt{-g}}$. The corresponding
equation in the model (\ref{toy-1}) is obtained from
(\ref{toy-1-gmunu}) by omitting the term with gradients of
$\sigma$. It follows from Eqs.(\ref{toy-1-phi}) and
(\ref{toy-1-gmunu}) that
\begin{equation}
\frac{1}{\chi}=\frac{M^{4}-\sigma}{2b\sigma^{2}}
\label{toy-constr}
\end{equation}
This result holds  in both model.

 In the model (\ref{toy-1}), variation of $\sigma$ results in
 $\frac{1}{\chi}=-\frac{1}{2b\sigma}$ which is consistent with
 Eq.(\ref{toy-constr}) only if the integration constant $M=0$.
 This means that through the classical mechanism displayed in TMT,
 {\it it is impossible to achieve
 spontaneous breakdown of the local scale invariance}  the first model
 possesses.

 Transition to the Einstein frame where the space-time becomes
 Riemannian is implemented by means of the conformal
 transformation $\tilde{g}_{\mu\nu}=\chi g_{\mu\nu}$. For the
 model (\ref{toy-1}) the gravitational equations in the Einstein frame
 read
\begin{equation}
\frac{1}{\kappa}G_{\mu\nu}(\tilde{g}_{\alpha\beta})=\frac{1}{8b}\tilde{g}_{\mu\nu}.
\label{toy-1-grav}
\end{equation}
This means that the model (\ref{toy-1}) with auxiliary
(nondynamical) field $\sigma$ intrinsically contains a constant
vacuum energy.

In the model (\ref{toy-2}), where $\sigma$ appears as a dynamical
field,  the gravitational equations in the Einstein frame results
from Eq.(\ref{toy-1-gmunu})
\begin{equation}
\frac{1}{\kappa}G_{\mu\nu}(\tilde{g}_{\alpha\beta})=
\frac{(M^{4}-\sigma)^{2}}{8b\sigma^{2}}\tilde{g}_{\mu\nu}
+\frac{1}{2}\left(\frac{\partial_{\mu}\sigma
\partial_{\nu}\sigma}{\sigma^{2}}-
\frac{1}{2}\tilde{g}_{\mu\nu}\tilde{g}^{\alpha\beta}\frac{\partial_{\alpha}\sigma
\partial_{\beta}\sigma}{\sigma^{2}}\right). \label{toy-2-grav}
\end{equation}
It is convenient to rewrite this equation in terms of the scalar
field $\ln\sigma\equiv\phi$:
\begin{equation}
\frac{1}{\kappa}G_{\mu\nu}(\tilde{g}_{\alpha\beta})=V_{eff}(\phi)
\tilde{g}_{\mu\nu} +\frac{1}{2}\left(\partial_{\mu}\phi
\partial_{\nu}\phi-
\frac{1}{2}\tilde{g}_{\mu\nu}\tilde{g}^{\alpha\beta}\partial_{\alpha}\phi
\partial_{\beta}\phi\right), \qquad where \qquad
V_{eff}(\phi)=\frac{1}{8b\sigma^{2}}(M^{4}e^{-\phi}-1)^{2}.
\label{toy-2-grav-phi}
\end{equation}
 The
$\phi$-equation reads $\Box\phi +V_{eff}^{\prime}(\phi)=0$.
Similar to the general discussion in the main text we see that if
the $\sigma$ field is dynamical then TMT provides the vacuum with
zero energy without fine tuning.

Hence the main difference between the TMT models with auxiliary
and dynamical scalar field consists in radically different results
concerning the cosmological constant problem.

However it is very unlikely that a nondynamical scalar field will
not acquire a kinetic term after quantum
corrections\cite{Gross-Neveau}. Then it  becomes dynamical which
restores the above results for the model (\ref{toy-2}). This is
why we have ignored the rather formal possibility of introducing
the nondynamical scalars into the fundamental action of the models
studied in this paper.

\end{document}